\documentclass[12pt]{article}
\usepackage{amscd, amsmath}
\usepackage{amsthm, amssymb}
\usepackage{mathtools}
\usepackage{slashed}
\usepackage{scalerel}
\usepackage{graphicx}
\usepackage[bottom,flushmargin]{footmisc}
\usepackage{mathrsfs}
\usepackage[format=plain, labelfont=it]{caption}
\usepackage{subcaption}
\usepackage{float}
\usepackage{array}
\usepackage{accents}
\usepackage{empheq}
\usepackage[dvipsnames]{xcolor}
\usepackage[most]{tcolorbox}
\usepackage[british]{babel}
\usepackage{tikz-cd} 
\usepackage{hyperref}

\renewcommand{\baselinestretch}{1.2}
\newcommand{\linebr}{\\[0.2em]}

\DeclareMathAlphabet{\mathpzc}{OT1}{pzc}{m}{it}

\definecolor{myblue1}{RGB}{0, 0, 139}

\hypersetup{
    linktocpage,
    colorlinks,
    citecolor=myblue1,
    filecolor=black,
    linkcolor=myblue1,
    urlcolor=black
}

\newtheorem*{assumption(A)}{Condition (A)}

\theoremstyle{remark}

\theoremstyle{remark}

\numberwithin{equation}{section}

\newcommand{\MM}{\mathcal M}
\newcommand{\MT}{P}
\newcommand{\ST}{S}

\newcommand{\CC}{\mathbb{C}}
\newcommand{\Lie}{\mathcal L}

\newcommand{\VV}{\mathscr V}
\newcommand{\HH}{\mathcal H}

\newcommand{\FF}{\mathcal F}
\newcommand{\tg}{\tilde{g}}
\newcommand{\tbeta}{\hat{g}}

\newcommand{\bg}{\bar{g}}
\newcommand{\bk}{\gamma}
\newcommand{\sD}{\slashed{D}}
\newcommand{\sL}{\slashed{L}}

\newcommand{\slLie}{\slashed{\mathcal{L}}}

\newcommand{\sM}{\slashed{M}}

\newcommand{\trho}{\tilde{\rho}}

\newcommand{\SU}{\mathrm{SU}(3)}
\newcommand{\su}{\mathfrak{su}(3)}
\newcommand{\Utwo}{\mathrm{U}(2)}
\newcommand{\utwo}{\mathfrak{u}(2)}
\newcommand{\sutwo}{\mathfrak{su}(2)}
\newcommand{\kk}{\mathfrak{k}}
\newcommand{\SO}{\mathrm{SO}}
\newcommand{\Spin}{\mathrm{Spin}}
\newcommand{\GL}{\mathrm{GL}}
\newcommand{\sSpin}{{\scaleto{\Spin}{5.9pt}}}
\newcommand{\sSO}{{\scaleto{\SO}{4.9pt}}}
\newcommand{\tGLp}{\widetilde{\GL}{\vphantom{\GL}}^{+}}

\newcommand{\dd}{{\mathrm d}}
\newcommand{\vol}{\mathrm{vol}}
\newcommand{\Vol}{\mathrm{Vol}}
\newcommand{\diag}{\mathrm{diag}}
\newcommand*\DAlembert{\mathop{}\!\mathbin\Box}
\newcommand{\Diff}{{\rm Diff}}
\newcommand{\Sym}{{\rm Sym}}

\DeclareMathOperator{\Tr}{Tr}
\DeclareMathOperator{\Ad}{Ad}
\DeclareMathOperator{\ad}{ad}

\DeclareMathOperator{\grad}{grad}
\DeclareMathOperator{\divergence}{div}

\newcommand{\beq}{\begin{equation}}
\newcommand{\eeq}{\end{equation}}
\newcommand{\bal}{\begin{align}}
\newcommand{\eal}{\end{align}}

\newcommand{\bmatr}{\begin{bmatrix}}
\newcommand{\ematr}{\end{bmatrix}}

 \newtcolorbox{empheqboxed}{
 opacityback=0,
 enhanced jigsaw,
 width=\textwidth,
 boxrule=.5pt,
 sharpish corners,
 left=0pt,
 right=2pt,
 top=-9pt, 
 bottom=3pt
}
\newcommand*\widefbox[1]{{\setlength\fboxsep{6pt}\fbox{\hspace{0.2em}#1\hspace{0.2em}}}}

\newcommand{\LL}{{\scaleto{L}{4.8pt}}}
\newcommand{\RR}{{\scaleto{R}{4.8pt}}}
\newcommand{\LLL}{{\scaleto{L}{3.8pt}}}
\newcommand{\RRR}{{\scaleto{R}{3.8pt}}}
\newcommand{\PPP}{{\scaleto{P}{3.1pt}}}
\newcommand{\MMM}{{\scaleto{M}{3.1pt}}}
\newcommand{\KKK}{{\scaleto{K}{3.1pt}}}
\newcommand{\Szero}{{\scaleto{0}{4.3pt}}}
\newcommand{\SSzero}{{\scaleto{0}{3.3pt}}}
\newcommand{\SSSzero}{{\scaleto{0}{2.7pt}}}

\makeatletter

\renewenvironment{bmatrix}
  {\left[\mkern3.5mu\env@matrix}
  {\endmatrix\mkern3.5mu\right]}

\def\blfootnote{\xdef\@thefnmark{}\@footnotetext}
\makeatother

\begin{document}

\begin{titlepage}

\title{\vspace{-.5cm} \bf Internal symmetries \\ in Kaluza-Klein models}


\author{{Jo\~ao Baptista}}  
\date{June 2023}

\maketitle

\thispagestyle{empty}
{\centerline{{\large \bf{Abstract}}}}
\noindent
The usual approach to Kaluza-Klein considers a spacetime of the form $M_4 \times K$ and identifies the isometry group of the internal vacuum metric, $g_K^\Szero$, with the gauge group in four dimensions. In these notes we discuss a variant approach where part of the gauge group does not come from full isometries of $g_K^\Szero$, but instead comes from weaker internal symmetries that only preserve the Einstein-Hilbert action on $K$. Then the weaker symmetries are spontaneously broken by the choice of vacuum metric and generate massive gauge bosons within the Kaluza-Klein framework, with no need to introduce ad hoc Higgs fields. Using the language of Riemannian submersions, the classical mass of a gauge boson is calculated in terms of the Lie derivatives of $g_K^\Szero$. These massive bosons can be arbitrarily light and seem able to evade the standard no-go arguments against chiral fermionic interactions in Kaluza-Klein. As a second main theme, we also question the traditional assumption of a Kaluza-Klein vacuum represented by a product Einstein metric. This should not be true when that metric is unstable. In fact, we argue that the unravelling of the Einstein metric along certain instabilities is a desirable feature of the model, since it generates inflation and allows some metric components to change length scale. In the case of the Lie group $K = \SU$, the unravelling of the bi-invariant metric along an unstable perturbation also breaks the isometry group from $(\SU \times \SU) / \mathbb{Z}_3$ down to $( \SU \mkern-1mu \times \mkern-1mu \mathrm{SU}(2) \mkern-1mu  \times \mkern-1mu \mathrm{U}(1) ) /  \mathbb{Z}_6$, the gauge group of the Standard Model. We briefly discuss possible ways to stabilize the internal metric after that first symmetry breaking and produce an electroweak symmetry breaking at a different mass scale.

\vspace{-1.0cm}
\let\thefootnote\relax\footnote{
\noindent
{\small {\sl \bf  Keywords:} Kaluza-Klein theories; Riemannian submersions; spontaneous symmetry breaking; \\ unstable Einstein metrics; Standard Model; inflation.}
}

\end{titlepage}

\renewcommand{\baselinestretch}{1.10}\normalsize
\tableofcontents
\renewcommand{\baselinestretch}{1.2}\normalsize

\newpage 

\section{Introduction and overview of results}
  
This paper describes a collection of geometrical observations about internal symmetries and mass generation in Kaluza-Klein models. It also explores the natural role that metric instabilities can play in symmetry breaking and scale change in those models.
  
 The starting point for the discussion is the observation that while the Einstein-Hilbert action on $M_4 \times K$ has a very large group of symmetries---including the whole group of diffeomorphisms of the internal space---the traditional Kaluza-Klein ansatz associates gauge fields only to the much smaller internal isometry group, i.e. to the symmetries that preserve a specific choice of vacuum metric on $K$. The main reason for this restricted focus is the sense that any other gauge fields should have bosons with masses in the Planck scale, so too heavy to be observed experimentally. This is not necessarily true, however. A small perturbation of an initial vacuum metric can reduce its isometry group, for example, and the gauge bosons associated to former isometries will acquire non-zero but arbitrarily small masses. To ignore these gauge fields because they do not preserve the new vacuum metric then seems unnatural. It also precludes the investigation of relevant phenomena within the Kaluza-Klein framework, such as spontaneous symmetry breaking or a possible dependence of the vacuum energy density on the internal geometry. 
Thus, the first purpose of these notes is to discuss the consequences of having a theory that associates gauge fields also to non-isometric diffeomorphisms of internal space, co-existing with gauge fields that come from full isometries. Since distinct geometric origins in higher-dimensions lead to gauge fields with different properties in four dimensions, a primary motivation for the study is the possibility that these variations could be useful to model the peculiarities of the weak force field in the Standard Model, when compared to the strong and electromagnetic fields.

In fact, some of the main difficulties associated with the traditional Kaluza-Klein framework come from shortfalls in modeling the peculiarities of the weak force. 
For instance, the standard viewpoint is that all observed gauge bosons correspond to Killing vector fields on the internal space and are classically massless. The weak bosons gain their experimental, extremely small masses (when compared to the Planck mass) through a quantum mechanism that is not fully understood within the framework. Fermionic masses, on the other hand, are determined by the eigenvalues of the Dirac operator on $K$, or of a natural deformation thereof generally called the internal mass operator. The need to consider deformations is imposed by the geometric fact that, due to the Schr{\"o}dinger-Lichnerowicz formula, the standard Dirac operator does not have zero modes on a compact internal space with positive scalar curvature, but a deformation may well have them. In any case, one expects the internal mass operator to be natural and commute with all isometries of internal space. However, as pointed out by Witten in \cite{Witten83}, a result of Atiyah and Hirzebruch implies that any gauge group that acts through isometries on the internal space cannot have complex (chiral) representations on the zero modes of the internal Dirac operator or, more generally, on the zero modes of such deformations. So the Kaluza-Klein picture seems to be inconsistent with the chiral nature of fermions when responding to the weak force.

Our main suggestion to address this difficulty follows from the calculation that, in a more complete Kaluza-Klein model, the gauge boson associated to a vector field $e_a$ on $K$ has a mass proportional to the norm of the Lie derivative of the internal vacuum metric:
\[
\left(\text{Mass} \ A_\mu^a \right)^2 \ \ \propto \ \ \frac{ \int_K  \; \left\langle \Lie_{e_a}\, g_K^\Szero,  \; \Lie_{e_a}\, g_K^\Szero \right\rangle  \, \vol_{g^\SSzero_\KKK} }{ 2 \int_K  \,  g_K^\Szero (e_a ,  e_a  ) \ \vol_{g_\KKK^\SSzero} } \ .
\]
So mass generation for gauge bosons should operate through a mechanism that perturbs the vacuum metric $g_K^\Szero$ in a way that $e_a$ is no longer an exact Killing field. If the perturbation is small enough, the new bosons will be light. But if a boson is no longer acting through isometries on the internal space, then it can also evade the Atiyah-Hirzebruch theorem, and at least part of the gauge group may well have chiral fermionic representations. In other words, any perturbation of the classical vacuum metric on $K$ that allows some gauge bosons to gain a small mass should also allow that same part of the gauge group to evade the Atiyah-Hirzebruch theorem. This could help to understand, within the Kaluza-Klein framework, the fact that the massive gauge fields in nature are precisely the ones that have chiral fermionic representations. Note that this observation does not exhibit an explicit  mechanism leading to the appearance of chiral representations. It merely points out a possible new way to circumvent the no-go arguments described in \cite{Witten83} that rule out such representations in Kaluza-Klein models. There are other ways to circumvent those arguments, some of them well-verified but perhaps  farther from the original Kaluza-Klein philosophy, such as adding to the theory gauge fields living on the higher-dimensional spacetime \cite{CS}.

Another indication that the weak force field may be better modeled by an association with non-Killing vector fields on the internal space, as suggested here, is the experimental fact that the weak field mixes fermions with different masses, while the strong and electromagnetic fields only mix fermions with the same mass. As previously mentioned, in the Kaluza-Klein picture fermionic masses in four dimensions are determined by the eigenvalues of a natural, Dirac-like mass operator acting on spinors over $K$. One expects this operator to commute with all isometries of internal space. So a gauge field associated with an isometry of $K$ should preserve the eigenspaces and eigenvalues of the mass operator, i.e. should not mix fermions with different masses. Since the weak field does not have this property, it is plausible that it is not associated with exact internal isometries.

 A second significant difficulty associated with the traditional Kaluza-Klein framework is finding classical solutions of the higher-dimensional equations of motion that can be regarded as good candidates for the vacuum configuration \cite{Bailin, Duff, CJBook, CFD}. If one takes the simplest option available and chooses the standard Einstein-Hilbert Lagrangian for the higher-dimensional action, possibly with a non-zero cosmological constant, then the classical solutions are the Einstein metrics on $M_4 \times K$. If one furthermore assumes that the classical vacuum should be represented by a simple product metric $g_M^e + g_K^e$, then the higher-dimensional Einstein equations impose that $g_M^e$ and $g_K^e$ are both Einstein metrics on the respective spaces with the {\it same} constant. In particular, if one takes $M_4$ to be flat Minkowski space or a nearly flat constant curvature space (corresponding to the observed very small value of the four-dimensional cosmological constant), then the internal metric $g_K^e$ would also need to be flat or nearly flat. But Ricci flat metrics on a compact space cannot have the continuous, non-abelian isometries that are needed in Kaluza-Klein to model the strong force gauge fields. This problem disappears if we take $g_K^e$ to have positive curvature, as there are plenty of examples of positive Einstein metrics with non-abelian isometry groups. Unfortunately, in all these examples, the inverse scaling relation between scalar curvature and Riemannian volume raises a new problem: one would need an Einstein internal space of unreasonably big volume in order to obtain the tiny scalar curvature of $g_K^e$ necessary to match the observed curvature of the spacetime vacuum metric $g_M^e$. This unreasonably big internal space would then affect the predicted values of the gauge coupling constants of the theory. Equivalently, a small internal Einstein space will have good gauge couplings but large internal curvature, so the higher-dimensional equations of motion force a way too large four-dimensional curvature and cosmological constant. Hence the traditional difficulty to reconcile the Kaluza-Klein picture with the smallness of the spacetime cosmological constant \cite[ch. V.1]{CFD}.

Our suggestion regarding this difficulty is based on the observation that Einstein metrics are often unstable, and when this happens they may not be the best candidate to describe the vacuum configuration. We will give two examples of such instabilities. Firstly, it is well-known that a product Einstein metric with positive curvature is always unstable under a rescaling of the relative size of $M$ and $K$, even when the metrics $g_M^e$ and $g_K^e$ are both stable on their respective spaces. If we take a function $\phi$ in $C^\infty (M)$ and contract the internal metric $g_K^e$ by a factor $e^{-\phi}$ while expanding the metric on $M$ by an appropriate power of $e^\phi$, a quadratic analysis of the Einstein-Hilbert action says that this is an unstable perturbation of the product solution. One can go beyond the quadratic analysis and flesh out the field $\phi$ in the full action and its equation of motion. This reveals a potential $V(\phi)$ and confirms that the Einstein metric $g_M^e + g_K^e$ should unravel along this rescaling direction, in a process akin to cosmological inflation controlled by the scalar field $\phi$. This process is studied in section \ref{UnstableModesInflation}. During this unravelling, the 4D spacetime curvature will become smaller. Thus, assuming that the deformation represented by $\phi$ stabilizes at a given value $\phi_\Szero$, presumably through a complex process akin to reheating, the final vacuum configuration can have a spacetime curvature much smaller than the internal curvature. In particular, after the unravelling of the product Einstein metric along the $\phi$-instability, the scale of the gauge couplings no longer has to coincide with the scale of the spacetime cosmological constant.

Besides the rescaling represented by $\phi$, there can be additional interesting instabilities of a product Einstein metric, such as the ones coming from TT-deformations of the internal metric $g_K^e$. Let us consider the example of the internal space $K = \SU$. It is well-known that the bi-invariant metric on this group is Einstein, has positive curvature but is unstable, meaning that it is a saddle point of the Einstein-Hilbert action under TT-deformations of the metric, not a local maximum. Most TT-deformations of the bi-invariant metric will reduce its scalar curvature, but  a particular deformation found in \cite{Jensen} takes the scalar curvature up to positive infinity. Since $ - R_{g_\KKK}$ plays the role of a potential for TT-deformations, we see that that particular mode is unstable and its equation of motion is governed by a potential unbounded from below. Now, instead of discarding this example as unphysical, let us entertain the possibility of a physical unravelling of the initial bi-invariant metric along the unstable direction, if that metric ever represented the internal configuration over an ancient region of spacetime.

Firstly, we note that when the bi-invariant metric on $\SU$ is deformed along that unstable direction its isometry group is broken from $(\SU \times \SU) / \mathbb{Z}_3$ down to the suggestive $( \SU \mkern-1mu \times \mkern-1mu \mathrm{SU}(2) \mkern-1mu  \times \mkern-1mu \mathrm{U}(1) ) /  \mathbb{Z}_6 $, the gauge group of the Standard Model. This is described in section \ref{DeformedMetricsSU(3)}. Identifying left-invariant metrics on $\SU$ with inner-products on the Lie algebra $\su$, one can use the vector space decomposition
\[
\su \ = \  \mathfrak{u}(1) \oplus \sutwo \oplus \CC^2  \ \ , \qquad   v \ =  \ v_Y\ +\ v_W \ + \ v''  \ \ ,
\]
to write the deformed internal metric as  
\beq \label{DefinitionTBeta2}
g_K (u,v) \ = \ e^{- \frac{\beta \phi}{5}} \; \frac{15}{2\, \Lambda} \; \Big[\,  e^{2 \varphi}\, \Tr(u_Y^\dag\, v_Y) \ + \ e^{-2 \varphi}\, \Tr(u_W^\dag\, v_W) \ + \ e^{\varphi} \,\Tr\big[(u'')^\dag \, v'' \big] \, \Big] \ .
\eeq
Here $\beta$ and $\Lambda$ are positive constants, $\phi$ is the scalar field on $M$ that controls the rescaling deformation, as before, while $\varphi$ is the scalar field that controls the TT-deformation and breaks the internal isometry group. The value $\varphi = 0$ corresponds to internal bi-invariant metrics. As $\varphi$ grows, four initially massless gauge bosons acquire a classical mass that increases with the extent of the TT-deformation, as calculated in section \ref{InternalSymmetryBreaking}. But the higher-dimensional Einstein-Hilbert action implies that the dynamics of the deformation fields are governed by a classical potential, denoted $V(\phi, \varphi)$, that is unbounded from below for large values of the fields. So will $\phi$ and $\varphi$  just grow indefinitely, increasing the internal curvature $R_{g_\KKK}$ and the bosons' masses across all orders of magnitude, as the internal space collapses to zero size or becomes infinitely deformed? Or will new physics kick in at some point, physics not contained in the Einstein-Hilbert action, and stabilize the deformation? 

Given that physical reality contains phenomena operating at different scales, we argue that the appearance of runaway instabilities in the metric derived from the simple Einstein-Hilbert action should be viewed as a desirable feature of the model. As an opportunity to allow some metric components to transition to other scales. The unbounded deformations just describe how those components will change by many orders of magnitude before encountering new physics at a different scale, physics hidden in a second part of the Lagrangian that only becomes relevant after the rescalings. In other words, without the instabilities and runaway deformations at the Einstein-Hilbert level, everything in the metric would be stuck at Planck scale. There would be no open door  for its components to reach the small scale of the cosmological constant, for instance. To describe phenomena at those scales we would then need to introduce new fields, besides the higher-dimensional metric, and this is undesirable in the Kaluza-Klein framework.

Embracing the existence of instabilities at the Einstein-Hilbert level, in section \ref{StabilizingInternalCurvature} we discuss possible ways to stabilize the runway deformations of the internal geometry.
One of these attempts, for example, is based on the intuitive notion that there should be an energy cost for increasing the gauge bosons' masses all the way up to infinity. This notion may very well be false in the classical picture of the vacuum, where Yang-Mills fields can be exactly zero. In the quantum picture the odds seem more favourable, though, with fields that are always fluctuating and never vanish entirely. In fact, although the calculation of the zero-point energy of a quantum field does not seem to be a settled matter, according to most calculations the renormalized vacuum energy density does increase with the mass $m$ of the field, for instance as $m^4  \log\big(\frac{m^2}{\mu^2}\big)$ \cite{Martin}. Here $\mu$ is a new mass scale, not necessary close and presumably smaller than the Planck mass. Adding this vacuum energy density to the Einstein-Hilbert action we obtain an effective potential that, one calculates, is now bounded from below as the metric deformation increases. So the unstable TT-deformation could be stabilized once the gauge bosons associated to the broken part of the isometry group reach masses that can balance the (Planck scale) contribution of $- R_{g_\KKK}$ to the effective potential. Those heavy gauge bosons would be unobservable at current experimental energies. Near the new equilibrium point the contributions to the effective potential of the classical term $R_{g_\KKK}$ and of the vacuum energy density would be comparable to each other. Hence the effective Lagrangian is no longer just the Einstein-Hilbert one, and the Kaluza-Klein vacuum will not be a classical Einstein metric on $M_4 \times K$.

In principle this game can be played with other models for the internal space, besides $K \mkern-1mu =  \mkern-1mu \SU$. One basically needs an Einstein metric $g_K^e$ whose isometry group contains $G_{\rm SM}$ and, simultaneously, has unstable TT-deformations that can break the isometry group down to $G_{\rm SM}$ at Planck scale. The electroweak symmetry breaking at a lighter mass scale needs separate arguments.

In any case, the overall message is that unstable Einstein metrics may deserve a second look in the Kaluza-Klein framework. The existence of unstable directions of perturbation could be more than a nuisance in the search for the vacuum configuration, more than a criterion to exclude candidates among possible internal geometries. Firstly, the internal symmetry breaking caused by unstable perturbations can carve out the peculiar $( \SU \mkern-1mu \times \mkern-1mu \mathrm{SU}(2) \mkern-1mu  \times \mkern-1mu \mathrm{U}(1) ) /  \mathbb{Z}_6 $ as a subgroup of larger, seemingly more natural gauge groups. Secondly, the rescaling of the components of the unstable metric that fall along steep potentials may be a good classical model for physical processes that require specific metric components to change by many orders of magnitude, before encountering new physics at a different scale. Classically stable Einstein metrics are too nice to have individual components breaking out of the Planck scale. That is why they are hard to reconcile with a tiny four-dimensional cosmological constant.

In the second part of this Introduction we will give a more detailed overview of the main calculations in this paper. For general reviews of Kaluza-Klein theory see \cite{Bailin, Bleecker, Bou, CJBook, WessonOverduin}, for example, and the comprehensive texts \cite{CFD, Duff} stressing the supergravity viewpoint. Here we do not work with supergravity. We also do not delve into the beautiful topic of black holes in Kaluza-Klein \cite{GW, HW}. Some of the early original references for Kaluza-Klein theory are \cite{Original}, with much more complete lists given in the mentioned reviews. Although some of the observations in these notes were suggested by the calculations in \cite{Ba1, Ba2}, here we try to provide a broader picture of the models, generalizing parts of those calculations along the way.

\subsubsection*{Decomposing the higher-dimensional scalar curvature} 

Consider the higher-dimensional space $P =  M_4 \times K$ viewed as a fibre-bundle over four-dimensional spacetime. The fibre over a spacetime point, also called the internal space over that point, is isomorphic to $K$. Let $g_P$ be a submersive metric on $P$. As in the usual Kaluza-Klein framework, it determines three more familiar objects: 
\begin{itemize}
\item[{\bf i)}]  through projection, a unique metric $g_M$ on the four-dimensional spacetime; 
\item[{\bf ii)}] through restriction to the fibres, a family of metrics $g_K$ on the internal spaces;
\item[{\bf iii)}] gauge fields on spacetime, which can be encapsulated in a one-form $A$ on $M_4$ with values in the Lie algebra of vector fields on $K$.
\end{itemize}
The equations linking these objects to the higher-dimensional metric $g_P$ are
\bal \label{MetricDecomposition}
g_P (U, V) \ &= \ g_K (U, V) \nonumber \\
g_P (X, V) \ &= \  - \ g_K \left(A (X), V \right) \nonumber \\
g_P (X, Y) \ &= \ g_M (X, Y) \ + \  g_K \left(A(X) , A(Y) \right) \ ,
\end{align}
for all tangent vectors $X,Y \in TM$ and vertical vectors $U, V \in TK$. These relations are usually called the Kaluza-Klein ansatz for $g_P$. They allow one to reconstruct the higher-dimensional metric from the more familiar data $(g_M , A, g_K)$. The correspondence between submersive metrics on $P$ and the data i), ii), iii) is a bijection. It will be described in more detail in section \ref{DecomposingScalarCurvature}.

Choosing a set $\{ e_a \}$ of independent vector fields on $K$, the one-form on spacetime can be decomposed as a sum 
\beq \label{GaugeFieldExpansion}
A(X) \ = \ \sum\nolimits_a \,A^a(X) \, e_a \ ,
\eeq
where the real-valued coefficients $A^a(X)$ are the traditional gauge fields on $M_4$. For general submersive metrics on $P$ this can be an infinite sum, with $\{ e_a \}$ being a basis for the full space of vector fields on $K$, which coincides with the Lie algebra of the diffeomorphism group ${\rm Diff} (K)$. Most often, however, the approach in Kaluza-Klein is to restrict the attention to special families of higher-dimensional submersive metrics, such as the ones obtained when $K = G/H$ is a homogeneous space, the fibre metrics $g_K$ are $G$-invariant, and the vector fields $\{ e_a \}$ on $K$ come from a basis of a finite-dimensional subspace of ${\rm Lie}(G)$. For the moment we will not make any such restrictions, so will consider the case of general submersive metrics on $P$.

 In these notes we investigate the scalar curvature of the metric $g_P$. We want to express it, as explicitly as possible, in terms of the equivalent data $(g_M , A, g_K)$. A standard result in Riemannian submersions \cite[ch. 9]{Besse} says that $R_{g_\PPP}$ can be decomposed as 
\beq   
R_{g_\PPP} \ = \ R_{g_\MMM} \ + \ R_{g_\KKK} \ -\  |\FF|^2 \ - \ |S|^2 \ - \ |N|^2 \ - \ 2\, \check{\delta} N  \nonumber  \ .
\eeq
Here $R_{g_\MMM}$ and $R_{g_\KKK}$ denote the scalar curvatures of the metrics $g_M$ and $g_K$, respectively; $|\FF|^2$ is the component that originates the Yang-Mills terms $|F_A|^2$ in the usual Kaluza-Klein calculation; the tensor $S$ is the second fundamental form of the fibres $K$, also called shape operator; the vector $N$ is the metric trace of $S$, usually called the mean curvature vector of the fibres. 

In the lowest-dimensional Kaluza-Klein model, where the internal space $K$ is just the circle $S^1$, the scalar curvature $R_{g_\KKK}$ vanishes and the tensors $S$ and $N$ merge into the same object. This essentially coincides with the gradient of the Brans-Dicke scalar field, measuring the variation of the size of the internal circle as one moves along $M_4$. For higher-dimensional $K$ the structure of $R_{g_\KKK}$ and $S$ is much richer. In section \ref{SectionScalarCurvature} we investigate that structure for a general $K$, extending the results described in \cite{Ba1} in the particular case where $K = \SU$ equipped with a special family of left-invariant metrics.

For a general Riemannian submersion, one calculates that the third term of the higher-dimensional scalar curvature can be expressed as
\bal \label{NormF3}
|\FF|^2 \ = \ \frac{1}{4} \; g_M^{\mu \nu} \; g_M^{\sigma \rho}   \  g_K (e_a ,  e_b  ) \ (F^a_{A})_{\mu \sigma} \ (F^b_{A})_{\nu \rho} \ ,
\end{align}
where the $(F^a_{A})_{\mu \sigma}$ are the components of the curvature of the gauge fields \eqref{GaugeFieldExpansion} on $M_4$. The right-hand side broadly coincides with the form of the Yang-Mills terms in the Standard Model Lagrangian. The fact that such terms can be obtained from the higher-dimensional scalar curvature $R_{g_\PPP}$ is a remarkable and very well-known result, sometimes dubbed the Kaluza-Klein miracle \cite{Original}. 

The terms $|S|^2$ and $|N|^2$ are much less studied in the Kaluza-Klein literature, where the assumption of totally geodesic fibres ($S=0$) or constant internal geometry are common.
Section \ref{SecondFundamentalForm} describes how the term $|S|^2$ is akin to the covariant derivative of a Higgs field. For a general submersion metric, it can be expressed in terms of the gauge fields and the Lie derivatives of $g_P$ as
\begin{multline}
\label{NormT3}
|S|^2 \ = \   \frac{1}{4} \ g_M^{\mu \nu}  \left\langle \Lie_{X_\mu}\, g_P,  \ \Lie_{X_\nu}\, g_P \right\rangle \ + \  \frac{1}{2} \ g_M^{\mu \nu} \ A^a_\mu  \left\langle \Lie_{e_a}\, g_P,  \ \Lie_{X_\nu}\, g_P \right\rangle \linebr + \   \frac{1}{4} \ g_M^{\mu \nu} \ A^a_\mu\  A^b_\nu  \left\langle \Lie_{e_a}\, g_P,  \ \Lie_{e_b}\, g_P \right\rangle \ .
\end{multline}
So the fibres' second fundamental form produces the quadratic terms in the gauge fields $A^a_\mu$ that are necessary for mass generation through spontaneous symmetry breaking. These terms are encoded in the higher-dimensional scalar curvature $R_{g_\PPP}$, with no need to introduce ad hoc Higgs fields, as in the traditional Brout-Englert-Higgs mechanism \cite{EBH}. 

The parallel between the different components of the curvature $R_{g_\PPP}$ and those of an Einstein-Yang-Mills-Higgs Lagrangian is made more explicit in section \ref{GaugedSigmaModels}. There we observe that a metric $g_P$ on the product $M\times K$ determines, by restriction, natural maps from $M$ to the space of Riemannian metrics and to the space of volume forms on $K$:
\begin{align*}
g_K &: \, M\,  \longrightarrow \, \MM(K) \ ,  \hspace{-2.5cm} &x \,&\mapsto\, g_{P_x}  \ ;   \linebr 
\vol_{g_\KKK} &: \, M\,  \longrightarrow \, \Omega^k(K, \mathbb{R}) \ , \hspace{-2.5cm}  &x \,&\mapsto\, \vol_{g_{\PPP_x}} \ . 
 \end{align*}
 Here $g_{P_x}$ denotes the restriction of the metric $g_P$ to the fibre $P_x \simeq K$ over the point $x$ in $M$. Since the diffeomorphism group $\text{Diff} (K)$ acts naturally on the target spaces $\MM(K)$ and $\Omega^k(K, \mathbb{R})$, it is possible to define $\text{Diff} (K)$-gauge transformations and covariant derivatives of these maps. Then one shows that the Einstein-Hilbert action for a submersive metric $g_P$ can be rewritten in terms of these maps as
 \beq \label{GaugedSigmaModelAction2}
\int_P \, R_{g_\PPP} \, \vol_{g_\PPP} \ = \ \int_P \, \Big[\, R_{g_\MMM} \, + \, R_{g_\KKK} \, - \, \frac{1}{4}\, |F_A|^2 \, - \,  \frac{1}{4}\,  |\dd^A g_K|^2  \ + \  |D^A \, (\vol_{g_\KKK})|^2 \, \Big] \,\vol_{g_\PPP} \ .  \nonumber
\eeq
This is essentially an Einstein-Yang-Mills-Higgs Lagrangian. The terms $|S|^2$ and $|N|^2$ become the norms of the covariant derivatives of the scalar fields that describe how the internal metric $g_K$ and its volume form vary from fibre to fibre. The opposite $-R_{g_\KKK}$ plays the role of a classical Higgs potential for those scalar fields. 

If instead of general submersive metrics on $P$ we only consider the homogeneous ones, namely submersions with homogeneous restrictions to an internal space of the form $K =G/H$, then the result is a simpler $G$-gauge theory, instead of a $\text{Diff} (K)$-gauge theory. That is the standard setting in the Kaluza-Klein literature (e.g. \cite{Bailin, CFD, CJBook, Duff}). The Higgs-like maps $g_K$ that we discuss here will then have values in the finite-dimensional space of homogeneous metrics on $G/H$, instead of the infinite-dimensional $\MM(K)$. 


\subsubsection*{Mass of the gauge fields} 

In the simplest case of vector fields $e_a$ on $K$ having vanishing divergence with respect to the vacuum metric, the mass of the associated four-dimensional gauge fields can be calculated to be
\begin{equation} \label{MassGaugeBosons2}
\left(\text{Mass} \ A_\mu^a \right)^2 \ \propto \ \frac{ \int_K  \; \left\langle \Lie_{e_a}\, g_K,  \; \Lie_{e_a}\, g_K \right\rangle  \, \vol_{g_\KKK} }{ 2 \int_K  \,  g_K (e_a ,  e_a  ) \, \vol_{g_\KKK} } \ .
\end{equation}
Here the internal metric should be taken to be constant at its vacuum value $g_K = g^{\Szero}_K$. Thus, the classical mass of a gauge field $A^a_\mu$ is determined by the geometrical properties of the vector field $e_a$ with respect to the vacuum metric on the internal space. Whenever $e_a$ is Killing, the Lie derivative vanishes and the fields $A^a_\mu$ will be massless.

The precise relation between the gauge bosons' classical mass and the Lie derivatives depends on whether the calculation is performed with a higher-dimensional Lagrangian in the Einstein frame or in the Jordan frame. In the latter case, the constant of proportionality in the relation above is the unity (section \ref{SectionMassGaugeBosons}). In the Einstein frame the constant of proportionality depends on the total volume $\Vol_{g^{\SSSzero}_\KKK}$ of the vacuum internal metric (appendix \ref{AppendixBosonsMassEinsteinFrame}). In general, the bosons' masses scale inversely with the size of the internal space, but depend on much more than that size. 

If the internal vector field $e_a$ is nearly Killing, but not exactly, then the mass of the boson associated to $A^a_\mu$ will be non-zero and very small. Thus, a natural way to generate light bosons within the Kaluza-Klein framework would be to start with a classical internal metric with a larger isometry group, for instance with $\SU \mkern-1mu \times \mkern-1mu \mathrm{SU}(2) \mkern-1mu  \times \mkern-1mu \mathrm{U}(1)$ isometries, and then suppose that a mechanism operating at a different scale, for example a quantum mechanism, slightly perturbs the initial metric in a way that some Lie derivatives $\Lie_{e_a}\, g^\Szero_K$ become non-zero but remain small. In particular, in a physical system with very light bosons the true internal vacuum metric should not be the classical metric---a solution of the Einstein equations at Planck scale. Instead, it should be the perturbed metric $g^\Szero_K$ with nearly Killing vector fields---presumably a solution of equations derived from the Einstein-Hilbert action added to some effective potential. Such scenarios will be discussed in section \ref{StabilizingInternalCurvature}.

There are two immediate advantages of generating the gauge bosons' mass through a perturbation of the classical internal metric, as opposed to using ad hoc Higgs fields. Firstly, it fits naturally with the Kaluza-Klein framework, where everything should come from the higher-dimensional metric. Secondly, if the physical internal vacuum metric $g^\Szero_K$ has a reduced isometry group, for instance $\SU \mkern-1mu   \times \mkern-1mu \mathrm{U}(1)$ instead of $\SU \mkern-1mu \times \mkern-1mu \mathrm{SU}(2) \mkern-1mu  \times \mkern-1mu \mathrm{U}(1)$, then it can evade the main no-go arguments against chiral fermions in Kaluza-Klein, since these arguments only rule out chiral interactions with the gauge fields associated to internal isometries. This will be further discussed in section \ref{NoGoArguments}.

\subsubsection*{Gauge couplings in the Einstein frame} 

A naive dimensional reduction of the Einstein-Hilbert action on $M \times K$ produces a Lagrangian in four dimensions that is not in the Einstein frame. This means that the gravity component of the 4D Lagrangian does not appear in the traditional guise $R_{g_\MMM} \sqrt{-g_M}$, but instead appears multiplied by a scalar field that depends on the spacetime coordinate. This field is a hallmark of Kaluza-Klein theories. It is related to the volume of the internal space over each spacetime point, which can vary along $M$. 

Since the Einstein frame is generally preferred for a good physical interpretation of the 4D theory \cite{FGN}, in section \ref{EinsteinFrame} we use a Weyl rescaling of the metric $g_M$ to transform our Lagrangian to the Einstein frame. This rescaling is a standard technique in many models.  To apply it to our general Kaluza-Klein model, we first calculate how the geometric terms $|\FF|^2$, $|S|^2$ and $|N|^2$ transform under separate Weyl rescalings of $g_M$ and $g_K$. This is done in appendix \ref{AppendixWeylRescalings} and, subsequently, leads to the general Lagrangian \eqref{ExpandedLagrangianScalarField}.

Using those results, in section \ref{GaugeCouplingConstants} we argue that the traditional formula relating the scale of the gauge couplings to the size of the internal space is unnecessarily strict. Starting from the Einstein-Hilbert action on $M\times K$, the traditional Kaluza-Klein argument assumes a constant vacuum metric $g^\Szero_K$ and performs the dimensional reduction
\[
\frac{1}{2 \, \kappa_{P}} \int_P \, (R_{g_\PPP} \, - \,  2 \Lambda )\, \vol_{g_\PPP} \ = \ \frac{1}{2 \, \kappa_{M}}  \int_M \, \Big[\, R_{g_\MMM} \, + \, R_{g^\SSzero_\KKK} \, - \, \frac{1}{4}\, |F_A|^2 \, - \, 2 \Lambda \, \Big] \,\vol_{g_\MMM} \ 
\]
by fixing the value $ \kappa_{P} := \kappa_{M} \Vol (K, g^\Szero_K)$ and integrating over the fibre $K$. Here $\kappa_M$ denotes the four-dimensional Einstein constant. Since gauge fields have values on the vector fields on $K$, the norm $|F_A|^2$ depends on the internal metric $g^\Szero_K$. This leads to gauge coupling constants $\alpha_{\rm{Wein}}$ of the order
\[
\alpha^2_{\rm{Wein}} \ \sim \ \frac{8\pi^2 \kappa_M}{l^2_{g_\KKK^\SSzero}} \ ,
\]
where $l_{g_\KKK^\SSzero}$ denotes the ``average'' length of the circumferences on $(K, g^\Szero_K)$ generated by the vector field associated to the gauge field. This was derived in a very general setting by Weinberg in \cite{WeinbergCharges}, with the slightly ajusted conventions described in section \ref{GaugeCouplingConstants}. A similar relation is stated in most reviews. 

Our point here is that fixing the constant $\kappa_P$ to a vacuum-dependent value does not sound like a general approach. For example, if the theory had several local vacua, associated to internal spaces of different sizes, the derivation of the respective couplings $\alpha^2_{\rm{Wein}}$ would run into an ambiguous choice when imposing the normalization of $\kappa_P$.
A more general approach to obtain a 4D action in the Einstein frame is to keep $\kappa_P$ unconstrained and, instead, redefine the physical metric $g_M$ through the Weyl rescaling of section \ref{EinsteinFrame}. This approach produces a Lagrangian in the Einstein frame even when the internal metric $g_K$ is not constant throughout the spacetime $M$. Thus, using the Lagrangian of section \ref{EinsteinFrame}, in section \ref{GaugeCouplingConstants} we adapt the derivation of  \cite{WeinbergCharges} to obtain a slightly more general formula for the scale of the gauge couplings in four dimensions:
\[
\alpha^2 \  \sim \  \frac{8\pi^2 \,\kappa_P}{l_{g_\KKK^\SSzero}^2 \, \big( \Vol_{g_\KKK^\SSzero} \big)}  \ .
\]
This reduces to the traditional expression when $\kappa_P = \kappa_M \, \Vol_{g_\KKK^\SSzero}$, but here $\kappa_P$ remains unconstrained. In particular, notice that the $\alpha^2$ derived from the full Lagrangian in the Einstein frame scales differently from $\alpha^2_{\rm{Wein}}$ when $(K, g_K^\Szero)$ changes size. This happens because, in the traditional derivation, the volume factor $(\Vol_{g_\KKK^\SSzero})^{-1}$ is hidden in the normalization of the constant $\kappa_P$, which in fact becomes a non-constant for a rescaling $g_K^\Szero$.

\subsubsection*{Unstable Einstein metrics and cosmological inflation}

Ideally, a classical vacuum configuration on $M\times K$ would be characterized by vanishing gauge fields, a flat or perhaps constant curvature four-dimensional metric $g_M$, and internal metrics $g_K$ that are constant across the different fibres and equal to a fixed metric on $K$. In other words, the ideal vacuum configuration of $g_P$ would be a product metric $g_M + g_K$. To satisfy the higher-dimensional equations of motion, both metrics $g_M$ and $g_K$ need to be Einstein on the respective spaces with the {\it same} constant.

But what happens when those Einstein products are unstable? It is well-known that a product Einstein metric with positive curvature is always unstable under a rescaling of the relative size of $M$ and $K$, even when $g_M$ and $g_K$ are both stable on their respective spaces \cite{Besse, Kro}. The metric $g_K$ may also be unstable under TT-deformations of the internal geometry. The initial dynamics of those deformations  can be studied by considering submersive metrics of the form
\[ 
g_P \ = \   e^{ \beta_1 \phi} \; g_M \  +  \ e^{- \beta_2 \phi}  \; g_K(\varphi) \ .
\]
Here $\phi$ and $\varphi$ are four-dimensional scalar fields that parametrize the unstable rescaling and the unstable internal TT-deformations, respectively, over each spacetime point. The $\beta_i$ are positive constants to be specified later. The gauge fields are taken to vanish in this approximation, so we are looking at the conditions close to the initial Einstein product metric $g_M + g_K$. Applied to such metrics $g_P$, the higher-dimensional Einstein-Hilbert action (converted to the Einstein frame) reduces to a four-dimensional action of the illustrative form
\[
\int_{P} \; \left( R_{g_\PPP} \, - \, 2\, \Lambda  \right) \vol_{g_\PPP} \ = \ \int_M \, \Big[ \, \frac{1}{2\, \kappa_M} \, R_{g_\MMM} \, - \, \frac{1}{2} \, |\dd \phi|^2_{g_\MMM}  \, - \, \frac{5}{2} \, |\dd \varphi|^2_{g_\MMM} \, - \, V(\phi, \varphi) \, \Big]\, \vol_{g_\MMM} \ , 
\]
after integration over the fibre $K$. So the deformation scalars are similar to Klein-Gordon fields governed by a potential $V(\phi, \varphi)$. 

The overall form of the dimensionally-reduced action is similar to the action found in scalar models of cosmological inflation. Thus, optimistically, Kaluza-Klein suggests how the ad hoc scalar fields of inflation could have their origin in the scalar components of the internal metric that unravel under unstable deformations of a primordial, product Einstein metric on $M\times K$. This interpretation suggests a new way to generate microscopically motivated, multi-field models of the initial stages of inflation. Experiment with different Einstein metrics on $K$, which will have different TT-instabilities and hence will generate distinct inflationary models in four dimensions.

In section \ref{UnstableModesInflation} we study the most basic example, where the internal metric has no unstable TT-deformations and the rescaling represented by $\phi$ is the only instability present. Then the dimensionally-reduced Einstein-Hilbert action is similar to the inflaton action in single-field models of  inflation. The potential $V(\phi)$ has a maximum at $\phi=0$ and generates inflation as $\phi$ rolls down from the maximum, during the initial stages of rescaling of the internal space. However, in this approximation (no TT-instabilities, no gauge fields, etc.), it does not satisfy the quantitative conditions of slow-roll inflation.

Section \ref{InternalSymmetryBreaking} illustrates the case where TT-instabilities are also present, working in the example $K=\SU$. The potential $V(\phi, \varphi)$ is seen to have a $\phi$-maximum and a $\varphi$-saddle point at the values $\phi=\varphi=0$, so the product Einstein metric is manifestly unstable. The additional instability represented by $\varphi$ breaks the internal symmetry, reducing the isometry group of $K$ from $(\SU \times \SU) / \mathbb{Z}_3$ down to $( \SU \mkern-1mu \times \mkern-1mu \mathrm{SU}(2) \mkern-1mu  \times \mkern-1mu \mathrm{U}(1)  ) /  \mathbb{Z}_6 $. The action defines a more evolved, two-field model of the initial stages of inflation. The potential $V(\phi, \varphi)$ is unbounded from below for large values of $\phi$ and $\varphi$. This suggests that these components of the higher-dimensional metric can change significantly in a classical dynamical process, perhaps by many orders of magnitude, before being stabilized at a different scale by new physics not contained in the Einstein-Hilbert action (if they can be stabilized at all). Of course at later times also the gauge fields can become non-zero, so even the classical dynamics is more complicated than what is represented by the simplified, four-dimensional action written above. 

If the unstable deformations of $g_P$ can be stabilized by an effective potential, then the resulting deformed metric should be a better candidate for the present-time vacuum configuration than the product Einstein metric. Due to the rescaling represented by $\phi$, in that vacuum configuration the scale of the gauge couplings no longer needs to coincide with the scale of the cosmological constant on $M$, as discussed previously in this Introduction.

\subsubsection*{Stabilizing the internal curvature} 

As described before, one expects the initial Einstein product metric on $M\times K$ to unravel along the unstable deformations represented by $\phi$ and $\varphi$. When this happens, the internal space goes through a rescaling and, simultaneously, through a slower TT-deformation that breaks the isometry group and increases the internal scalar curvature. But will these deformations of the internal geometry increase indefinitely, across all orders of magnitude of size and curvature, as indicated by the classical potential $V(\phi, \varphi)$? 

Generally speaking, it is reasonable to expect that new physics may start to be relevant at other scales of size or curvature. One should not be able to confine quantum particles in arbitrarily small internal directions, for example. Meanwhile, it is probably a tall order to ask the (higher-dimensional) Einstein-Hilbert action to cover the phenomena in all those scales. So the question becomes how to complement that action and model mathematically the micro-scale effects. For example, is there a natural addition to the Einstein-Hilbert action that becomes relevant for small internal spaces and prevents a total collapse of $K$? Or an effective potential that increases with the internal curvature and prevents gauge bosons with infinite mass? If they exist, should those additional terms to the Lagrangian be regarded as purely ad hoc, as in most cosmological models of inflation and reheating, or can they be justified as having origins in micro-scale physics? These speculative matters are discussed in section \ref{StabilizingInternalCurvature}.

One possible ideia would be to introduce an effective potential inspired by the QFT vacuum energy density. Well-known calculations suggest that the renormalized, four-dimensional vacuum energy density increases with the masses of the quantum fields. Although the precise formula does not seem to be consensual, the contribution of a gauge field of mass $m$ to the vacuum energy density should increase as the power $m^4$. The calculations in \cite{Martin}, for instance, suggest a contribution proportional to 
\beq 
m^4 \, \log\Big(\frac{m^2}{\mu^2} \Big) \ ,
\eeq
where $\mu$ is a new mass scale. But in a Kaluza-Klein model the (classical) mass of a gauge field depends on the vacuum internal metric, as expressed by formulae such as \eqref{MassGaugeBosons} or \eqref{MassGaugeBosons5}. So we have a non-trivial dependence $m^2 = m^2 (g_K)$, and this means that a deformation of the internal metric will affect the vacuum energy density:
\[
\rho_{\rm vac}^{4D} \ = \ \rho_{\rm vac}^{4D} (g_K) \ .
\]
Adding this density to the classical Einstein-Hilbert Lagrangian creates an effective potential for the internal deformations that, in a favourable setting, could contribute to stabilize them around a local minimum. In the simplified example of section \ref{StabilizingInternalCurvature}, and only taking into account the contributions of the natural gauge bosons, we have something like 
\[
V_{{\rm eff}} (\phi, \varphi) \ = \ V(\phi, \varphi)  \ + \   \, m^4(\varphi, \phi) \, \log \Big[\, \frac{m^2(\varphi, \phi) }{\mu^2} \, \Big]  \ ,
\]
with a dependence $m^2(\phi,\varphi)$ established in \eqref{MassBosonsTTDeformation}. An inspection of that formula shows that the power $m^4(\phi,\varphi)$ grows as $e^{2 \beta \phi} \, e^{4\varphi}$ for large, positive values of the deformation fields. So the second term in $V_{{\rm eff}}$ will dominate the initial potential $V(\phi, \varphi)$ in this regime, which decreases as $- e^{\beta \phi} \, e^{2\varphi}$ for large values of the same fields. This suggests that the unstable $(\phi,\varphi)$-deformations of the internal metric could in principle be stabilized by contributions coming from the vacuum energy density. This is just a rough argument, nonetheless, since the true vacuum energy density is a complex quantity that should depend on all gauge fields and all fermionic fields.

The appearance of a new mass scale $\mu$ in the effective potential is also an interesting feature. In favourable conditions, some classically massless gauge fields may acquire a mass dependent on the new scale, which can be distinct from the Planck scale. In general, adding a new potential term to the classical Einstein-Hilbert action implies the introduction of new constants and scales in the model. This could help to model the physical electroweak symmetry breaking within the Kaluza-Klein framework, although we do not pursue that line here.

\subsubsection*{Comments on fermions} 

Having gauge fields associated to non-isometric diffeomorphisms of the internal space creates new opportunities---such as the possibility to generate mass for the gauge bosons within the Kaluza-Klein framework, or the possibility to evade the Atiyah-Hirzebruch theorem---but also additional theoretical challenges---such as understanding how fermions should transform under diffeomorphisms that do not preserve the metric. This challenge exists in any gravity theory, not just in Kaluza-Klein.

In section \ref{NoGoArguments} we discuss the opportunity to circumvent the traditional no-go arguments against chiral fermions in Kaluza-Klein. These include the general argument derived from the Atiyah-Hirzebruch theorem, ruling out chiral fermions in all dimensions, as well as other arguments applicable only to certain dimensions. 
For the rest of section  \ref{CommentsFermions} we describe three possible geometric approaches to the question of how fermionic fields should transform under non-isometric diffeomorphisms of the underlying space. This implies thinking about the definition of spinor itself. More precisely, how to extend this definition to objects that are not tied down to a fixed background metric, i.e. to objects that have a natural action of the double cover of the diffeomorphism group. The approaches described include using $\tGLp_k$-representations, instead of $\Spin_k$ representations; using what we call extended spinors; or using more general universal spinors. Some of these approaches already exist in the literature, in different forms, but do not seem to be widely explored.

\newpage

\section{Scalar curvature of submersive metrics on \texorpdfstring{$M_4 \times K$}{MK}}     
\label{SectionScalarCurvature}

\subsection{Decomposing the higher-dimensional scalar curvature}
\label{DecomposingScalarCurvature}

The purpose of this section is to recall the notion of Riemannian submersion on the higher-dimensional manifold $P = M \times K$. More details can be found in \cite{ONeill, Besse}. A submersive metric on $P$, denoted by $g_P$, is the classical field of a general version of the Kaluza-Klein ansatz. It can be fed into an action functional $\mathcal{E} (g_P) $ such as the Einstein-Hilbert action. We will spend a few paragraphs recalling the formula for the scalar curvature of a Riemannian submersion and establishing the associated notation.

Let $\pi$ denote the natural projection $\pi: P \to M$. The inverse image $\pi^{-1}(x)$ of a given point $x$ in $M$ is called the fibre of $P$ above $x$, or the internal space above $x$. It is sometimes denoted by $P_x$ or $K_x$, and it is of course isomorphic to $K$. The tangent space to $P$ at any given point $p=(x,h)$ has a distinguished subspace $\VV_p$ defined by the kernel of the derivative map $\pi_\ast: T_p P \to T_x M$. This is called the vertical subspace of the projection $\pi$ and is just the tangent space to the fibre $P_x$. When $P$ is a simple product, it can be identified with the tangent space $T_h K$. Given an arbitrary Riemannian metric $g_P$ on $P$, the orthogonal complement to $\VV_p$ is called the horizontal subspace $\HH_p$ of the tangent space $T_pP$. Then we have a decomposition
\beq \label{HorizontalDistribution}
T_p P \; =\; \HH_p \oplus \VV_p   \, \qquad \quad  E \; = \;  E^\HH  \; +  \; E^\VV \ ,
\eeq
and every tangent vector $E\in T_p P$ can be written as a sum of the respective components. The map $\pi$ is called a submersion if the derivative $\pi_\ast: T_pP \rightarrow T_xM$ is surjective for every $p\in P$. In this case the derivative induces an isomorphism of vector spaces $\HH_p \simeq T_x M$. This is always true in the case of a product manifold $P = M \times K$. The pair $(\pi, \, g_P)$ is said to define a {\it Riemannian} submersion if it satisfies the property
\beq \label{RiemannianSubmersion}
\pi_\ast \big(E_1^\HH\big) \ = \ \pi_\ast \big(E_2^\HH \big) \qquad  \implies  \qquad  g_P \big(E_1^\HH, \, E_1^\HH \big) \ = \ g_P \big(E_2^\HH, \, E_2^\HH \big)
\eeq
for any horizontal vectors $E_1^\HH$ and $E_2^\HH$ on $TP$. In words, horizontal vectors that project down to the same vector on the base $M$ must have the same $g_P$-norm on $P$, even if they are tangent to different points of the same fibre. This is a restriction on the metric $g_P$. 

Any Riemannian submersion $(\pi, \, g_P)$ defines a metric $g_M$ on the base by projection. To be more precise, given a vector $X \in T_x M$, let $p$ be a point on the fibre above $x$ and let $X^\HH$ be the unique horizontal vector in $\HH_p$ such that $\pi_\ast \big(X^\HH\big) = X$. Then one defines
\[
(g_M)_x (X, X) \ := \ g_P(X^\HH , X^\HH ) \ .
\]
This definition is independent of the choice of the point $p$ on the fibre $\pi^{-1}(x)$ because of property \eqref{RiemannianSubmersion}. 

A Riemannian submersion $(\pi, \, g_P)$ also defines a natural one-form $A$ on $M$ with values on the space of vertical vector fields. In fact, given point $p=(x,h)$ on $P$ and a tangent vector $X \in T_x M$, the identification $T_p P = T_x M \oplus T_h K$ allows us to regard $X$ also as vector in $T_p P$. Using this identification and decomposition \eqref{HorizontalDistribution} one simply defines
\beq  \label{DefinitionGaugeFields}
A(X) \, |_p \ := \ - \, X^\VV \ , 
\eeq
where $X$ is regarded as a vector in $T_x M$ on the left-hand side and as a vector in $T_p P$ on the right-hand side. This definition implies that 
\[
(g_P)_p (A(X), V) \ = \ -\, (g_P)_p (X, V)
\]
for every vector $V \in \VV_p$. The information contained in the one-form $A$ on $M$ is equivalent to the information contained in the horizontal distribution $\HH \subset TP$ associated to $g_P$. In fact, if a vector field $E$ on $P$ is written as a sum $E = E_M + E_K$ according to the decompositon $TP = TM \oplus TK$, then we have the identities
\beq \label{DefinitionHorizontalDistribution}
E^\VV  \ = \   E_K \; - \;  A (E_M)    \quad  \qquad \qquad  E^\HH  \ = \  E_M \; + \;   A (E_M)   \ . 
\eeq
The one-form $A$ on $M$ is called the connection form, or the Yang-Mills form, of the submersion. It can also be regarded as the one-form on the base defined by a connection on a (trivial) $\text{Diff} (K)$-principal bundle over $M$. This works because the Lie algebra of the diffeomorphism group $\text{Diff} (K)$ is the space of vector fields on $K$, so precisely the space where $A$ has its values. In this view $P=M\times K$  should be regarded as the bundle associated to that principal bundle by the natural $\text{Diff} (K)$-action on $K$.

Besides determining a metric $g_M$ and a one-form $A$, a Riemannian submersion $(\pi, \, g_P)$ also defines a family of metrics $g_K$ on the internal space by restricting $g_P$ to the different fibres, all isomorphic to $K$. In other words, $g_P$ defines a map from $M$ to the space of Riemannian metric on $K$, $M \rightarrow \MM(K)$, through the natural restriction $x\mapsto g_P |_{\pi^{-1}(x)}$. This point of view will be further discussed in section \ref{GaugedSigmaModels}.

As described in the Introduction, using \eqref{MetricDecomposition} one can fully reconstruct the submersive metric $g_P$ from the equivalent data $(g_M, A, g_K)$. So all the natural quantities associated to $g_P$ can also be expressed in terms of that data. For example, it follows from groundwork in \cite{ONeill} that the scalar curvature of a submersive metric $g_P$ can be written as a sum of components
\beq \label{DecompositionScalarCurvature}
R_{g_\PPP} \ = \ R_{g_\MMM} \, + \, R_{g_\KKK} \, - \, |\mathcal{F}|^2 \, - \, |S|^2  \ - \  |N|^2 \, -\, 2\,\check{\delta} N \ . 
\eeq
Here $R_{g_\MMM}$ and $R_{g_\KKK}$ denote the scalar curvatures of $g_M$ and $g_K$, respectively, and $\mathcal{F}$, $S$ and $N$ are the tensors on $P$ that we will now describe (see chapter 9 of \cite{Besse}\footnote{The notation here differs from that in \cite{ONeill, Besse} in the following points: the tensor called $A$ in \cite{ONeill, Besse} is called here $\FF$, to avoid confusion with the gauge fields; the tensor called $T$ in \cite{ONeill, Besse} is called here $S$, to avoid confusion with the energy-momentum tensor.}). 

Let $\nabla$ denote the Levi-Civita connection of the metric $g_P$; let $U$ and $V$ denote vertical vector fields on $P$; let $W$ and $Z$ denote horizontal vector fields on $P$. Then $S$ denotes the linear map $\VV \times \VV \to \HH$ that extracts the horizontal component of the covariant derivative of vertical fields,
\beq \label{tensorT}
S_U V  \ := \   (\nabla_U V)^\HH  \ .
\eeq
Since $U$ and $V$ are tangent vectors to the fibre $K$, the map $S$ can be identified with the second fundamental form of the fibres immersed in $P$. When $S$ vanishes, all the fibres are geodesic submanifolds of $P$ and are isometric to each other \cite{Hermann, Besse}. 

On its turn, $\FF$ is the linear map $\HH \times \HH \to \VV$ that extracts the vertical component of the covariant derivative of horizontal fields,
\beq \label{tensorF}
\mathcal{F}_W\, Z \ := \ (\nabla_{\!W} Z)^\VV \ =  \ \frac{1}{2} \,  [W, Z]^\VV  \ .
\eeq
The second equality is a standard result for torsionless connections \cite{ONeill, Besse}. When $\FF$ vanishes, the Lie bracket of horizontal fields is also horizontal, and hence $\HH$ is an integrable distribution. It is clear from the respective definitions that both $S$ and $\FF$ are $C^\infty$-linear when their arguments are multiplied by smooth functions on $P$. 

The vector field $N$, perpendicular to the fibres of the submersion, is defined as the metric trace
\beq \label{vectorN}
N \ := \ \sum\nolimits_j \, S_{v_j} v_j  \ ,
\eeq
where $\{v_j\}$ denotes a $g_K$-orthonormal basis of the vertical space. So $N$ can be identified with the mean curvature vector of the fibres of $P$. The norms of all these objects are defined by
\bal \label{NormsTensors}
|\FF|^2 \ &:= \ \sum_{\mu,\nu}\  g_K \big(\, \FF_{X_\mu^\HH}\, X_\nu^\HH  \, , \, \FF_{X_\mu^\HH} \, X_\nu^\HH \, \big)  \linebr
|S|^2 \ &:= \ \sum_{i,j }\  g_P \big(\, S_{v_i} v_j  \, , \, S_{v_i} v_j \, \big) \ = \ \sum_{i, j,\mu}  \ g_P (S_{v_i} v_j\, , X_\mu) \  g_P (S_{v_i} v_j \, , X_\mu)  \nonumber  \linebr
|N|^2 \ &:= \ g_P (N, N) \ = \ \sum_\mu  \ g_P (N, X_\mu) \  g_P (N, X_\mu) \nonumber \ ,
\end{align}
where $\{X_\mu\}$ stands for a $g_M$-orthonormal basis of $TM$, which lifts to a $g_P$-orthonormal basis $\{X_\mu^\HH\}$ of the horizontal subspaces inside $TP$. Finally, the scalar $\check{\delta} N$ is defined as the negative trace
\beq \label{DivergenceN2}
\check{\delta} N \ = \ - \sum\nolimits_\mu \, g_P\big(\, \nabla_{X_\mu^\HH}\, N,\, X_\mu^\HH \, \big) \ . 
\eeq
It is useful to note that the scalar $\check{\delta} N$ can also be expressed as a the combination of the norm $|N|^2$ and a total divergence on $P$. In effect, using the fact that $\{X_\mu^\HH, v_j \}$ is a $g_P$-orthonormal basis of the tangent space to $P$, we have that
\bal \label{DivergenceN}
\check{\delta} N \ &= \ -\; \divergence_{g_\PPP} (N) \ + \ \sum_j \,  g_P\big(\nabla_{v_j} N, v_j \big) \nonumber \linebr 
&=  \ -\; \divergence_{g_\PPP} (N) \ + \ \sum_j \, \Lie_{v_j} \Big[   g_P\big( N, v_j \big)  \Big] \ - \ g_P\big( N, \nabla_{v_j} v_j \big) \nonumber \linebr 
&= \  -\; \divergence_{g_\PPP} (N) \ - \ \sum_j \, g_P\big( N, S_{v_j} v_j \big)  \ = \    -\; \divergence_{g_\PPP} (N) \; - \;  |N|^2  \ .
\end{align}
Observe that $S$ and $N$ are not independent tensors, as one is the metric trace of the other. To work with independent degrees of freedom it is convenient to isolate the traceless part of the fibres' second fundamental form as 
\beq \label{TracelessFundamentalForm}
\mathring{S} (U,V)\ := \ S(U,V) \ - \ \frac{1}{k} \, \, g_P(U,V) \, N \ ,
\eeq
where $k$ denotes the dimension of $K$. In this case we have the usual identity of norms
\bal \label{NormTracelessFundamentalForm}
|\mathring{S} |^2 \ &:= \  \sum_{i,j }\  g_P \big(\, \mathring{S}_{v_i} v_j  \, , \, \mathring{S}_{v_i} v_j \, \big)  \   \nonumber \linebr
 &= \  |S|^2 \; + \;  \frac{1}{k^2} \, \, |N|^2 \, \sum_{i,j } g_P (v_i, v_j)^2  \; - \;  \frac{2}{k} \, \sum_{i,j }\  g_P \big(\, S_{v_i} v_j  \, , \, N \, g_P (v_i, v_j) \, \big)  \   \nonumber \linebr
 &= \  |S|^2 \; + \;  \frac{1}{k}  \, \, |N|^2 \; - \; \frac{2}{k} \, \,  \sum_{i,j }\  g_P \big(\, S_{v_j} v_j  \, , \, N  \, \big)  \   
= \  |S|^2 \; - \; \frac{1}{k} \, \, |N|^2  \ .
\end{align}
The purpose of the next few sections will be to calculate more explicitly these different components of the higher-dimensional scalar curvature $R_{g_\PPP}$. This will lead to a better understanding of their role in terms of four-dimensional physics.

\subsection{Yang-Mills terms on \texorpdfstring{$M_4$}{M}}

The content of the famous Kaluza calculation, progressively generalized in \cite{Original}, is the verification that the Yang-Mills kinetic terms for the gauge fields on Minkowski space can be obtained from the higher-dimensional Einstein-Hilbert action, more precisely from the component $| \FF |^2$ of the higher-dimensional scalar curvature. In this section we will verify how this works for general submersive metrics $g_P$ on $P = M \times K$, as determined by the metrics $g_M$ and $g_K$ and by the one-form $A$ defined in \eqref{DefinitionGaugeFields}. Everything develops as expected, so this is mostly a training exercise in the language of Riemannian submersions.

Let $X$ be a tangent vector to $M$. From the identification $TP = TM \oplus TK$, it can be regarded also as a tangent vector to $P$ satisfying $\pi_\ast X = X$. From \eqref{DefinitionHorizontalDistribution} we can write the horizontal component of $X$ as
\bal \label{HorizontalVector}
X^\HH  \ &:= \  X \; + \; A(X)  \ = \ X \; + \; \sum\nolimits_a \; A^a (X) \, e_a    \ ,
\end{align}
where $A$ is the one-form on $M$ with values in the Lie algebra of vector fields on $K$ and, in the rightmost sum,  we have picked a basis $\{ e_a\}$ for that algebra. The same identification  $TP = TM \oplus TK$ also allows us to think of $A$ as having values on the vertical vector fields on $P$. Then an application of  \eqref{HorizontalVector} to the tensor $\FF$ of \eqref{tensorF} leads to
\bal 
2\, \, \FF_{X^\HH} Y^\HH \ &= \ [X^\HH, \, Y^\HH]^\VV       \nonumber   \linebr
&= \ \big\{ \, [X, \, Y]^\HH \ - \ A([X, Y]) \ + \ [A(X), Y] \ + \ [X, A(Y)] \ + \ [A(X), A(Y)] \, \big\}^\VV   \nonumber   \linebr
&= \ \big\{  - \, A([X, Y]) \  - \ \Lie_Y [A^a(X)]\, e_a \ + \ \Lie_X [A^a(Y)]\, e_a \ + \ [A(X), A(Y)] \, \big\}^\VV   \nonumber   \linebr
&= \ \big\{  (\dd_M  A) (X, Y) \ + \ [A(X), A(Y)] \, \big\}^\VV  \nonumber  \ .
\end{align}
Here we have used the Einstein summation convention and the standard identity for the exterior derivative of a one-form:
\beq
\dd \omega (u, v)  \ = \ \Lie_u [\, \omega(v) \, ] \ - \ \Lie_v [\, \omega (u) \, ] \ - \ \omega ([u, v]) \ .
\eeq
Standard properties of the Lie bracket of vector fields on $P$ also allow us to write
\bal 
[A(X), A(Y)] \; &= \; A^a (X)\, A^b (Y) \, [e_a, e_b]  \, - \, A^b(Y)\, \dd[A^a(X)]  (e_b) \, e_a  \,  + \, A^a (X) \, \dd[A^b(Y)]  (e_a)\, e_b        \nonumber   \linebr
&= \; A^a (X)\, A^b (Y) \, [e_a, e_b]   \nonumber  \ , 
\end{align}
since the $A^a(X)$ are functions on $M$ with constant value along the vertical fibres $K$. Thus
\bal \label{ExplicitTensorF}
2\, \, \FF_{X^\HH} Y^\HH \ = \ (\dd_M A^a) (X, Y) \, e_a  \ + \ A^a (X)\, A^b (Y) \, [e_a, e_b]  \  =: \ F_{A} (X, Y)  \ 
\end{align}
has constant values along the fibres and defines a 2-form on the base $M$ with values on the vertical vector fields of $P$. This is of course the curvature of the connection one-form $A$.
To explicitly write down the norm $|\FF|^2$, let $\{X_\mu\}$ denote a basis for the tangent space $TM$. It follows from \eqref{ExplicitTensorF} combined with \eqref{NormsTensors} that
\bal \label{NormF}
{\setlength{\fboxsep}{3\fboxsep} \boxed{
|\FF|^2 \ = \ \frac{1}{4} \; g_M^{\mu \nu} \; g_M^{\sigma \rho}   \  g_K (e_a ,  e_b  ) \ (F^a_{A})_{\mu \sigma} \ (F^b_{A})_{\nu \rho} \ 
}}
\end{align}
as a scalar function of $M\times K$. Even though the metric $g_M$ and the curvature coefficients $F^a_{A}$ only depend on the coordinate $x \in M$, the norm of $\FF$ is not a constant function along $K$, since the inner-products $g_K (e_a , e_b)$ in general do depend on the coordinates on $K$. 

This expression for $|\FF|^2$ can be integrated over the internal space $(K, \vol_{g_\KKK})$ to define a scalar function on the base $M$:
\begin{equation} \label{IntegralNormF}
\int_K \, |\FF|^2 \ \vol_{g_\KKK} \ = \ \frac{1}{4} \, g_M^{\mu \nu} \, g_M^{\sigma \rho}   \ (F^a_{A})_{\mu \sigma} \ (F^b_{A})_{\nu \rho}  \ \int_K   g_K (e_a ,  e_b  ) \ \vol_{g_\KKK} \ .
\end{equation}
Observe how the coefficients in front of the curvature components depend solely on the $L^2$ inner-product of the vector fields $e_a$ on $K$. This scalar density on $M$ broadly coincides with the form of the Yang-Mills terms in the Standard Model Lagrangian. This very well-known fact is sometimes dubbed the Kaluza-Klein miracle.

\subsection{Fibres' second fundamental form}
\label{SecondFundamentalForm}

A submersive metric $g_P$ on the higher-dimensional $M\times K$, together with a metric connection on the tangent bundle, determines the tensor $S$ defined in \eqref{tensorT}. This tensor can be identified with the fibres' second fundamental form in the case of the Levi-Civita connection. The purpose of this section is to understand $S$ and its norm $|S|^2$ in terms of the data $(g_M, g_K, A)$, equivalent to the initial metric $g_P$. In the Lagrangian density of a Kaluza-Klein model, the component $|S|^2$ of the higher-dimensional curvature encapsulates the kinetic terms for the scalar fields describing how the internal metric $g_K$ varies from fibre to fibre. These scalar fields are the analog of the Higgs field in the Kaluza-Klein framework. In particular, it is the component $|S|^2$ of the Lagrangian that contains the quadratic terms in the gauge fields that can lead to mass generation through spontaneous symmetry breaking.

Let $U$ and $V$ be vertical vector fields on the total space of a submersion $\pi : P \to M$ and let $\nabla$ be a metric connection on the tangent bundle $TP$. In a submersion, the Lie bracket of vertical fields is always vertical \cite{Besse}. So for torsionless connections it is clear that $S_U V$, as defined in \eqref{tensorT}, is symmetric:
\beq
S_U V \ = \    (\nabla_U V)^\HH   \ = \  \big( \ \nabla_V U + [U, V] + {\mathrm{Tor}}^{\nabla} (U, V) \ \big)^\HH \  = \  S_V U \ .
\eeq
It is not strictly necessary to start with a torsionless connection in order to obtain a symmetric $S$. It is enough to demand that ${\mathrm{Tor}}^{\nabla} (U, V)$ be a vertical vector field whenever $U$ and $V$ are vertical. This will be the case when ${\mathrm{Tor}}^{\nabla} (U, V)$ is proportional to the bracket $[U, V]$, for instance. Be that as it may, in the calculations ahead we will not explore this variant and will simply assume that $\nabla$ is the Levi-Civita connection. 

Using the definition of $S_U V$ and the properties of the Levi-Civita connection, one can write for every vector $X \in TM \subset TP$:
\bal
g_P \big(\,S_U V, \,  X\, \big) \ &= \ g_P \big(\,\nabla_U V, \, X^\HH \, \big)  \ 
= \ \Lie_U \big[\, g_P ( V, \, X^\HH) \,\big] \, - \, g_P \big(\,V, \,  \nabla_U  X^\HH \,) \nonumber \linebr
&= \ - \ g_P \big( \, V, \, \nabla_{X^\HH} U \, + \, [U, X^\HH] \, \big) \nonumber \ .
\end{align}
But $S_U V$ is symmetric in $U$ and $V$, so using again that $\nabla$ is a metric connection,
\bal \label{ExplicitTensorS}
2\, g_P \big(\, S_U V, \,  X \, \big) \ &=   \ g_P \big(\, S_U V, \,  X \, \big) \ + \ g_P \big(\, S_V U, \,  X \, \big)                  \nonumber \linebr
&= \ - \ \Lie_{X^\HH} \big[\, g_P ( U, \, V) \,\big] \, - \,  g_P \big( \, V, \, [U, X^\HH] \, \big) \, - \, g_P \big( \, U, \, [V, X^\HH] \, \big) \qquad \nonumber  \linebr
&= \ - \ (\Lie_{X^\HH} \,g_P ) \big(U, \, V  \big) \ , 
\end{align}
where the last equality is a standard identity for Lie derivatives of 2-tensors. This formula provides a concise relation between the tensor $S_UV$ and the horizontal Lie derivatives of the submersion metric $g_P$. Using expression \eqref{HorizontalVector} for the horizontal lift of $X$ and noting that the functions $A^a(X)$ are constant along the fibres, so have vanishing Lie derivative along the vertical fields $U$ and $V$, we can also write 
\begin{equation} \label{ExplicitTensorS2}
{\setlength{\fboxsep}{3\fboxsep} \boxed{
2\, g_P \big(\, S_U V, \,  X \, \big) \ = \ - \ (\Lie_{X} \, g_P ) (U, \, V )  \ - \  A^a(X) \; (\Lie_{e_a} \, g_P ) (U, \, V ) \  
}}
\end{equation}
as a scalar function of $M\times K$. The combination of \eqref{NormsTensors} and \eqref{ExplicitTensorS} allows us to express the squared-norm of the fibres' second fundamental form as
 \begin{equation} \label{ExplicitTensorS3}
| S |^2  \ = \    \frac{1}{4}  \;  \sum_{\mu , \nu}\; \sum_{i, j } \  (g_M)^{\mu \nu} \  (\Lie_{X_\mu^\HH} \, g_P ) (v_i, \, v_j )  \   (\Lie_{X_\nu^\HH} \, g_P ) (v_i, \, v_j ) \ ,
\end{equation}
where $\{v_j\}$ is any $g_K$-orthonormal basis of the vertical space. For a general submersive metric $g_P$ the norm $|S|^2$ is not a constant function along the fibres. If we use \eqref{ExplicitTensorS2} instead of \eqref{ExplicitTensorS}, the formula for $|S|^2$ becomes longer but perhaps more suggestive:
 \begin{multline}
\label{NormT2}
|S|^2 \ = \   \frac{1}{4} \;  \sum\nolimits_{\mu , \nu}\; g_M^{\mu \nu}  \left\langle \Lie_{X_\mu}\, g_P,  \ \Lie_{X_\nu}\, g_P \right\rangle \ + \  2 \, g_M^{\mu \nu} \ A^a (X_\mu)  \left\langle \Lie_{e_a}\, g_P,  \ \Lie_{X_\nu}\, g_P \right\rangle \linebr  
+  \ g_M^{\mu \nu} \ A^a (X_\mu)\  A^b (X_\nu)   \left\langle \Lie_{e_a}\, g_P,  \ \Lie_{e_b}\, g_P \right\rangle \ .
\end{multline}
Here $\left\langle \cdot \, , \, \cdot \right\rangle$ denotes the inner-product on the space of vertical, symmetric 2-tensors determined by the metric $g_P$, or by its restriction $g_K$ to the fibres.
 Explicitly, the inner-product of two symmetric 2-tensors on the internal space $K$ is defined as
\begin{equation} \label{DefinitionInnerProduct}
\left\langle h_1, \, h_2 \right\rangle_{g_K} \ := \ \sum_{i,\, j} \, h_1(v_i, v_j) \ h_2(v_i, v_j) \ , 
\end{equation}
where $\{ v_j \}$ is any local, $g_K$-orthonormal trivialization of $TK$. 

Expression \eqref{NormT2} shows how the fibres' second fundamental form gives rise to the quadratic terms in the gauge fields $A^a$ that are essential to mass generation through spontaneous symmetry breaking. Quite naturally, the coefficients of those quadratic terms are determined by the Lie derivative of the fibres' metric along the associated internal vector field $e_a$. So the components $A^a$ associated to Killing vector fields will disappear entirely from $|S|^2$, and thus correspond to massless bosons.

\subsection{Mean curvature of the fibres}

Among the six components of the higher-dimensional scalar curvature $R_{g_\PPP}$ in formula \eqref{DecompositionScalarCurvature}, only the terms involving the mean curvature vector of the fibres---denoted by $N$ in that formula---have not yet been calculated in terms of the data $(g_M, g_K, A)$. That is the purpose of this section.

Let $p = (x, h)$ be a fixed point in the product manifold $P = M \times K$. Let $\{ u_j\}$ denote a trivialization of $TK$ in a neighbourhood of $h$ that is orthonormal with respect to the 
restriction of the metric $g_P$ to the fibre $P_x = \{ x\} \times K$. This restriction is denoted by $g_{P_x}$. If we are talking about a general fibre of $P$, the vertical restriction of $g_P$ is denoted simply by $g_K$. Having in mind definition \eqref{vectorN} of the horizontal field $N$, we start by taking the trace of \eqref{ExplicitTensorS} to write 
\bal \label{AuxTraceS} 
2\, g_P (  N, \,  X) \ = \ 2\, g_P \big( S_{u_j} u_j, \,  X  \big) \ = \ - \ \Lie_{X^\HH} \big[\, g_P ( u_j, \, u_j) \,\big] \, - \,  2 \; g_P \big( \, u_j, \, [u_j, X^\HH] \, \big)  \ .
\end{align}
The orthonormality of the vector fields $\{ u_j\}$ with respect to $g_{P_x}$ implies that the functions $g_P ( u_j, \, u_j)$ have vanishing Lie derivatives in the vertical directions, so 
\[
\Lie_{X^\HH} \left[\, g_P ( u_j, \, u_j) \,\right] \  =\   \Lie_{X} \left[\, g_P ( u_j, \, u_j) \,\right]  \ 
\]
over the slice $\{ x\} \times K$. On the other hand, since $X$ is a vector field on $M$ and $u_j$ is a field on $K$, the bracket $[X, u_j]$ vanishes over the product $M\times K$. So
\bal 
 g_P \big( \, u_j, \, [u_j, X^\HH] \, \big) \ &= \ \sum_{j, \, a}  \ g_P \big( \; u_j, \, \big[ u_j, \, A^a (X) \, e_a  \big] \;  \big) \ = \  \sum_{j, \, a} \, A^a (X) \ g_P \big(  u_j, \, [u_j, \, e_a]  \big) \nonumber \ , 
\end{align}
where the last equality uses that the coefficients $A^a (X)$ are constant along the fibres. This expression can be further simplified by making use of the properties of the Levi-Civita connection $\nabla$ on the fibre $P_x$. In fact, for any vertical vector field $V$ on $K$ we can write
\bal \label{IdentityDivergence}
 \sum_{j}  \ g_{P_x} \big(  u_j, \, [u_j, \, V]  \big) \ &= \  \sum_{j}  \  g_{P_x} \big(  u_j, \, \nabla_{u_j} V  -  \nabla_V\,  u_j   \big)     \nonumber  \linebr 
 &= \ \sum_j \  g_{P_x} \big(  u_j, \, \nabla_{u_j} V  \big)   - \,  \frac{1}{2} \; \Lie_V \Big[ g_{P_x} (  u_j, \, u_j) \Big]   \nonumber  \linebr 
 &= \  \divergence_{g_{\PPP_x}} (V) \ \ ,
\end{align}
where the last equality uses that the local vector fields $\{ u_j \}$ are orthonormal, besides the definition of divergence of a vector field. Applying this simplification to \eqref{AuxTraceS} we get that, for any vector $X \in TM \subset TP$,
\beq \label{ExplicitVectorN1}
2\, g_P (  N, \,  X) \ =  \  - \; \sum_j  \; \Lie_{X} \big[ \, g_P ( u_j, \, u_j) \,\big]   \ - \ 2 \, \sum_a \;  A^a (X) \, \divergence_{g_\KKK} (e_a) \ .
\eeq
When reading this expression it is important to keep in mind that the vertical restrictions of the submersive metric $g_P$, generically denoted by $g_K$, may vary across different fibres, while the basis $\{ u_j\}$ was defined to be $g_K$-orthonormal only at the fibre $P_x$. In particular, the functions $g_P ( u_j, \, u_j)$ have value 1 on $P_x$ but need not be constant when moving across fibres along the flow of $X$. 

Using the Jacobi formula that relates the derivative of the determinant of an invertible matrix with the derivative of the trace the matrix, one can also write locally
\[
\Lie_{X} \Big[ \, \sum\nolimits_j\ g_P ( u_j, \, u_j) \,\Big] \ = \ 2 \; \Lie_{X} \left[ \; \log \big(\sqrt{|g_K |} \big) \; \right]  \ .
\]
Here the symbol $|g_K |$ denotes the determinant of the square matrix $g_P(u_i, u_j)$ that represents the fibre metric in the local, fixed, $g_{P_x}$-orthonormal trivialization $\{ u_j\}$ of $TK$. So $\sqrt{|g_K |}$ is the scalar density of the volume forms $\vol_{g_\KKK} $ in this trivialization. Denoting by $\vol_{g_{\PPP_x}}$ the fixed volume form on $K$ associated to the metric $g_{P_x}$, we can write
\[
\vol_{g_\KKK} \ = \ \sqrt{|g_K |} \ \vol_{g_{\PPP_x}} \ . 
\]
This expression shows that the function  $\sqrt{|g_K |}$ is globally defined on the product $M \times K$. Now let $\vol_{g_\SSzero}$ be any fixed volume form on $K$. It can be written as $f_0  \cdot \vol_{g_{\PPP_x}} $ for some function $f_0$ on $K$ that does not depend on $M$. Since $X$ is a vector field on $M$, we have
\bal
\Lie_{X} \left[ \; \log \big(\sqrt{|g_K |} \big) \; \right] \ &= \  \Lie_{X} \left[ \; \log \Big(\, \frac{\vol_{g_\KKK}}{\vol_{g_{P_x}}  }  \, \Big)  \; \right]  \ = \ \Lie_{X} \left[ \; \log \Big(\, \frac{\vol_{g_\KKK}}{\vol_{g_0} }  \, \Big)    +    \log (f_0) \; \right]  \nonumber \linebr
 &= \  \Lie_{X} \left[ \; \log \Big(\, \frac{\vol_{g_\KKK}}{\vol_{g_\SSzero} }  \, \Big)  \; \right]  \nonumber
\end{align}
as a function on $P = M\times K$. So we can finally write
\begin{equation} \label{ExplicitN2}
{\setlength{\fboxsep}{3\fboxsep} \boxed{
g_P (  N, \,  X) \ = \  - \; \Lie_{X} \Big[  \log \Big(\, \frac{\vol_{g_\KKK}}{\vol_{g_\SSzero} }  \, \Big)  \Big]  \ - \  \sum\nolimits_a \;  A^a (X) \, \divergence_{g_\KKK} (e_a)  \ .
} }
\end{equation}
This expression makes it manifest that the mean curvature vector $N$ depends on the internal metric only through its volume form $\vol_{g_\KKK}$. More precisely, it measures how $\vol_{g_\KKK}$ varies along the flows of $X$ in the base $M$ and along the vertical flows of $e_a$. The function $g_P (  N, \,  X)$ in general is not constant along the fibres of $P$, as both $\vol_{g_\KKK}$ and the vertical divergence of $e_a$ vary with the metric $g_P$ along both $K$ and $M$.

 The previous expression for $g_P (  N, \,  X) $ leads to a remarkably simple integral relation between both sides of the equation. This happens because the coefficients $A^a (X)$ are constant along $K$ and the fibre-integral of the divergence $\divergence_{g_\KKK} (e_a)$ always vanishes. Using also that the volume form $\vol_{g_\SSzero}$ is fixed on K and does not change with the flow of $X$ along $M$, we get simply
\bal \label{PropertyMeanCurvature}
\int_K \ g_P (  N, \,  X) \  \vol_{g_\KKK} \ &= \  - \; \int_K  \Lie_{X} \Big[  \log \Big(\, \frac{\vol_{g_\KKK}}{\vol_{g_\SSzero} }  \, \Big)  \Big]  \  \vol_{g_\KKK} \ = \  - \;  \int_K   \frac{ \Lie_{X}  (\vol_{g_\KKK})}{\vol_{g_\KKK} }   \ \vol_{g_\KKK} \nonumber \linebr
&= \  - \;  \int_K    \Lie_{X}  (\vol_{g_\KKK}) \ = \ -  \, \Lie_{X}  \Big( \Vol_{g_\KKK} \Big) \ .
\end{align}
as a function on the base $M$. Here $\Vol_{g_\KKK}$ denotes the total volume of the fibre, and the last equality follows from Leibniz rule for the exchange the order of integrals and derivatives. The resulting expression is a well-known formula for the first variation of the volume of the fibres as one moves along the base of a Riemannian submersion (e.g. \cite{Besson}\footnote{This reference uses the opposite sign convention in the definition of divergence of a vector field.}).

The norm $|N|^2$ of the mean curvature vector can be expressed in terms of the data $(g_M, \vol_K, A)$ by substituting \eqref{ExplicitN2} in the definition \eqref{NormsTensors} of the squared-norm. Here we will not write it down. We will, however, give a slightly simplified expression for the fibre-integral of the resulting expression, $\int_K |N|^2 \ \vol_{g_\KKK}$. To this end, note that using again Leibniz integral rule and the fact that $\divergence_{g_\KKK} (e_a)$ integrates to zero over the fibre, we have
\bal
\int_K \divergence_{g_\KKK} (e_a) \ \Lie_{X_\mu} \Big[  \log \Big(\, \frac{\vol_{g_\KKK}}{\vol_{g_\SSzero} }  \, \Big)&  \Big]  \vol_{g_\KKK} \ = \ \int_K  \divergence_{g_\KKK} (e_a) \   \Lie_{X_\mu} (\vol_{g_\KKK} ) \nonumber \linebr
&= \  \Lie_{X_\mu} \int_K  \divergence_{g_\KKK} (e_a) \ \vol_{g_\KKK}    \  - \, \int_K  \Lie_{X_\mu}  \big[\divergence_{g_\KKK} (e_a) \big] \ \vol_{g_\KKK} \nonumber \linebr
&= \ - \int_K  \Lie_{X_\mu}  \big[\divergence_{g_\KKK} (e_a) \big] \ \vol_{g_\KKK} \ .
\end{align}
So the fibre-integral of the full norm of the mean curvature vector can be expressed as
\bal
\int_K |N|^2 \ \vol_{g_\KKK} \ = \ &\sum_{\mu, \,\nu} \, g_M^{\mu \nu} \int_K \left\{  \Lie_{X_\mu} \Big[  \log \Big(\, \frac{\vol_{g_\KKK}}{\vol_{g_\SSzero} }  \, \Big)   \Big]  \right\} \left\{  \Lie_{X_\nu} \Big[  \log \Big(\, \frac{\vol_{g_\KKK}}{\vol_{g_\SSzero} }  \, \Big)   \Big]  \right\} \, \vol_{g_\KKK} \nonumber \linebr
&- \ 2\; \sum_{\mu,\, \nu, \, a} \, g_M^{\mu \nu}  \; A^a (X_\mu) \, \int_K  \Lie_{X_\nu}  \big[\divergence_{g_\KKK} (e_a) \big] \ \vol_{g_\KKK} \nonumber \linebr
&+ \  \sum_{\mu,\, \nu, \, a, \, b} \, g_M^{\mu \nu} \; A^a (X_\mu) \, A^b (X_\nu) \, \int_K  \divergence_{g_\KKK} (e_a) \;   \divergence_{g_\KKK} (e_b) \ \vol_{g_\KKK} 
\end{align}
as a function on the base $M$. This formula displays explicitly the contribution of the gauge fields $A^a (X_\mu)$ to this component of the scalar density on $M$.


\subsection{Submersive metrics and gauged sigma-models}
\label{GaugedSigmaModels}

The field theory consisting of a submersive metric on the product $M\times K$ governed by the higher-dimensional Einstein-Hilbert action has an alternative interpretation as four-dimensional gravity plus a gauged sigma-model. This sigma-model is a theory for maps $M \rightarrow \MM(K)$, where the target is the infinite-dimensional space of Riemannian metrics on $K$ acted upon by the gauge group $\text{Diff} (K)$, the group of diffeomorphisms of $K$. The nub of this view is the interpretation of the tensor $S$ of the Riemannian submersion as a covariant derivative of the fibres' metric as one moves along the base $M$. This is the Kaluza-Klein analog of the covariant derivative of Higgs fields. 

The purpose of this section is to describe this alternative viewpoint. The general idea is partially present in the mathematical literature on Riemannian submersions, where one studies the Ehresmann connection and holonomy groups associated to a submersion  \cite{Besse, Hermann}. However, to highlight the correspondence with the objects present in traditional Yang-Mills-Higgs models, we find it helpful to describe Riemannian submersions also in terms of explicit $\text{Diff} (K)$-connections, $\text{Diff} (K)$-gauge transformations and covariant derivatives expressed by local formulae such as \eqref{CovariantDerivativeFibreMetricCoordinates}. This seems to be new. Reading this section is not an essential requisite to follow the arguments in the subsequent sections. The presentation here is streamlined and we do not write down all the calculations and  proofs.

\subsubsection*{Covariant derivative of the fibre metric}

A higher-dimensional metric $g_P$ on the product $P= M\times K$ determines by restriction a natural map 
\beq \label{DefinitionMapToMetrics}
g_K : \, M\,  \longrightarrow \, \MM(K) \ , \qquad x \,\mapsto\, g_{P_x}
 \eeq
 where $g_{P_x}$ denotes the restriction of the metric $g_P$ to the fibre $P_x \simeq K$ over the point $x$ in $M$. Now let $X$ be a vector field on $M$ and let $U$ and $V$ be vertical fields on $P$. We define the covariant derivative of the map $g_K$ by the expression
\beq \label{DefinitionCovariantDerivative}
(\dd^A g_K)_X (U, V) \ := \  (\Lie_{X}\, g_P) (U, V) \ + \ A^a (X) \; (\Lie_{e_a}\, g_P) (U, V) \ ,
\eeq
where we are using the Einstein convention and summing over a basis $\{e_a\}$ of the space of vector fields on $K$. The right-hand side of this equation is $C^\infty (M)$-linear in the $X$ entry and is $C^\infty (M \! \times \! K)$-linear in the $U$ and $V$ entries, so $\dd^A g_K$ is a section of the bundle $T^\ast M \otimes {\text{Sym}}^2 (\VV^\ast)$ over $M\times K$. Here $\VV^\ast$ denotes the dual of the vertical sub-bundle of $TP$. Of course by \eqref{ExplicitTensorS2} we also have that
\beq \label{RelationCovariantDerivativeS}
(\dd^A g_K)_X (U, V) \ = \  -\, 2 \, g_P (S_U V, X) \ ,
\eeq
so this covariant derivative is essentially the same object as $S$ in a different notation. To recognize that the notation is justified, i.e. to recognize that the right-hand side of \eqref{DefinitionCovariantDerivative} measures how the fibres' metric $g_K$ changes as one moves along the flow of $X$ on the base $M$, pick a local coordinate system  $\{x^\mu , y^j \}$ on the product $M \! \times \! K$ and the associated trivializations of the tangent and cotangent bundles. Then a short calculation shows that the covariant derivative can be locally written as
\[
\dd^A g_K \ = \ (\dd^A g_K)_{\mu ij} \, \dd x^\mu \otimes  \dd y^i \otimes  \dd y^j
\]
with coefficients given by 
\beq \label{CovariantDerivativeFibreMetricCoordinates}
(\dd^A g_K)_{\mu ij} \ := \  \partial_\mu \, (g_K)_{ij}  \, + \,  A^a_\mu \, \left[ \,  (e_a)^l \,   \partial_l \, (g_K)_{ij}  \,+ \,  (g_K)_{li} \, \partial_j  (e_a)^l \, + \, (g_K)_{lj} \, \partial_i  (e_a)^l\, \right] 
\eeq
as functions on $M\! \times\! K$.  The first term is just the derivative of the fibres' metric coefficients $g_K \big(\frac{\partial}{ \partial y^i} ,\, \frac{\partial}{ \partial y^j} \big)$ along the four-dimensional $\frac{\partial}{ \partial x^\mu}$ direction, which supports the interpretation of $\dd^A g_K$, and hence $S$, as a covariant derivative of the fibre metric. From \eqref{RelationCovariantDerivativeS} it is clear that the map $g_K$ will be ``covariantly constant'' along $M$ if and only if the fibres of $P$ are totally geodesic ($S=0$). This is consistent with the well-known result that in a submersion with totally geodesic fibres the metrics of the different fibres are the image of each other under parallel transport \cite{Hermann, Besse}. 
Note as well that $e_a =  (e_a)^l \, \frac{\partial}{ \partial y^l}$ was defined as vector field on $K$, so the local coefficients $(e_a)^l$ only depend on the coordinates $y^j$. The coefficients $(g_K)_{ij}$ depend both on $x$ and $y$, because the fibres' metric may change as one moves along the base $M$. The gauge fields $A^a_\mu$ only depend on $x$.

To justify the designation of ``covariant derivative" one should show that $\dd^A g_K$ has some kind of covariance property with respect to gauge transformations. Here we are dealing with general Riemannian submersions, and this means that expansion \eqref{GaugeFieldExpansion} can be taken over a basis $\{e_a\}$ of the full Lie algebra of vector fields on $K$, which as a space coincides with the Lie algebra of the infinite-dimensional diffeomorphism group $\text{Diff} (K)$. So we will consider the full group of $\text{Diff} (K)$-gauge transformations on the bundle $P \rightarrow M$.

\subsubsection*{$\text{Diff} (K)$-gauge transformations}

A $\text{Diff} (K)$-gauge transformation is defined simply as a diffeomorphism $\varphi: P \rightarrow P$ that projects to the identity map on $M$ through the bundle projection $\pi: P \rightarrow M$. So we have $\pi \circ \varphi = \pi$. Such a transformation defines a family of diffeomorphisms $\varphi_x: K \rightarrow K$, parametrized by the points in the base $x \in M$, through the natural identity 
\[
\varphi (x, y)\ = \  (x, \varphi_x(y))
\]
for all $(x,y)$ in $M\times K$.
Gauge transformations act on the right on the space of metrics on $P$ through the pullback operation $g_P \rightarrow \varphi^\ast g_P$. Since $\pi \circ \varphi = \pi$ this action preserves the vertical distribution in $TP$, the Riemannian submersion property and the metric $g_M$ on the base determined by $g_P$. However, a gauge transformation in general will change the horizontal distribution determined by $g_P$, so it will have a nontrivial action on the one-forms $A(X) = \sum_a A^a(X) \, e_a$ defined in \eqref{DefinitionGaugeFields}. To write down the transformation rule of $A$ under gauge transformations, start by defining canonical one-forms $\Theta_\varphi$ on $M$ with values in the vertical fields on $P$ by
\beq
\Theta_\varphi (X) \ |_{(x,y)} \ := \ \frac{\dd}{\dd t} \ \left(x, \ \varphi_{\Phi^X_t (x)} \circ \varphi_x^{-1} (y) \right) \ |_{t=0}
\eeq
where $ \Phi^X_t :M \rightarrow M$ is the flow on $M$ of the vector field $X$. It is clear that for the identity gauge transformation $\varphi_x = \mathrm{id}$ the one-form $\Theta_{\text{id}}$ is identically zero. Moreover, one can show that under composition $\varphi_1 \circ \varphi_2$ of two gauge transformations the one-forms $\Theta$ transform as
\[
\Theta_{\varphi_1 \circ \varphi_2} (X) \ = \ \Theta_{\varphi_1} (X) \ + \  (\varphi_1)_\ast \left[\, \Theta_{\varphi_2} (X) \, \right]  \ ,
\]
where $(\varphi_1)_\ast$ denotes the push-forward of vertical vector fields on $P$.
The transformation rule $A \rightarrow A_\varphi$ of the connection one-form under a gauge tranformation $\varphi$ is defined by
\[
A_\varphi (X) \ := \  \varphi^{-1}_\ast \left[\,  A(X) -  \Theta_\varphi (X)\,  \right] \ = \  \varphi^{-1}_\ast \left[\,  A(X) \,  \right] \ + \ \Theta_{\varphi^{-1}} (X)
\]
as vertical vector fields on $P$, for all vector fields $X$ on the base $M$. With this rule one can show that, for any gauge transformation $\varphi$, the covariant derivative \eqref{DefinitionCovariantDerivative} transforms according to
\beq
{\setlength{\fboxsep}{3\fboxsep} \boxed{
\varphi^\ast \left[ (\dd^A g_K)_X \right] \ = \  \left[ \dd^{A_\varphi} ( \varphi^\ast g_K)  \right]_X \
}}
\eeq
as a section of the bundle ${\text{Sym}}^2 (\VV^\ast)$ over $M\times K$, for any vector field $X$ on $M$. Here $\varphi^\ast g_K$ denotes the map $M \rightarrow \MM(K)$ determined by the pullback $\varphi^\ast g_P$ of the higher-dimensional metric, as in \eqref{DefinitionMapToMetrics}. Moreover, one can also verify that if a submersive metric $g_P$ is represented by the data $(g_M, A, g_K)$, in the spirit of \eqref{MetricDecomposition}, then the pullback metric $\varphi^\ast g_P$ is represented by the gauge-transformed data:
\beq \label{GaugeTransformationData}
g_P \ \sim \ (g_M, \, A, \, g_K) \qquad  \implies \qquad  \varphi^\ast g_P \ \sim \ (g_M, \, A_\varphi, \, \varphi^\ast g_K) \ .
\eeq
Since the covariant derivative $\dd^A g_K$ is just the fibres' second fundamental form in another guise, one can wonder how this transformation rule looks like in terms of the tensor $S$. The answer is given by the following identity, valid for all vertical vector fields $U$ and $V$ on $P$, all vector fields $X$ on $M$ and all $\text{Diff} (K)$-gauge transformations:
\beq \label{GaugeTransformationS}
 (\varphi^\ast g_P) \big( \, (S^{\, \varphi^\ast g_\PPP})_{U} V,  \, X \, \big) \  = \  \varphi^\ast  \left[ \, \, g_P \big( \, (S^{\, g_\PPP})_{\varphi_\ast U} \, \varphi_\ast V \, ,\,  X\, \big)  \, \,  \right]  \ .
\eeq
This is an identity of functions on $P$. We have also made explicit the dependence of $S$ on the higher-dimensional metric $g_P$. 

If $\{ v_j \}$ is a local trivialization of the vertical bundle $\VV \rightarrow M\times K$ that is orthonormal with respect to the pull-back metric $\varphi^\ast g_P$, then the push-forward fields $\{ \varphi_\ast v_j \}$ are orthonormal with respect to $g_P$. Thus, taking the metric trace of both sides of \eqref{GaugeTransformationS} we get the transformation rule for the mean curvature vector:
\beq \label{GaugeTransformationN}
 (\varphi^\ast g_P) \big( \, N^{\, \varphi^\ast g_\PPP} ,  \, X \, \big) \  = \ \varphi^\ast \left[ \, g_P \big(  N^{g_\PPP} ,\,  X \big) \,  \right] \ .
\eeq
These rules imply that under gauge transformations the norms of these tensors change as
\beq \label{TransformationSDiffeo}
|  S^{\, \varphi^\ast g_\PPP} |^2_{\varphi^\ast g_\PPP} =   \varphi^\ast\, |  S^{\, g_\PPP} |_{g_\PPP}^2  \qquad   \qquad  \quad |  N^{\, \varphi^\ast g_\PPP} |^2_{\varphi^\ast g_\PPP} =   \varphi^\ast\, |  N^{\, g_\PPP} |_{g_\PPP}^2    \  \ , 
\eeq
where the notation signals that $S$, $N$ and the norms all depend on the metric on $P$. The standard invariance of integrals under diffeomorphisms then leads to, for instance,
\beq \label{TransformationIntSDiffeo}
\int_P\,  \big| \, S^{\, \varphi^\ast g_\PPP} \big|^2_{\varphi^\ast g_\PPP} \ \vol_{\varphi^\ast g_\PPP} \ = \  \int_P  \, \varphi^\ast \Big( \big| \, S^{\, g_\PPP} \big|^2_{g_\PPP} \, \vol_{g_\PPP} \Big) \ = \  \int_P  \, \big| \, S^{\, g_\PPP} \big|^2_{g_\PPP} \ \vol_{g_\PPP} \ .
\eeq
This of course is very natural and outlines how the different components of the Einstein-Hilbert action, not just the total action, are invariant under diffeomorphisms of $P$ that descend to the identity on $M$.

\subsubsection*{Covariant derivative of the fibre's volume form}

A Riemannian metric on $K$ defines a volume form on the manifold. So composing \eqref{DefinitionMapToMetrics} with this correspondence we get that each higher-dimensional metric $g_P$ determines a map
 \beq \label{DefinitionMapToVolumeForms}
\vol_{g_\KKK} : \, M\,  \longrightarrow \, \Omega^k(K, \mathbb{R}) \ , \qquad x \,\mapsto\, \vol_{g_{\PPP_x}} \ .
 \eeq
For any vector field $X$ on $M$ and any point $x$ on that manifold we define the covariant derivative of the map $\vol_{g_\KKK}$ by the expression
\beq \label{CovariantDerivativeVolumeForm}
D^A_X \, (\vol_{g_\KKK}) \ |_x  \ := \  \dd  (\vol_{g_\KKK})_x (X) \ + \ A^a (X) \; \Lie_{e_a} [\vol_{g_\KKK} (x)]  \ .
\eeq
This expression is $C^\infty (M)$-linear in the $X$ entry and $D^A_X \, (\vol_{g_\KKK})$ is just another map $ M\,  \rightarrow \, \Omega^k(K)$. 
 Not by accident, this definition of covariant derivative of the map $\vol_{g_\KKK}$ determined by the submersive metric $g_P$ is very much related to the mean curvature vector $N$ of the submersion. Using \eqref{ExplicitN2} it is not difficult to verify the simple identity
\beq \label{RelationCovariantDerivativeToN}
D^A_X \, (\vol_{g_\KKK})  \ = \  -\,   g_P(N, X)  \  \vol_{g_\KKK}  \ .
\eeq
The diffeomorphism group $\text{Diff} (K)$ acts on the right on the space of volume forms, $\Omega^k(K, \mathbb{R})$, so $\text{Diff} (K)$-gauge transformations $\varphi$ act on maps $ M\,  \rightarrow \, \Omega^k(K)$. It is clear that the $\varphi^\ast \vol_{g_\KKK}$ coincides with the map determined through \eqref{DefinitionMapToVolumeForms} by the higher-dimensional, pullback metric $\varphi^\ast g_P$. Thus, combining \eqref{RelationCovariantDerivativeToN} with rules \eqref{GaugeTransformationN} and \eqref{GaugeTransformationData} for the gauge transformations of $N$ and $A$, we get that 
\bal
\varphi^\ast \big[ D^A_X \, (\vol_{g_\KKK}) \big] \ &= \ - \, \varphi^\ast \big[ g_P(N, X)\big] \ \varphi^\ast \vol_{g_\KKK}  \nonumber \linebr
&= \ (\varphi^\ast g_P) \big( \, N^{\, \varphi^\ast g_\PPP} ,  \, X \, \big) \ \varphi^\ast \vol_{g_\KKK}  \nonumber \linebr
&= \ D^{A_\varphi}_X \, ( \varphi^\ast \vol_{g_\KKK})  \ .
\end{align}
So $D^A_X (\vol_{g_\KKK})$ deserves its designation as a covariant derivative under $\text{Diff} (K)$-gauge transformations.

\subsubsection*{Einstein-Hilbert action in desguise}
 
 Combining the definitions and results of this section with decomposition \eqref{DecompositionScalarCurvature} of the higher-dimensional scalar curvature, we can write the Einstein-Hilbert action of a submersive metric $g_P$ as
 \beq \label{GaugedSigmaModelAction}
\int_P \, R_{g_\PPP} \, \vol_{g_\PPP} \ = \ \int_P \, \Big[\, R_{g_\MMM} \, + \, R_{g_\KKK} \, - \, \frac{1}{4}\, |F_A|^2 \, - \,  \frac{1}{4}\,  |\dd^A g_K|^2  \ + \  |D^A \, (\vol_{g_\KKK})|^2 \, \Big] \,\vol_{g_\PPP} \ . 
\eeq
To derive this expression we have also used \eqref{DivergenceN} and ignored the integral over $P$ of the total divergence term. The definition of the squared-norms of the covariant derivatives are the natural ones, given the definition of the covariant derivatives themselves presented before. The right-hand side of \eqref{GaugedSigmaModelAction} broadly resembles the action of a four-dimensional gauged sigma-model for the maps $g_K$ with a potential term $-R_{g_\KKK}$ and a gravity component $R_{g_\MMM}$. The difference is the presence of the additional term $|D^A \, (\vol_{g_\KKK})|^2$ and the fact that the integral is over $P$, not over $M$. These differences can disappear in special situations where we consider a smaller subset of submersive metrics as the domain of the action functional. For instance, when $K$ is an unimodular Lie group and we define the functional only on the set of submersive metrics on $M\times K$ that are invariant under left group multiplication on $K$ and have fibres of unit volume. In this case the gauge group of the sigma-model can be taken to be $K$, instead of $\text{Diff} (K)$; the maps $g_K$ have values in the finite-dimensional space of left-invariant metrics of unit volume on $K$, instead of having values on the whole $\MM(K)$; and after dimensional reduction by fibre integration the functional  \eqref{GaugedSigmaModelAction} becomes the traditional action of four-dimensional gravity plus a gauged sigma-model on $M$.

\newpage

\section{Dynamical models on \texorpdfstring{$M_4 \times K$}{MxK}}       
\label{SectionDynamicalModels}

\subsection{Lagrangian densities}

The purpose of this short section is to bring together previous work and write down the Lagrangian density whose dynamics will be studied subsequently. According to  \eqref{DecompositionScalarCurvature}, in a Riemannian submersion the higher-dimensional scalar curvature can be written as  
\[   
R_{g_\PPP} \ = \ R_{g_\MMM} \ + \ R_{g_\KKK} \ -\  |\FF|^2 \ - \ |S|^2 \ - \ |N|^2 \ - \ 2\, \check{\delta} N    \ .
\]
From \eqref{DivergenceN}, the function $\check{\delta} N$ can be expressed as combination of the norm $|N|^2$ and the divergence $\divergence_{g_\PPP}(N)$, so an alternative formula for the scalar curvature on $P$ is
\bal \label{DecompositionScalarCurvature2}
R_{g_\PPP} \ &= \ R_{g_\MMM} \, + \, R_{g_\KKK} \, - \, |\mathcal{F}|^2 \, - \, |S|^2  \, + \,  |N|^2 \, +\, 2\,\divergence_{g_\PPP}(N)  \nonumber \linebr
&= \ R_{g_\MMM} \, + \, R_{g_\KKK} \, - \, |\mathcal{F}|^2 \, - \, |\mathring{S}|^2  \, + \, \Big(\,1 - \frac{1}{k} \, \Big) \, |N|^2 \, +\, 2\,\divergence_{g_\PPP}(N)  \ .
\end{align}
The last equality used identity \eqref{NormTracelessFundamentalForm} between the squared-norm of $S$ and that of its traceless part $\mathring{S}$. When included in the higher-dimensional Einstein-Hilbert action, each one of these components will have a different role, or interpretation, in terms of four-dimensional physics. The term $R_{g_\MMM}$ generates four-dimensional gravity; $|\mathcal{F}|^2$ is the kinetic term for the Yang-Mills fields; $|\mathring{S}|^2$ and $|N|^2$ are the kinetic terms for the scalar fields that describe how the internal metric $g_K$ varies from fibre to fibre, with $|\mathring{S}|^2$ associated to the variations that preserve the internal volume form and $|N|^2$ associated to the variations of $\vol_{g_\KKK}$; the opposite $-R_{g_\KKK}$ plays the role of a classical potential for those scalar fields; finally, the total divergence $\divergence_{g_\PPP}(N)$ can be ignored in the action if standard boundary conditions are assumed on $M$.

One point that should be made is that, once the higher-dimensional action $\mathcal{E} (g_P)$ is restricted to the domain of submersive metrics on $M \times K$, instead of all possible metrics on this space, each of the previously described components gets a life of its own, independent of the higher-dimensional scalar curvature. By this we mean that $R_{g_\KKK}$, $|\mathcal{F}|^2$, $|\mathring{S}|^2$ and the remaining terms are all natural geometric functions on $M \times K$ that can be included or not in an action functional for $g_P$. In the gauged sigma-model interpretation of submersive metrics, each one of those terms is $\text{Diff} (K)$-gauge equivariant, as exemplified in \eqref{TransformationSDiffeo}, and so produces a $\text{Diff} (K)$-gauge invariant term in the functional $\mathcal{E} (g_P)$, as exemplified in \eqref{TransformationIntSDiffeo}. Thus, once the domain of the functionals is restricted to submersive metrics, there are natural generalizations of the Einstein-Hilbert action on $M \times K$, namely any linear combination of the components $R_{g_\MMM}$, $|\mathcal{F}|^2$, $ |\mathring{S}|^2$, etc., not necessarily with the coefficients that appear in the curvature \eqref{DecompositionScalarCurvature2}.

In these notes we will not analyze this general action functional for submersive metrics. Instead, we will restrict ourselves to studying a modest generalization of the Einstein-Hilbert action by allowing an arbitrary coefficient of the mean curvature term $|N|^2$. This is the kinetic term for the fields that control the volume of internal space, which are directly related to the gauge couplings and play an important role in Kaluza-Klein models. Thus, here we will consider the higher-dimensional action
\bal     \label{ActionP}
\mathcal{E} (g_P) \ &:= \  \frac{1}{2\, \kappa_P}\, \int_P \, \left( R_{g_\PPP} \, - \, \lambda \, |N|^2 \, - \, 2\, \Lambda \right) \ \vol_{g_\PPP}  \linebr
&= \  \frac{1}{2\, \kappa_P}\, \int_P \Big(\,R_{g_\MMM} \, + \, R_{g_\KKK} \, - \, |\mathcal{F}|^2 \, - \, |\mathring{S}|^2  \, - \, \Big(\, \lambda - 1 + \frac{1}{k} \, \Big) \, |N|^2 \, - \, 2\, \Lambda \,  \Big) \ \vol_{g_\PPP}  \nonumber   \ ,
\end{align}
where $\kappa_P$, $\lambda$ and $\Lambda$ are real constants and in the last equality we have ignored the total divergence term. When $\lambda=0$ this definition reduces to the traditional Einstein-Hilbert action, which remains the most natural one because it is also defined for non-submersive metrics on $M \times K$.

In a somewhat unrelated drift, before ending this section we note that there is a nice combination of the scalar curvature $R_{g_\PPP}$ with the two functions $|N|^2$ and $\divergence_{g_\PPP}(N)$ that satisfies a particularly simple rule of transformation under Weyl rescalings of the submersive metric. The combination is 
\beq   \label{DensityP2}
W({g_P}) \ :=  \  R_{g_\PPP} \ - \ \frac{(m+k-1)(m+k-2)}{k^2} \; \big| N \big|^2_{g_\PPP} \, - \, 2\, \frac{(m+k-1)}{k} \ \divergence_{g_\PPP}  N  \ ,
\eeq
where $m$ and $k$ denote the dimensions of $M$ and $K$, respectively. Indeed, if $\Omega: \, P \to \mathbb{R}^+$ is any positive function with constant values on the fibres and $\tilde{g}_P :=  \Omega^2\, g_P$ is the corresponding Weyl transformation, it is shown in appendix \ref{AppendixWeylRescalings} that the function $W$ calculated for the rescaled metric satisfies the simple relation
\[
W({\tilde{g}_P}) \ = \ \Omega^{-2} \ W({g_P}) \ .
\]
This contrasts with the complicated behaviour of the curvature $R_{g_\PPP}$ under the same Weyl transformation. 
It is clear from the results of appendix \ref{AppendixWeylRescalings} that this simple transformation rule will also hold for any linear combination of $W({g_P})$ with the scalar functions $|\mathring{S}_{g_\PPP}|^2$ and $R_{g_\KKK}$ on the higher-dimensional space $P$.

\subsection{Mass of the gauge bosons}    
\label{SectionMassGaugeBosons}

The calculations leading to a mass formula for the fields $A^a_\mu$ on $M$ mimic, in every essential way, the calculations usually performed in the case of the electroweak gauge fields of the Standard Model \cite{Weinberg67, Weinberg, Hamilton}. One works in the approximation where the gauge fields $A^a_\mu$ are small, close to their vanishing vacuum value, and where the internal metric $g_K$ (the restriction of $g_P$ to the fibres $K$) is constant as one moves across the fibres and equal to the vacuum metric $g_K^\Szero$. 

Start by combining definition \eqref{ActionP} of the action $\mathcal{E} (g_P)$ with the results of section \ref{SectionScalarCurvature}, which express the different components of $R_{g_\PPP}$ in terms of the data $(g_M, A, g_K)$, equivalent to the submersive metric $g_P$. After fibre-integration, one finds that the terms of the four-dimensional Lagrangian density that depend on the field $A^a_\mu$ are proportional to
\bal  \label{GaugeTerms}
\frac{1}{4} \; g_M^{\mu \nu} \, g_M^{\sigma \rho}   \ (F^a_{A})_{\mu \sigma} \ (F^b_{A})_{\nu \rho}  \ B_{ab} \ + \ g_M^{\mu \nu} \ A^a_\mu \; D_{\nu a}     \ + \ 
 \ g_M^{\mu \nu} \ A^a_\mu \  A^b_\nu   \ C_{ab}  \ .
\end{align}
The coefficients $B_{ab}$, $C_{ab}$ and $D_{\nu a}$ depend on the metric on the internal space and are given by the fibre-integrals
\bal \label{MassCalculationCoefficients}
B_{ab}  \ &:= \ \int_K   g_K^\Szero (e_a ,  e_b  ) \ \vol_{g_\KKK^\SSzero}  \nonumber \linebr
D_{\nu a} \ &:= \   \frac{1}{2} \; \int_K  \left[\; \left\langle \Lie_{e_a}\, g_K^\Szero,  \ \Lie_{X_\nu}\, g_P^\Szero \right\rangle \ +\ 4 \, (1-\lambda)\; \Lie_{X_\nu}  (\divergence_{g_\KKK^\SSzero} e_a)  \; \right]  \, \vol_{g_\KKK^\SSzero}  \nonumber \linebr
C_{ab}  \ &:= \   \frac{1}{4} \int_K \left[ \; \left\langle \Lie_{e_a}\, g_K^\Szero,  \ \Lie_{e_b}\, g_K^\Szero \right\rangle  \ + \ 4\, (\lambda - 1)\, ( \divergence_{g_\KKK^\SSzero} e_a) \,   (\divergence_{g_\KKK^\SSzero} e_b) \; \right]  \, \vol_{g_\KKK^\SSzero}  \ .
\end{align}
Working with the Levi-Civita connection $\nabla^M$ on $M$ and ignoring total derivatives, the first variation of \eqref{GaugeTerms} with respect to $\delta (A^a)^\mu$ leads to the equations of motion
\[
g_M^{\mu \nu} \, g_M^{\sigma \rho} \; (\nabla^M_\nu F^a_{A})_{\mu \sigma}  \; B_{ab} \ - \ 2 \;  g_M^{\mu \rho} \;  A^a_\mu \; C_{ab}  \ = \ 0 \ .
\]
When the vector fields $e_a$ on the internal space $K$ are chosen to simultaneously diagonalize the quadratic forms $B_{ab}$ and $C_{ab}$, the equations of motion reduce to
\beq  \label{EquationMotionGaugeField}
g_M^{\mu \nu} \; (\nabla^M_\nu F^a_{A})_{\mu \sigma} \ - \ 2 \; \frac{C_{aa} }{B_{aa}}\; A^a_\sigma  \ = \ 0 \ .
\eeq
 The usual arguments using the Lorenz condition $\partial^\mu A_\mu^a = 0$ (e.g. see \cite[section 2.7]{MS}) then say that, to first order in the fields, these equations can be simplified to the Klein-Gordon equation for fields on $M_4$ of squared-mass equal to $2 \; C_{aa}  / B_{aa}$. So we get a formula for the classical mass of the gauge fields:
\begin{equation} \label{MassGaugeBosons}
{\setlength{\fboxsep}{3\fboxsep} \boxed{
\left(\text{Mass} \ A_\mu^a \right)^2 \ = \ \frac{ \int_K \left[ \; \left\langle \Lie_{e_a}\, g_K^\Szero,  \; \Lie_{e_a}\, g_K^\Szero \right\rangle  \, + \, 4\, (\lambda -1)\, (\divergence_{g_\KKK^\SSzero} e_a)^2 \; \right] \vol_{g_\KKK^\SSzero} }{ 2 \int_K   g_K^\Szero (e_a ,  e_a  ) \ \vol_{g_\KKK^\SSzero} } \ .
}}
\end{equation}
Thus, the classical mass of the field $A^a_\mu$ is determined by the geometrical properties of the vector field $e_a$ with respect to the vacuum metric $g^{\Szero}_K$ on the internal space. The relevant quantities are the $L^2$-norm of $e_a$, the divergence of $e_a$ and the Lie derivative of the vacuum metric in the direction of $e_a$. For instance, if the vector field $e_a$ is Killing with respect to $g^{\Szero}_K$, then both the divergence and the Lie derivative vanish, so the fields $A^a_\mu$ will be massless for this particular value of the index $a$. In a sense, the classical mass of a gauge boson is a measure of how much the internal metric changes along the flow generated by the corresponding internal vector field. 

The right-hand side of the mass formula is manifestly non-negative for all vector fields $e_a$ with vanishing divergence on $(K, g_K^\Szero)$, i.e. for all vector fields that preserve the vacuum volume form $\vol_{g_K^\SSzero}$ on the internal space. However, if we associate gauge fields also to vector fields with non-vanishing divergence on $K$, we could end up with negative masses, unless the freedom in the parameter $\lambda$ of the Lagrangian is used to prevent this.

In general, on a Riemannian manifold the Lie derivative $\Lie_{V}g$ can be decomposed as
\[
\Lie_{V}g  \ = \ \frac{2}{n} \; \divergence_g (V) \; g \ + \ \eta_V  \ ,
\]
where $n$ is the dimension of the manifold and $ \eta_V$ is a symmetric 2-tensor satisfying the traceless condition $\sum_j \eta_V (v_j , v_j) = 0$, where $\{ v_j\}$ denotes a local, $g$-orthonormal trivialization of the tangent bundle.  Taking the norm of both sides of the equation, a calculation analogous to \eqref{TracelessFundamentalForm} and \eqref{NormTracelessFundamentalForm} leads to
\begin{equation} \label{NormInequality}
\left\langle \Lie_{V}\, g,  \; \Lie_{V}\, g \right\rangle  \ - \  \frac{4}{n} \, ( \divergence_g  V )^2 \ = \ \left\langle \eta_V,  \, \eta_V \right\rangle  \  \geq \ 0 \ ,
\end{equation}
with equality only if $\eta_V = 0$, i.e. only if $V$ is a conformal Killing vector field with respect to $g$. Applying \eqref{NormInequality} to the numerator of the mass formula we obtain that
\begin{equation*} \label{NormInequality2}
\left\langle \Lie_{e_a}\, g_K^\Szero,  \; \Lie_{e_a}\, g_K^\Szero \right\rangle  \, + \, 4\, (\lambda -1)\, (\divergence_{g_\KKK^\SSzero} e_a)^2  \ = \ \left\langle \eta_{e_a},  \, \eta_{e_a} \right\rangle \,  + \, 4\, \Big(\lambda -1 + \frac{1}{k} \Big) (\divergence_{g_\KKK^\SSzero} e_a)^2 \  .
\end{equation*}
Thus when $\lambda  >  1 -  \frac{1}{k}$ it is clear that all gauge fields have non-negative mass, with the massless case occurring precisely when the $A^a_\mu$ are associated to Killing vector fields on the internal space. When the parameter $\lambda$ is equal to $1  -   \frac{1}{k}$ all the masses are still non-negative, but the massless case occurs in the slightly more general situation of gauge fields associated to conformal Killing vector fields on $K$. For $\lambda  <  1 -  \frac{1}{k}$ all the gauge fields associated with zero divergence vector fields still have non-negative mass, but for example any gauge field associated with a conformal Killing vector field that is not Killing on $(K, g_K^\Szero)$ (if such fields exist) will have negative mass.

The mass formula for $A^a_\mu$ is manifestly invariant under a rescaling of the internal vector field $e_a$ by a positive constant. This is equivalent to a rescaling of the gauge field $A^a_\mu$. On the other hand, if one considers a constant rescaling of the internal vacuum metric, $\tg_K^\Szero = \omega^2\, g_K^\Szero$ with $\omega^2$ in $\mathbb{R}^+$, it is easy to check that \eqref{MassGaugeBosons} rescales as
\beq \label{NaiveMassScaling}
\left(\text{Mass} \ A_\mu^a \right)^2_{\tg_K^\SSzero} \  = \  \omega^{-2} \left(\text{Mass} \ A_\mu^a \right)^2_{g_K^\SSzero} \ .
\eeq
This follows from the observation that, for any vector field $V$ on $K$, the quantities $\left\langle \Lie_{V}\, g,  \; \Lie_{V}\, g \right\rangle_g$ and $\divergence_g V$ are both invariant under a constant rescaling of $g$, which itself follows from definition \eqref{DefinitionInnerProduct} of the inner-product and formula \eqref{IdentityDivergence} for the divergence. Thus, reducing the size of the vacuum internal space leads to an increase of the classical masses of all the massive gauge bosons.

We would like to end this section with two cautionary comments. Firstly, the traditional arguments that lead to the identification of the coefficient $2 \; C_{aa}  / B_{aa}$ in equation \eqref{EquationMotionGaugeField} as the mass of the gauge field are well-justified only in the case where $(M, g_M)$ is Minkowski space. In curved backgrounds the concept of mass of a gauge field is less clear-cut and should be used with caution. Secondly, note that the mass formula \eqref{MassGaugeBosons} was derived from the terms \eqref{GaugeTerms} of the four-dimensional Lagrangian density, which themselves were obtained by dimensional reduction (through fibre-integration) of the corresponding terms in the higher-dimensional action $\mathcal{E} (g_P)$, written in \eqref{ActionP}. However, a more complete of analysis of that dimensional reduction shows that it leads to a gravity Lagrangian for $g_M$ that is not in the Einstein frame, but is in the alternative Jordan frame. Since Lagrangians in the Jordan frame are generally disfavoured for a good physical interpretation of four-dimensional theories \cite{FGN}, further ahead we will redefine the fields in order to obtain a higher-dimensional action that produces a 4D Lagrangian in the Einstein frame, after dimensional reduction. After this process is complete the mass formula for the gauge fields will be slightly changed, with the appearance of an overall factor related to the total volume of $(K, g_K^\Szero)$. The new mass formula is calculated in appendix \ref{AppendixBosonsMassEinsteinFrame} and the result is stated in \eqref{MassGaugeBosons5}. It has a scaling behaviour under constant rescalings of $g_K^\Szero$ distinct from \eqref{NaiveMassScaling}. Namely, formula \eqref{MassGaugeBosons5} implies that the bosons' squared-mass scales more strongly as $\omega^{-2- 2k/(m-2)}$, where $k$ and $m$ are the dimensions of $K$ and $M$, respectively, instead of scaling as $\omega^{-2}$. In contrast, the {\it relative} masses of the different gauge bosons are the same in \eqref{MassGaugeBosons} and \eqref{MassGaugeBosons5}.

\subsection{Stability of product Einstein solutions}
\label{StabilityEinsteinSolutions}

Submersive metrics on $M\times K$ have degrees of freedom associated to the spacetime metric $g_M$, the gauge fields and the internal metric $g_K$. Decomposing the higher-dimensional scalar curvature, we recall, allowed us to define a generalized Einstein-Hilbert action
\bal \label{ActionP2}
\mathcal{E} (g_P) \; = \;  \frac{1}{2\, \kappa_P}\,  \int_P  \Big( R_{g_\MMM}  +  R_{g_\KKK}  -  |\mathcal{F}|^2  -  |\mathring{S}|^2   +    \Big(1 - \lambda - \frac{1}{k} \Big)  \, |N|^2  -  2\, \Lambda\,  \Big) \, \vol_{g_\PPP} \ .  
\end{align}
In this section we study configurations with vanishing gauge fields, so vanishing tensor $\mathcal{F}$.  Starting with a product Einstein metric $g^e_M + g^e_K$ on $P$, which is solution of the higher-dimensional equations of motion, we study the second variation of the action functional under small perturbations of the internal metric $g_K$. The perturbations can be both through Weyl rescalings or TT-deformations. The purpose is to calculate the mass of the four-dimensional perturbation fields and better understand the stability properties of the initial product Einstein solution. The calculations replicate very standard methods used to study the stability of general Einstein metrics under the Einstein-Hilbert action, described for instance in \cite{Besse, Kro}, applied to the case of the action \eqref{ActionP2} for submersive metrics near product Einstein solutions. The main takeaway is that while there is at most a finite number of unstable TT-deformations of $g^e_K$, the number of negative-mass perturbation modes associated to Weyl recalings of $g^e_K$ depends on the value of the constant $\lambda$. For small $\lambda$ there is just one such mode, corresponding to Weyl rescalings that are constant in the $K$-direction. For $\lambda > 1 - 1/k$ there will be infinite unstable modes.

Let us perturb the product Einstein metric $g^e_M + g^e_K$ by a variation with parameter $t$:
\beq \label{MetricPerturbation}
g_P^t \ := \ g_M^e \,+\, g_K^t \ =\  g_M^e \,+\, g_K^e\, + \, t\, \big(\, h \,+\, f g_K^e \, +\,  \Lie_E \, g^e_K  \, \big) \ .
\eeq
Here $h$ is a transverse, traceless variation of the fibre metrics in the vertical direction, so that $h$ restricted to each fibre is a TT-tensor on $K$, although this tensor may vary from fibre to fibre; $f$ is a real smooth function on $M\times K$; the term $ \Lie_E \, g^e_K$ denotes the Lie derivative of the Einstein metric along some vertical vector field $E$ on $P$, so is a variation of the metric through an infinitesimal diffeomorphism of the fibres. This represents the most general variation of the fibre metrics $g_K$, as follows from standard results \cite{Besse}.

The variation of $g_M^e \,+\, g_K^e$ by fibre diffeomorphisms does not change the total Einstein-Hilbert action nor the integral of its components $|S|^2$ and $|N|^2$. This was observed in \eqref{TransformationSDiffeo} and \eqref{TransformationIntSDiffeo}, for example. So we may as well take $E=0$ and ignore the term $\Lie_E \, g^e_K$ when calculating the second variation of the total action.

For vanishing gauge fields, formula \eqref{ExplicitTensorS2} applied to the variation $g_P^t$ implies that the traceless part of the fibres' second fundamental form satisfies
\begin{equation*}
 2\,\, g_P^t \big(\, (\mathring{S}^{\, g_\PPP^t})_U V, \,  X \, \big) \ |_{E=0} \ = \ - \ t\, (\Lie_{X} \, h ) (U, \, V ) 
\end{equation*}
for all vertical vector fields $U$ and $V$ on $P$ and for all fields $X$ in $M$. Similarly, from \eqref{ExplicitN2} it follows that for vanishing gauge fields
\begin{equation*}
 2\,\, g_P^t (  N^{\, g_\PPP^t}, \,  X) \ |_{E=0} \ = \  - \; t\, k\,\, \Lie_{X}  f \ .
\end{equation*}
Taken together these formulae imply that the total variation of the squared-norms is
\bal
\int_P \, |\mathring{S}|^2_{g_\PPP^t}  \; \vol_{g_\PPP^t}  \ &= \ \frac{t^2}{4}\, \int_P\,  g_M^{\mu\nu} \,  \langle\, \Lie_{X_\mu} h ,  \, \Lie_{X_\nu} h \,  \rangle_{g_\KKK^e}  \ \vol_{g_\PPP^e}  \ + \ O(t^3) \nonumber \linebr
\int_P \, |N|^2_{g_\PPP^t }\; \vol_{g_\PPP^t} \ &= \  \frac{t^2 \, k^2}{4}\, \int_P \,   g_M^{\mu\nu} \, (\partial_\mu f)  \, (\partial_\nu f )  \ \vol_{g_\PPP^e}  \ + \ O(t^3) \ .
\end{align}
Note in passing that these last formulae are valid even for non-vanishing $E$, because integration kills all variations through diffeomorphisms. In any case we conclude that the expansion of the components $|\mathring{S}|^2$ and $|N|^2$ of the higher-dimensional scalar curvature produces dynamical terms in the action for the perturbation fields $h$ and $f$, respectively, involving their derivatives in the four-dimensional  spacetime directions. 

In contrast, the fibres' scalar curvature $R_{g^t_\KKK}$ depends on the deforming tensors $h$, $f$ and their derivatives in the vertical, $K$-directions, with no derivatives in the spacetime directions. So one recognizes that the fibre integral
\[
\int_K R_{g^t_\KKK} (h, \, f) \; \vol_{g_\KKK^t} \ \ , 
\]
which is a scalar function on $M$, plays the role of a spacetime potential for the perturbation fields. The integral of the internal scalar curvature is invariant under the group of diffeomorphisms of $K$, so the variations of the vacuum metric in the directions tangent to diffeomorphisms preserve this potential. The fields $h$ and $f$, on the other hand, represent variations of $g_K^e$ that are transverse to diffeomorphisms \cite{Besse} and may change the value of the potential.

The classical masses of the perturbation fields are essentially the eigenvalues of the Hessian of the potential evaluated at the initial solution. They can also be extracted from the linearized equations of motion of the fields. So we should start by considering the potential-like components of the higher-dimensional Einstein-Hilbert action, 
\[
\int_P  \Big( R_{g_M^e} \, + \, R_{g^t_\KKK} \, - \, 2\, \Lambda\,  \Big) \, \vol_{g_\PPP^t} \ = \ \int_M   \mathcal{V}(t)   \ \vol_{g^e_M}  \ ,
\]
and calculate the second variation of the spacetime potential
\[
 \mathcal{V}(t) \ := \  \int_K  \Big( \, R_{g^t_\KKK} \, - \, 2\,  \Lambda\, + R_{g_M^e} \  \Big) \, \vol_{g_\KKK^t} \ = \ \mathcal{V}(0) \; + \; \frac{t^2}{2}\; \mathcal{V}''(0)  \; + \; O(t^3) \ .
\]
There is no linear term in $t$ because $g_M^e + g_K^e$ is a solution of the higher-dimensional equations of motion and $g_K^e$ is an Einstein metric on $K$. The second variation of the Einstein-Hilbert functional $\mathcal{V}(t)$ is very well-known \cite{Besse, Kro}. It can be written as
\[
\mathcal{V}''(0) \ = \ \frac{1}{2}\, \int_K  \left\{ - \, \langle\, h,\,  \nabla^\ast \nabla h -  2\, \mathring{R}_{g_\KKK^e} \, h \, \rangle_{g_\KKK^e} \; + \;  (k-2) \Big[ (k-1)\, f\, \Delta_{g^e_\KKK} f \; - \; R_{g_\KKK^e} \, f^2   \Big]   \right\}  \vol_{g_\KKK^e} \ .
\]
Here $\Delta_{g^e_\KKK}$ denotes the positive Laplacian on $K$ and, in the case of an Einstein manifold, the operator in the first term essentially coincides with the Lichnerowicz Laplacian $\Delta^{g_\KKK^e}_L$ acting on symmetric 2-tensors on $K$,
\[
\nabla^\ast \nabla h \ - \ 2\, \mathring{R}_{g_\KKK^e} \, h \ = \  \Delta^{g_\KKK^e}_L\, h \; - \; \frac{2}{k} \, R_{g_\KKK^e}  \, h \ .
\]
We have also used the relations 
\beq \label{CurvatureEinsteinProduct}
R_{g_\KKK^e} \ = \ \frac{k}{m} \, R_{g_M^e}   \ = \    \frac{2\,  \Lambda\, k}{m + k -2} \ 
\eeq
to eliminate $R_{g_\MMM^e}$ and $\Lambda$ from the end result and write them in terms of $R_{g_\KKK^e}$. These relations follow from the fact that $g_M^e + g^e_K$ is an Einstein metric on $M\times K$ with higher-dimensional cosmological constant $\Lambda$. The letters $m$ and $k$ denote the dimension of $M$ and $K$, respectively.
In summary, up to second order in $t$ the action \eqref{ActionP2} is
\[
\mathcal{E} \big(\, g_M^e +  g_K^t \big)  \; = \; \mathcal{E} \big(\, g_M^e +  g_K^e \big) \ + \ \frac{t^2}{2}\, \left( I_f \; + \; I_h  \right)  \; + \; O(t^3)
\]
with
\bal \label{PerturbativeLagrangians}
I_h \ &= \  \frac{1}{2}\,  \int_P \left\{ - \, g_M^{\mu \nu}\,  \left\langle\Lie_{X_\mu}  h ,  \, \Lie_{X_\nu}  h \right\rangle_{g_\KKK^e} \, - \,  \frac{1}{k}  \left\langle \, h,\,  k\,  \Delta^{g_\KKK^e}_L\, h \; - \; 2 \, R_{g_\KKK^e}  \, h  \, \right\rangle_{g_\KKK^e} \right\}  \vol_{g_\PPP^e}         \linebr
I_f \ &= \ \frac{1}{2}\, \int_P \left\{ k(k - \lambda k - 1)  \; g_M^{\mu \nu} \,  (\partial_\mu f)  \, (\partial_\nu f )  \; + \;  (k-2)  \Big[ (k-1)\, \Delta_{g^e_\KKK} f \; - \; R_{g_\KKK^e} \, f \Big] \, f \right\} \vol_{g_\PPP^e}    \nonumber 
\end{align}
Through variation it leads to the linearized equations of motion of the perturbation fields:
\bal \label{EquationsMotionPerturbation}
&g_M^{\mu\nu} \, \nabla_\mu (\Lie_{X_\nu}  h) \ - \ \Delta^{g_\KKK^e}_L\, h \ + \ \frac{2}{k} \, R_{g_\KKK^e}  \, h \  =  \ 0    \linebr
& k\,(k - \lambda k - 1)\, g_M^{\mu\nu} \, \nabla_\mu (\partial_\nu  f) \; - \; (k-2 ) \Big[ (k-1)\, \Delta_{g^e_\KKK} f \; - \;  R_{g_\KKK^e} \, \, f \Big] \ = \ 0 \ . \nonumber 
\end{align}
These are equations on the total space $M \times K$. Recall that the fields $f$ and $h$ represent deformations of the higher-dimensional metric around a classical solution $g_M^e + g^e_K$. These deformations change the internal metric $g_K$ in directions transverse to diffeomorphisms, while leaving the spacetime metric and the Riemannian product structure unchanged. For vanishing $\lambda$, the equations above are just the linearization of the Einstein equations on $M \times K$ in the direction of these deformations. There is no equation for the vector field $E$ of \eqref{MetricPerturbation} because diffeomorphisms preserve the Einstein equations, so the linearization is trivial in that direction. 

To obtain equations of motion in four dimensions from equations \eqref{EquationsMotionPerturbation} on $M\times K$, one should decompose the function $f$ as a sum of eigenfunctions $\omega_n$ of the scalar Laplacian $\Delta_{g^e_\KKK}$ on $K$. In parallel, one should decompose the field $h$ of TT-tensors as a sum of eigentensors $\eta_n$ of the Lichnerowicz Laplacian $ \Delta^{g_\KKK^e}_L$. Both these Laplacians are elliptic, self-adjoint operators on the compact $(K, g_K^e)$ with respect to the $L^2$-inner-product of scalar functions and TT-tensors, respectively. They have finite-dimensional eigenspaces with real eigenvalues that form a discrete, increasing sequence in the real line that accumulates only at $+\infty$ \cite{Bourguignon, Kro}. We write the decompositions as
\bal \label{eigenexpansion}
h(x,y) \ &=\ \sum\nolimits_{n= 0}^{\infty} \ h^n(x) \; \eta_n (y)    \qquad {\rm with} \quad \Delta^{g_\KKK^e}_L  \eta_n \, = \,\sigma_n \,  \eta_n  \nonumber \linebr
f(x,y) \ &=\ \sum\nolimits_{n=0}^{\infty} \ f^n(x) \; \omega_n(y)  \qquad {\rm with} \quad \Delta_{g^e_\KKK} \omega_n \, = \,   \tau_n \,  \omega_n \  \ .
\end{align}
Here $x$ denotes the coordinates on $M$ and $y$ denotes the coordinates on $K$, so the coefficients of the expansions are real functions on four-dimensional spacetime. We also assume that the real eigenvalues $\sigma_n$ and $\tau_n$ are ordered in non-decreasing sequences. Substituting \eqref{eigenexpansion} into the equations of motion of $h$ and $f$ we obtain Klein-Gordon equations on $M$ for the coefficient fields $ h^n(x)$ and $f^n(x)$,
\bal \label{KGEquationModes1}
&g_M^{\mu\nu} \, \nabla_\mu (\partial_\nu  \, h^n) \  +\ \Big(\, \frac{2}{k} \, R_{g_\KKK^e} \,- \, \sigma_n \, \Big)\, h^n  \  =  \ 0    \linebr
& k\,(k - \lambda k - 1)\, g_M^{\mu\nu} \, \nabla_\mu (\partial_\nu \, f^n) \; - \; (k-2 )  \left[ (k-1)\, \tau_n \, - \, R_{g_\KKK^e}  \right]  f^n  \ = \ 0 \ . \label{KGEquationModes2}
\end{align}
The squared-mass parameters of the fields are then:
 \begin{empheq}[box=\widefbox]{align} \label{MassPerturbationModes}
\left(\text{Mass} \ h^n \right)^2  \  &= \   \sigma_n \; - \;  \frac{2}{k} \, R_{g_\KKK^e} \nonumber  \linebr
\left(\text{Mass} \ f^n \right)^2  \  &= \   \frac{(k-2 )  \left[ (k-1)\, \tau_n \, - \, R_{g_\KKK^e}  \right] }{k\, (k - \lambda k - 1)} \ .
\end{empheq}
Since the eigenvalues of the Lichnerowicz Laplacian form a discrete, increasing sequence that accumulates only at positive infinity, we know that the number of unstable modes $h^n$ will always be finite, if non-zero. Furthermore, for an Einstein metric $g_K^e$ with positive scalar curvature that is not a standard sphere, the non-zero eigenvalues of the scalar Laplacian always satisfy \cite{Obata, Kro}
\[
(k-1) \, \tau_n \; > \; R_{g_\KKK^e}  \qquad {\rm{for}}  \ \  n \,  \geq \, 1 \ .
\]
This means that for $k>2$ and $n  \geq 1$ the squared-mass of the modes $f^n$ has the same sign as the factor $(k - \lambda k - 1)$. On the other hand, since the scalar Laplacian has $\tau_0 = 0$, for $k> 2$ and $R_{g_\KKK^e} > 0$ we recognize that the squared-mass of the mode $f^\Szero$ has the opposite sign to the factor $(k - \lambda k - 1)$. So for an Einstein metric $g_K^e$ with positive scalar curvature it is not possible to have non-negative squared-masses for all the modes $f^n$ unless $k=2$, in which case all the modes will be massless, or unless $\lambda= (k - 1)/k$, in which case all these modes are non-dynamical on spacetime and equation \eqref{KGEquationModes2} only has the zero solution. 

Thus, when the constant $\lambda$ is zero or small, corresponding to the traditional Einstein-Hilbert action on $M\times K$, only the mode $f^\Szero$ has negative squared-mass among all the perturbation modes $f^n$. 

Observe also that the kinetic term for the field $f$ in the second integral of \eqref{PerturbativeLagrangians} appears with a flipped, positive sign unless $\lambda \geq (k - 1)/k$. Such a positive sign would violate the requirements of the weak energy principle for coupling the scalar curvature $R_{g_M}$ to matter fields in the four-dimensional Lagrangian \cite{HE}. However, \eqref{PerturbativeLagrangians} is a truncation of the full higher-dimensional Lagrangian and, after fibre integration, it produces a four-dimensional Lagrangian in the Jordan frame, not in the standard Einstein frame where the weak energy principle is stated. A transformation to the Einstein frame can change the sign of scalar kinetic terms, as will be seen in detail in the next section. So a definite evocation of the weak energy principle would require additional care in these circumstances.

\subsection{Lagrangian in the Einstein frame}
\label{EinsteinFrame}

Dimensional reduction of our action \eqref{ActionP} produces a Lagrangian in four dimensions that is not in the Einstein frame, but is in the alternative Jordan frame. This means that the gravity component of the Lagrangian does not appear in the traditional guise $R_{g_\MMM} \sqrt{-g_M}$, but instead appears multiplied by a scalar factor that depends on the spacetime coordinate. This scalar factor is a hallmark of Kaluza-Klein theories. It is related to the volume of the internal space over each spacetime point, which in general can vary along $M$.

Since Lagrangians in the Jordan frame are generally disfavoured for a good physical interpretation of the four-dimensional theory \cite{FGN}, in this section we will use a Weyl rescaling to redefine what is understood to be the physical, four-dimensional metric $g_M$. After this redefinition, dimensional reduction does produce a 4D Lagrangian in the Einstein frame. This technique to transform a Lagrangian from the Jordan to the Einstein frame is very standard \cite{WessonOverduin, FGN}. In the case of action \eqref{ActionP} the computations are somewhat evolved because one first has to obtain the transformation rules of the norms $|\FF|^2$, $|S|^2$ and $|N|^2$ under the rescaling, which can then be combined with the standard transformation of the curvature $R_{g_\MMM}$. These calculations are summarized in appendix \ref{AppendixWeylRescalings}. 

In order to study the dynamics of the rescaling field $\phi$---which, recall, measures the volume of the internal space---we also want its kinetic term to appear with the standard Klein-Gordon normalization in the dimensionally-reduced Lagrangian. This requires additional care in the definition of $\phi$. The main output of this section is the expression of the higher-dimensional action in the Einstein frame registered in \eqref{ActionScalarField} and \eqref{ExpandedLagrangianScalarField}.

To describe that result, start by recalling that a submersive metric $g_P$ on the bundle $M\times K \rightarrow M $ defines a metric $g_M$ on the base, gauge fields on the base and a family of metrics $g_K$ on the internal spaces $K$, one metric for each fibre. A distinctive degree of freedom of the internal metrics is their overall scale, and how it varies from fibre to fibre. This is the variable of interest in this section, so let us autonomize it by explicitly writing 
\beq \label{InternalRescaling}
g_K \ =\  a_1 \, e^{- b_1 \phi} \, \bg_K \ ,
\eeq
where the metrics $\bg_K$ on the fibres are arbitrary but have fixed volume over $K$, normalized to a given value $\bk$; the real function $\phi$ on $M$ determines the scale of the fibres; $a_1$ and $b_1$ are positive constants that we will specify later on. So the field $\phi$ is similar to the mode $f^\Szero$ of last section except for the constants. 

Let us also declare that the ``physical" metric $g_M$ over spacetime is not the simple projection of $g_P$ down to $M$, but is instead the rescaled version of the projection that produces a spacetime Lagrangian in the Einstein frame after dimensional reduction. In other words, produces a Lagrangian with a gravitational component that coincides with that of traditional GR. This can be achieved be a redefinition of the form $g_M \rightarrow a_2\, e^{b_2 \phi} \, g_M$. Thus, when writing the submersive metric $g_P$ as a triple in the spirit of \eqref{MetricDecomposition}, the components can be written in terms of the redefined fields as
\beq   \label{MetricPerturbation2}
g_P \ \simeq \ \big( \, a_2\, e^{b_2 \phi} \, g_M \, , \, A \, , \, a_1\, e^{- b_1 \phi} \, \bg_K \, \big) \ .
\eeq
To determine the appropriate value of the constants and to study the dynamics of the field $\phi$ we should write action \eqref{ActionP} in terms of the rescaled variables, that is, we should write
\beq \label{ActionScalarField}
\mathcal{E} (g_P) \ = \ \frac{1}{2\, \kappa_P}\,   \int_P \; \mathcal{L} (\phi, g_M, A, \bg_K ) \;  \vol_{\bg_\PPP} \ , 
\eeq
where $\bg_P$ denotes the unscaled submersive metric on $P$,
\beq  \label{UnscaledMetric}
\bg_P\  \simeq \ \big( \, g_M \, , \, A \, , \, \bg_K\, \big)  \ .
\eeq 
The calculation of the expanded Lagrangian density $\mathcal{L} (\phi, g_M, A, \bg_K )$ is a slightly lengthy one, using the results of appendix \ref{AppendixWeylRescalings}. To write the result in a readable format let us first introduce some notation. The fixed normalization of the metrics $\bg_K$ is defined by a dimensionless, positive constant $\bk \in \mathbb{R}^+$ through the relation
\beq \label{ActionConstants3}
\Vol(K, \, \bg_K)  \  = \  \kappa_M^{-1}\,\kappa_P \, \bk  \ ,
\eeq
where $\Vol(K, \, \bg_K)$ denotes to total volume of the manifold and $\kappa_M$ stands for the Einstein gravitational constant, so $8\pi G c^{-4}$ in the four-dimensional spacetime. Let us also 
define the positive constant
\beq \label{ActionConstants}
\beta \ := \ 2\, \sqrt{\kappa_M} \,  \left[ \frac{m+k-2}{(m-2)k} \right]^{1/2}   \,  \left[1 + \lambda\, k\,  \frac{m-2}{m + k -2} \right]^{-1/2}   \ , 
\eeq
where $m$ and $k$ denote the dimensions of $M$ and $K$, respectively. Here we are assuming that 
\beq \label{condition_lambda}
\lambda \ > \ -\; \frac{m+k-2}{k(m-2)} \ ,
\eeq
otherwise the constant $\beta$ is not well defined. Then with the choices 
\begin{align} \label{ActionConstants2}
a_2 \ &= \ a_1 \ = \ \bk^{ \frac{-2}{m+k-2}}     \linebr
b_2 \ &= \  \frac{k}{m-2} \, b_1 \ = \  \frac{k \, \beta}{m+k-2}  \nonumber 
\end{align}
the expanded Lagrangian can be written as
 \begin{empheqboxed}
 \begin{align} \label{ExpandedLagrangianScalarField}  
 \mathcal{L} (\phi, g_M, A, \bg_K ) \ := \ \frac{1}{\bk} \;  \Big\{&  R_{g_\MMM} \, - \, \kappa_M \, | \dd \phi |^2_{g_\MMM} \, + \, R_{ \bg_\KKK} \, \, e^{\beta \phi} \, - \, 2 \, \Lambda \, a_2\,  e^{b_2 \phi} \, -  \,|\, \FF\,|^2_{ \bg_\PPP} \,  \, e^{- \beta \phi}
 \nonumber \\
 & -\,    |\,\mathring{S}_{ \bg_\PPP}\,|^2_{ \bg_\PPP} \, +\, \Big(1 - \lambda - \frac{1}{k}  \Big)\, |N_{ \bg_\PPP}|^2_{ \bg_\PPP}   \;  \Big\} \ .
\end{align}
\end{empheqboxed}
Now, by definition all the fibres of $(P, \bg_P)$ have the same fixed volume, equal to the normalized volume of $\bg_K$, so after fibre-integration we do get
\[
\frac{1}{2\,\kappa_P\, \bk} \;  \int_P \; \Big( \;  R_{g_\MMM} \, - \, \kappa_M \, | \dd \phi |^2_{g_\MMM} \; \Big) \; \vol_{\bg_\PPP} \ = \  \int_M \; \Big( \;  \frac{1}{2\, \kappa_M} \; R_{g_\MMM} \, - \, \frac{1}{2}  \, | \dd \phi |^2_{g_\MMM}   \; \Big) \; \vol_{g_\MMM} \  .
\]
This means that the field redefinitions \eqref{MetricPerturbation2}, with constants \eqref{ActionConstants2}, produce a dimensionally reduced Lagrangian with a gravitational component in the Einstein frame, as required, and with the kinetic term of the scalar field appearing in the canonical normalization. This is the main result of this section.

If condition \eqref{condition_lambda} on $\lambda$ were not satisfied, the kinetic term $| \dd \phi |^2_{g_\MMM}$ would appear with the wrong sign in the Lagrangian. If $\lambda$ takes the critical value 
\[
\lambda_0 \ = \ -\; \frac{m+k-2}{k(m-2)} 
\]
the kinetic term $| \dd \phi |^2_{g_\MMM} $ disappears altogether, so the equation of motion of $\phi$ becomes just a constraint.

The detailed calculation of Lagrangian \eqref{ExpandedLagrangianScalarField} uses the rescaling identities \eqref{B.10} of appendix \ref{AppendixWeylRescalings} in the special case where $2f= -b_1 \phi + \log a_1$ and $2\omega = b_2 \phi + \log a_2$. The metrics $\tg_P$ and $g_P$ in that appendix are taken to stand for the metrics $g_P$ and $\bg_P$ of this section, respectively. Besides employing these identities, one also needs to use that 
\[
 \int_P \, |N|^2_{g_\PPP} \, \vol_{g_\PPP} \ = \ \frac{1}{\bk} \, \int_P\,  \bigg[ \;  \frac{(k \, b_1)^2}{4} \, \, | \dd \phi |^2_{ \bg_\PPP}  \, + \, |N_{ \bg_\PPP}|^2_{ \bg_\PPP}  \;   \bigg] \; \vol_{\bg_\PPP}
\]
with no cross terms involving both $\dd \phi$ and  $N_{ \bg_\PPP}$. This result follows from the last identity in \eqref{B.10} after observing that the cross term is proportional to
\bal
\int_K \,  \bg_P \big(\, N_{ \bg_\PPP} \, , \, \grad_{\bg_\PPP} \phi \, \big) \;\vol_{\bg_\KKK} \ &= \ \int_K \,  \bg_P \big(\, N_{ \bg_\PPP} \, , \, \grad_{g_\MMM} \phi \, \big) \;\vol_{\bg_\KKK}  \linebr
&=\  -\; \Lie_{\,\grad_{g_\MMM} \phi} \, \big[ \, \Vol(K, \bg_K) \, ] \ = \ 0 \ .  \nonumber
\end{align}
Here we have used formula \eqref{PropertyMeanCurvature} and the constancy of the fibre-volume $\Vol(K, \bg_K)$ on $M$ for the normalized metrics $\bg_K$. We have also used that the gradient $\grad_{\bg_\PPP} \phi$, with $\phi$ regarded as a function on $P$ that is constant along the fibres, is the horizontal lift to $P$ of the gradient in the base $\grad_{g_\MMM} \phi$, with $\phi$ regarded as a function on $M$. 

For a clearer interpretation of the re-definition $g_M \rightarrow a_2\, e^{b_2 \phi} \, g_M$ of the physical metric on $M$, as carried out in \eqref{MetricPerturbation2}, it is also worth noting that
\beq \label{Identitya2b2}
a_2\, e^{b_2 \phi} \ = \ \Big[ \, \kappa_P^{-1}\,\kappa_M \, \Vol(K, \, g_K) \, \Big]^{2/(2-m)}
\eeq
as functions on the base $M$. This follows from \eqref{InternalRescaling} combined with \eqref{ActionConstants3} and \eqref{ActionConstants2}.

To end this section we will extract the equation of motion of the scalar field $\phi$ and compare it with the analogous, simpler equation often discussed in the setting of five-dimensional Kaluza-Klein. Varying Lagrangian \eqref{ExpandedLagrangianScalarField} with respect to $\phi$ and ignoring total derivatives yields the equation:
\beq \label{EquationMotionScalarField}
\kappa_M \, (g_M)^{\mu \nu}\, \nabla^M_\mu (\partial_\nu \phi ) \ + \ \beta \, |\FF\,|^2_{ \bg_\PPP} \,  \, e^{- \beta \phi}  \ + \ \beta \, R_{ \bg_\KKK} \, \, e^{\beta \phi} \ - \ 2 \, \Lambda \, a_2 \,b_2 \, e^{b_2 \phi}
 \ = \ 0 \  .
\eeq
In the traditional, five-dimensional Kaluza-Klein model $\Lambda$ is zero and the internal space is the circle $S^1$. So the internal scalar curvature $R_{ \bg_\KKK}$ vanishes. Moreover, the usual first-order analysis of this model takes all fields to be independent of the internal coordinate, so the only internal degree of freedom is the scaling factor $\phi$. This means that the normalized internal metric $\bg_K$ is trivial, the same constant number everywhere, and the tensors $S_{ \bg_\PPP}$ and $N_{ \bg_\PPP}$ necessarily vanish. Taking $m=4$ and $k=1$ we get that $\beta = \sqrt{6 \kappa_M} $ and the previous equation of motion reduces simply to $\sqrt{\kappa_M} (\nabla^M)^\mu (\partial_\mu \phi) = - \sqrt{6}\, |\FF\,|^2_{ \bg_\PPP}  \, e^{- \sqrt{6\, \kappa_\MMM} \phi}$. Thus, the squared-norm $|\FF\,|^2$ should vanish in any region of spacetime where $\phi$ is approximately constant, as required for the good interpretation of the gauge coupling constants at present time. This constrains the electromagnetic field and is often cited as a difficulty of the five-dimensional Kaluza-Klein model. This difficulty appears to be less blatant in the higher-dimensional models represented by equation \eqref{EquationMotionScalarField}, since in regions where $\phi$ is approximately constant we are left with the more evolved constraint
\beq
|\FF\,|^2_{ \bg_\PPP} \ +\ R_{ \bg_\KKK} \, \, e^{2 \beta \phi} \ - \  \frac{2 \, \Lambda \, k\, a_2 }{m+k-2} \,\, e^{(b_2 +\beta) \phi}   
\ \simeq \ 0 \ \ .
\eeq
So the squared-norm $|\FF\,|^2$ of the Yang-Mills curvature can be non-zero and vary in this region, as long as those changes are compensated by appropriate spacetime variations of the normalized internal metric $\bg_K$, which determines the function $R_{ \bg_\KKK}$. Moreover, these equations of motion are derived solely from the traditional Einstein-Hilbert action on $P$, which may not tell the whole story in a realistic model operating at different scales.

\subsection{Gauge coupling constants in the Einstein frame}
\label{GaugeCouplingConstants}

The purpose of this section is to write down a formula for the scale of the gauge couplings as extracted from the Lagrangian in the Einstein frame. We start from the Einstein-Hilbert action on $M\times K$, assume a vacuum internal metric $g_K^\Szero$ that is constant around present time and adapt a general result of Weinberg \cite{WeinbergCharges}.

In the traditional Kaluza-Klein setting, the result \cite[eq. 16]{WeinbergCharges} says that the gauge coupling constants are of the order
\[
\alpha^2_{\rm{Wein}} \ = \ \frac{8\pi^2 \kappa_M}{N\, l^2_{g_\KKK^\SSzero}}  \ \sim \ \frac{8\pi^2 \kappa_M}{l^2_{g_\KKK^\SSzero}} \ , 
\]
where $N$ is a positive integer and $l_{g_\KKK^\SSzero}$ denotes the average length (root-mean-square) of the circumferences on the internal space $(K, g^\Szero_K)$ generated by the vector field associated to the gauge field. The conventions in \cite{WeinbergCharges} differ from the ones used here, however. Firstly, the symbols $\kappa^2$, $\bar{\kappa}^2$ and $g_e$ used there correspond to what we denote here by $2 \kappa_M$, $2 \kappa_P$ and $\alpha_{\rm{Wein}}$, respectively. More importantly, \cite{WeinbergCharges} does not perform the field redefinitions to obtain the general, dimensionally-reduced Lagrangian in the Einstein frame, as was done here on the way to \eqref{ExpandedLagrangianScalarField}. Instead, along the lines of the traditional Kaluza-Klein ansatz, \cite{WeinbergCharges} fixes the value of the constant $\kappa_P$ to be $\kappa_M \, \Vol_{g_K^\SSzero}$, which produces a 4D Lagrangian in the Einstein frame only in the case of a constant internal metric, $g_K = g_K^\Szero$. So the relevant terms of the dimensionally-reduced action considered in \cite{WeinbergCharges} are
\beq \label{WeinbergLagrangian}
\int_M \left[ \, \frac{1}{2 \kappa_M}\, R_{g_\MMM}  \ - \  \frac{1}{2 \,\kappa_M\,  \Vol_{g_\KKK^\SSSzero}} \, \left(\, \int_K |\FF|_{g_\MMM + g_\KKK^\SSzero}^2 \ \vol_{g_\KKK^\SSzero}  \, \right) \, \right] \vol_{g_\MMM} \ .
\eeq
In contrast, in the present notes the constant $\kappa_P$ is not constrained {\it {a priori}} because the metric $g_M$ was redefined in order to obtain a Lagrangian in the Einstein frame in all circumstances, even when the internal metric is not constant throughout spacetime. In the case of a constant internal metric in the vacuum configuration, the relevant terms of our Lagrangian \eqref{ExpandedLagrangianScalarField} are
\beq \label{RelevantTerms}
\int_M \left[ \, \frac{1}{2 \kappa_M}\, R_{g_\MMM}  \ - \  \frac{1}{2 \,\kappa_P \, \bk } \, \left(\, \int_K |\FF|_{g_\MMM + \bg^\SSzero_\KKK}^2 \  e^{- \beta \phi_\SSzero} \, \vol_{\bg^\SSzero_\KKK}  \, \right) \, \right] \vol_{g_\MMM} \ .
\eeq
Here we have used \eqref{UnscaledMetric} and the notation $ |\FF|_{g_\MMM + \bg^\SSzero_\KKK}^2$ means that the norm \eqref{NormF} should be evaluated using the metrics $g_M$ and $\bg^\Szero_K$. Now we want to write this relation in terms of the internal vacuum metric  $g_K^\Szero$ instead of the normalized metric $\bg_K^\Szero$. From \eqref{InternalRescaling} and definition \eqref{NormF} it is clear that
\[
 |\FF|_{g_\MMM + \bg^\SSzero_\KKK}^2 \ \vol_{\bg^\SSzero_\KKK} \ = \  \big(\,  a_1\,  e^{-b_1\phi} \, \big)^{-1-k/2}  \ |\FF|_{g_\MMM + g^\SSzero_\KKK}^2 \ \vol_{g^\SSzero_\KKK} \ . 
\]
Using definitions \eqref{ActionConstants2} one calculates that
\[
\big(\,  a_1\,  e^{-b_1\phi_\SSzero} \, \big)^{-1-k/2} \ = \  \bk^{\frac{k+2}{m+k-2}} \ e^{\frac{(m-2)(k+2) \beta \phi_\SSzero}{2(m+k-2)}} \ = \ \bk \ e^{\beta \phi_\SSzero} \  \big(\,  a_2\,  e^{b_2 \, \phi_\SSzero} \, \big)^{\frac{m-4}{2}}   \ .
\]
Applying identity \eqref{Identitya2b2} we obtain that the terms \eqref{RelevantTerms} can be re-written as 
\beq \label{VariationWeinbergLagrangian}
\int_M \bigg[ \, \frac{1}{2 \kappa_M}\, R_{g_\MMM}  \ - \  \frac{\big( \kappa_P^{-1} \kappa_M \, \Vol_{g_\KKK^\SSzero} \big)^{\frac{m-4}{2-m}}}{2 \,\kappa_P} \, \left(\, \int_K |\FF|_{g_\KKK^\SSzero}^2 \, \vol_{g_\KKK^\SSzero}  \, \right) \, \bigg]  \, \vol_{g_\MMM} \ .
\eeq
Since our constant $\kappa_P$ is unconstrained, we are considering here a slight variation of the action \eqref{WeinbergLagrangian} of \cite{WeinbergCharges}. If we use our action in the calculations of \cite{WeinbergCharges}, the end result for the scale $\alpha^2 $ of the gauge coupling constants is instead
\beq \label{RelAux3}
\alpha^2 \ =  \ \big( \kappa_P^{-1} \kappa_M \, \Vol_{g_\KKK^\SSzero} \big)^{\frac{-2}{m-2}} \  \alpha^2_{\rm{Wein}}  \ .
\eeq
This is true because the denominators in front of the Yang-Mills term in \eqref{WeinbergLagrangian} and \eqref{VariationWeinbergLagrangian} differ by
\[
2 \, \kappa_P \,  \big( \kappa_P^{-1} \kappa_M \, \Vol_{g_\KKK^\SSzero} \big)^{\frac{m-4}{m-2}} \ =  \ \big( \kappa_P^{-1} \kappa_M \, \Vol_{g_\KKK^\SSzero} \big)^{\frac{-2}{m-2}} \  2 \, \kappa_M \, \Vol_{g_\KKK^\SSzero}  \ ,
\]
and, as is well-known, a rescaling of those denominators is equivalent to a rescaling of the squared gauge coupling constants by the same amount. Thus, in the Einstein frame we obtain
\beq \label{ScaleGaugeCouplings}
{\setlength{\fboxsep}{3\fboxsep} \boxed{
\alpha^2 \ \sim \  \frac{8\pi^2 \,\kappa_M}{l_{g_\KKK^\SSzero}^2 \, \big( \kappa_P^{-1} \kappa_M \, \Vol_{g_\KKK^\SSzero} \big)^{\frac{2}{m-2}} }  \ \ .
}}
\eeq
This reduces to the traditional expression when $\kappa_P = \kappa_M \, \Vol_{g_\KKK^\SSzero}$, but here $\kappa_P$ remains unconstrained. Notice also that the gauge couplings calculated in the Einstein frame scale exactly as the gauge bosons' masses of \eqref{MassGaugeBosons5} when the vacuum metric $g_K^\Szero$ changes by a constant factor. This scaling behaviour is different from that of $\alpha_{\rm{Wein}}$. This happens because, in the traditional derivation, a volume factor  is hidden in the normalization of the constant $\kappa_P$, which in fact becomes a non-constant for a rescaling $g_K^\Szero$.

\subsection{Unstable modes and scalar field inflation}
\label{UnstableModesInflation}

\subsubsection*{Early dynamics of the unstable mode $\phi$} 

The purpose of this section is to study the dynamics of the scaling field $\phi$ in the approximation where all gauge fields are zero and the other internal scalars are static. This is a coarse approximation, but could describe the initial stages of the unravelling of an unstable, higher-dimensional Einstein product metric on $M\times K$. In this regime, the dynamics of the scaling field is controlled by a simple potential $V(\phi)$ that will be written down and cursorily investigated. When the Einstein metric on $K$ also has TT-instabilities the initial dynamics of $\phi$ will be more evolved, since it will develop concurrently with internal symmetry breaking. Such an example will be described further ahead, in section \ref{InternalSymmetryBreaking}, in the case $K=\SU$.

Let us suppose that, over an ancient region of spacetime, the metric on $M\times K$ can be approximated by an Einstein product metric $g_P^e = g_M^e + g_K^e$ with positive constant. The scalar curvatures necessarily satisfy
\beq \label{CurvatureEinsteinProduct2}
R_{g_\KKK^e} \ = \  \frac{2\,  \Lambda\, k}{m + k -2} \  \qquad  \qquad  R_{g_\MMM^e}   \ = \   \frac{2\,  \Lambda\, m}{m + k -2}  \ .   
\eeq
This Einstein product  is an unstable solution of the higher-dimensional equations of motion. In section \ref{StabilityEinsteinSolutions} we saw that the number of unstable modes of perturbation is finite, at least for the internal metric. They include the fibre-rescaling mode represented by the real field $\phi$ and, depending on the details of $g_K^e$,  a possibly non-zero but finite number of perturbations by fields of TT-tensors on $K$. Let us further assume that the most unstable direction of perturbation (i.e. the direction with the most negative eigenvalue of the Hessian) is the rescaling mode represented by $\phi$. This is true in the example $K=\SU$ studied further ahead. Then one expects that the metric $g_M^e + g_K^e$ will first unravel in the $\phi$-direction. Thus, in a slightly larger region of spacetime, the metric on $M\times K$ would be well-approximated by a rescaled, warped product metric of the form  \eqref{MetricPerturbation2}, with no gauge fields and a normalized internal metric $\bg_K$ that is constant throughout the region and proportional to the Einstein solution. So let us take the normalized internal metric of \eqref{MetricPerturbation2} to be
\beq  \label{NormalizedMetricInflation}
\bg_K \ := \ \big( \kappa_P^{-1} \kappa_M \, \Vol_{g_\KKK^e} \big)^{\frac{2}{m-2}} \, g^e_K \ \  ,
\eeq
where $\Vol_{g_\KKK^e}$ denotes the total volume of $g_K^e$ and the (so far arbitrary) constant of proportionality has been chosen to simplify the formulae ahead. Standard identities then imply that the volume forms and scalar curvatures are related by
\bal \label{NormalizedMetricInflation2}
\vol_{\bg_\KKK} \ &= \ \big(\kappa_P^{-1} \kappa_M \, \Vol_{g_\KKK^e} \big)^{\frac{k}{m-2}} \, \vol_{g^e_\KKK}   \linebr
R_{\bg_\KKK} \ &= \ \big( \kappa_P^{-1} \kappa_M \, \Vol_{g_\KKK^e} \big)^{\frac{-2}{m-2}} \, R_{g^e_\KKK}  \nonumber  \linebr
\bk \ &= \ \kappa_P^{-1} \kappa_M \,  \Vol_{\bg_\KKK} \ = \ \big( \kappa_P^{-1} \kappa_M \, \Vol_{g_\KKK^e} \big)^{1+ \frac{k}{m-2}}    \nonumber   \ .
\end{align}
In particular, the higher-dimensional metric \eqref{MetricPerturbation2} can be written as
\beq   \label{MetricPerturbation3}
g_P \ = \  \big( \kappa_P^{-1} \kappa_M \, \Vol_{g_\KKK^e} \big)^{\frac{-2}{m-2}}  \; e^{ \frac{k \beta \phi}{m+k-2}} \; g_M \  +  \ e^{- \frac{(m-2) \beta \phi}{ m+k-2}}  \; g^e_K \ ,
\eeq
after taking into account the definitions \eqref{ActionConstants3} and \eqref{ActionConstants2} of the various constants. So $g_P$ is a warped product of the Einstein metric $g_K^e$ and the physical spacetime metric $g_M$. The warping factor is determined by the field $\phi$. By assumption $\bg_K$ is constant across the fibres, so the tensors $\mathring{S}$ and $N_{ \bg_\PPP}$ vanish in this region, just as $\FF$ does. Therefore, after fibre-integration, the full action \eqref{ActionScalarField} becomes simply
\beq \label{SimplifiedAction}
\int_M\, \Big[ \;  \frac{1}{2\, \kappa_M} \, R_{g_\MMM} \; - \; \frac{1}{2} \, |\, \dd \phi \,|^2_{g_\MMM} \; - \; V(\phi ) \; \Big] \, \vol_{g_\MMM}
\eeq
with a potential 
\beq \label{ScalarFieldPotential}
{\setlength{\fboxsep}{3\fboxsep} \boxed{
V(\phi ) \ = \    \frac{\Lambda}{\kappa_M \, ( \kappa_P^{-1} \kappa_M \, \Vol_{g_\KKK^e})^{\frac{2}{m-2}} } \,   \, e^{\frac{k \beta \phi}{m+ k-2}} \, \Big(\, 1  \, - \, \frac{k}{m+k-2}\; e^{\frac{(m-2) \beta \phi}{m+k-2}} \, \Big) \ .
}}
\eeq
Using relation \eqref{CurvatureEinsteinProduct2} between the constant $\Lambda$ and the scalar curvature of $g_K^e$, in the physical case $m=4$ the expression for the potential can also be written as
\beq \label{ScalarFieldPotential3}
V(\phi ) \ = \    \frac{  \kappa_P \; R_{g^e_\KKK} }{2\, \kappa_M^2 \; \Vol_{g_\KKK^e} } \, \,  e^{\frac{k \beta \phi}{k+2}} \,  \Big(\, 1 \,+\, \frac{2}{k} \, - \, e^{\frac{2 \beta \phi}{k+2}} \, \Big) \ .
\eeq
So the potential depends on the primordial Einstein metric in internal space only through a multiplicative factor that coincides with the simple ratio of its scalar curvature and Riemannian volume.
A plot of $V(\phi)$ for the illustrative values $m=4$, $k=8$ and $\lambda =0$ looks like this:
\begin{figure}[H]
\centering
\includegraphics[scale=0.35]{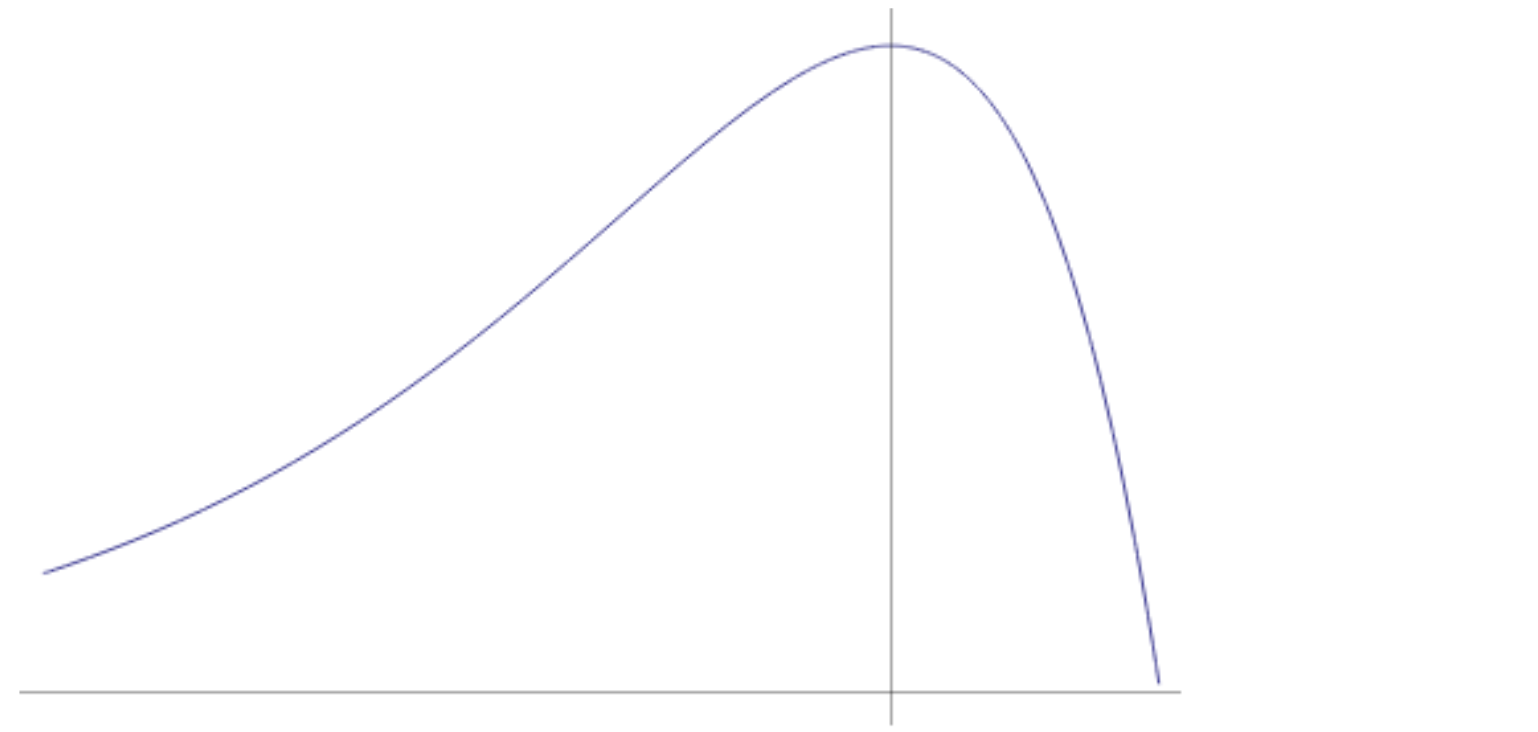}   
\vspace*{-.3cm}
\caption{Potential $V(\phi)$. \protect\footnotemark}
\label{fig:HillTopPotential}
\end{figure}
\footnotetext{All plots were generated with the free online version of Wolfram Alpha.}
The pure Einstein product configuration of the metric $g_P$ corresponds to $\phi= 0$ and defines a global maximum of the potential, which takes the value 
\[
V(0) \ = \ \frac{(m-2)\, \Lambda}{(m+k-2) \, \kappa_M \, ( \kappa_P^{-1} \kappa_M \, \Vol_{g_\KKK^e})^{\frac{2}{m-2}}}  \  \ .
\]
In this region spacetime is de Sitter. For non-zero $\phi$ the potential decreases. It is unbounded from below for positive values of $\phi$, as the internal space contracts, and it is bounded by zero for negative $\phi$, as the internal space expands from its primordial size $\Vol_{g_\KKK^e}$.

An inspection of the full Kaluza-Klein Lagrangian \eqref{ExpandedLagrangianScalarField}, however, also makes it clear that this simplified potential cannot be used when the tensor $\FF$ is not negligible, or when the scalar curvature of $\bg_K$ departs too much from its initial value. The last condition can take hold pretty quickly, one expects, if the primordial Einstein metric $g^e_K$ also has unstable TT-deformations, besides the $\phi$-instability. Even though the TT-instabilities can be milder than the rescaling instability, once the metric starts to unravel in that direction the curvature $R_{ \bg_\KKK}$ in Lagrangian \eqref{ExpandedLagrangianScalarField} does change the effective potential felt by $\phi$. All in all, one should not expect the potential $V(\phi)$ depicted in figure \ref{fig:HillTopPotential} to represent the true, effective potential for large values of $\phi$.

In section \ref{InternalSymmetryBreaking} we will discuss a more complete potential $V(\phi, \varphi)$ in the particular case $K=\SU$. This potential controls the early, joint dynamics of two unstable modes of the bi-invariant metric on $K$: the scaling mode represented by $\phi$ and a TT-deformation represented by $\varphi$. It would be interesting to analyze in detail how the presence of the $\varphi$-instability affects the early dynamics of $\phi$. In section \ref{StabilizingInternalCurvature} we will argue that the rescaling of the internal space induced by a changing $\phi$ may not proceed indefinitely, across all orders of magnitude. At some scale new physics may contribute to form a minimum of the potential $V(\phi, \varphi)$ and stabilize the rescaling deformation at a finite value $\phi = \phi_0$.

\subsubsection*{A model with scalar inflation} 

The simplified action \eqref{SimplifiedAction} that controls the early dynamics of the unstable mode $\phi$ is similar to the inflaton action that appears in single field models of cosmological inflation \cite{LLBook}. The general profile of the potential $V(\phi)$ depicted in figure \ref{fig:HillTopPotential} belongs to the family of hilltop potentials in small-field models. In these models inflation is generated by the scalar field slowly rolling down from the local maximum of $V$  \cite{BL, KLL}. The particular potential \eqref{ScalarFieldPotential}---a combination of two exponentials in $\phi$---belongs to a subfamily of hilltop models studied in \cite{CFM}. 

In the next few paragraphs we will verify that the simplified potential \eqref{ScalarFieldPotential} does indeed generate four-dimensional inflation, as the Einstein product metric starts to unravel along the scaling deformation. In principle, the scalar field can roll down to both sides of the maximum, leading to increasingly positive or negative values of $\phi$. Inflation can occur in both situations.
However, we will also see that it never satisfies the slow-roll conditions in this approximation. This happens because the hilltop of $V(\phi)$ is not flat enough, so the roll down of the scalar field from the global maximum will not be slow enough. When the field rolls down to the  $\phi < 0$ direction, the profile of the potential in figure \ref{fig:HillTopPotential} is qualitatively similar to the potentials found in models of quintessential inflation. 

Start by observing that the $n$th derivative of $V$ with respect to $\phi$ is
\[
V^{(n)}(\phi ) \ = \    \frac{k\, \beta^n\, \Lambda}{(m+ k-2)\, \kappa_M \, ( \kappa_P^{-1} \kappa_M \, \Vol_{g_\KKK^e})^{\frac{2}{m-2}} } \,   \, e^{\frac{k \beta \phi}{m+ k-2}} \, \bigg[\, \Big(\, \frac{k}{m+k-2}\, \Big)^{n-1}  \, - \,  e^{\frac{(m-2) \beta \phi}{m+k-2}} \, \bigg] \ .
\]
Using this expression one calculates that in the physical, $m=4$ case, the slow-roll parameters associated to the potential are
\bal
\epsilon (\phi) \ &:= \ \frac{1}{2 \, \kappa_M} \; \Big (\, \frac{V'}{V} \, \Big)^2 \ = \   \frac{ \beta^2}{2\, \kappa_M}\ \bigg( \, \frac{1 - e^{b_1 \, \phi}}{1\, + \, \frac{2}{k} \, -\, e^{b_1 \phi}}  \, \bigg)^{2}  \linebr
\eta (\phi) \ &:= \ \frac{1}{\kappa_M} \;  \frac{V''}{V} \ - \ \epsilon (\phi) \ = \   \frac{ \beta^2}{2\, \kappa_M\, \big(1\, + \, \frac{2}{k} \, -\, e^{b_1 \phi}\big)^2}\ 
\bigg[ \, (1 - e^{b_1  \phi})^2\, - \,   \frac{8\, e^{b_1  \phi}}{k\, (k+2)}  \, \bigg]  \nonumber \ .
\end{align}
Here $b_1 = 2 \,\beta /(k+2)$, as in definition \eqref{ActionConstants2} with $m=4$. Inflation should occur for values of the field between $\phi = 0$ and $\epsilon (\phi) = 1$. The last condition is satisfied for 
\[
\phi_{\mp} \ := \  -\, \frac{k+2}{2\, \beta} \ \log\bigg[\, \frac{k}{k+2} \, \Big(\, 1\, \pm \,  \frac{\beta}{\sqrt{2\, \kappa_M}} \, \Big)\, \bigg] \ .
\]
Since $\beta$ is positive by definition, the $\phi_{-}$ solution is always well defined. The $\phi_{+}$ solution exists only when $\frac{\beta}{\sqrt{2\, \kappa_M}} < 1$, and in this case $\phi_{+}$ is positive. However, the value of the second slow-roll parameter at $\phi= 0$ is
\[
\eta (0) \ = \ \frac{-2}{ 1\, + \, \frac{2 \lambda k}{ k+2}}  \ .
\]
So it does not satisfy the slow-roll condition $|\eta | \ll 1$ unless the constant $\lambda$ is large. At the same time, we do not want $\lambda$ to be large because, according to \eqref{MassPerturbationModes}, that would create many more unstable deformations of the primordial Einstein metric. Thus, the simplified potential $V(\phi)$ does not seem to be compatible with slow-roll inflation.

At the end of last section we described the limitations of the approximation leading to $V(\phi)$. Especially at later times, when the gauge field strength $\FF$ may become non-zero, or if the internal metric starts to unravel along TT-instabilities and the scalar curvature of $\bg_K$ departs from its initial value. The question is how these additional effects will modify the description of the inflationary period derived from the simplified potential $V(\phi)$. In principle it is possible that the dynamics of $\phi$ coming from the full Kaluza-Klein Lagrangian \eqref{ExpandedLagrangianScalarField}, if properly analyzed, could reveal a more realistic model of the inflationary process. This would be a multiple-field, microscopic model of inflation, where the scaling field $\phi$ dominates the initial stages of the process but soon enough, for instance after the start of internal symmetry breaking, other internal scalars (other coefficients of the internal metric) and gauge fields need to be considered to extend the inflationary period and, eventually, participate in the reheating period. 

A good starting point to investigate the corrections to $V(\phi)$ induced by TT-instabilities may be the example $K=\SU$ described in section \ref{InternalSymmetryBreaking}. There we write down the explicit two-field action \eqref{ActionTTDeformation} and associated potential $V(\phi, \varphi)$ that control the joint dynamics of the unstable deformations $\phi$ and $\varphi$. It would be interesting to understand the properties of an inflationary period described by this two-field model. Even though this classical Kaluza-Klein picture still ignores fermions and all quantum effects. In section \ref{StabilizingInternalCurvature} we also discuss the stabilization of the rescaling deformation at a finite value $\phi = \phi_0$. That is relevant to start thinking about the formation of present-time vacuum conditions.

Important efforts to link the Kaluza-Klein picture to FRW cosmological models are well-known in the literature. See \cite[ch. 6]{Bailin} for a review. These works generally adapt the FRW model by assuming two time-dependent scale factors---one for the three-dimensional space and one for the internal space---on an otherwise static space and internal geometry. Then they study the equations of motion of the scale factors as derived from the Einstein-Hilbert action or variations thereof. For example, the initial studies  \cite{CD, Freund} worked with a flat internal space, and with Kasner-type solutions of the higher-dimensional Einstein equations, to present models for the expansion of the three-dimensional space and simultaneous contraction of the internal space. In our language, these conditions lead to a trivial potential \eqref{ScalarFieldPotential} (since $\Lambda = 0$) and do not generate inflation. The article \cite{SW} considers non-standard curvature terms in the higher-dimensional action to obtain a sufficiently long period of inflation. It works with general internal geometries and a non-minimal gravity action with suitable parametres. These approaches differ from the one discussed in section \ref{InternalSymmetryBreaking} because they assume a static internal metric during the entire inflationary period, apart form the dynamical scaling factors (i.e. no room for TT-instabilities or internal symmetry breaking).

\subsection{Left-invariant metrics on \texorpdfstring{$\SU$}{SU}}
\label{DeformedMetricsSU(3)}

In the following two sections we will take as internal space the simple, eight-dimensional Lie group $K\! =\! \SU$. We want to study the behaviour of the Einstein-Hilbert action when the internal metric varies within a particular family of left-invariant metrics on $K$. These metrics are deformations of the bi-invariant, Cartan-Killing metric on $\SU$, which is known to have an unstable TT-perturbation lying within that family.

The discussion starts with a quick reminder of general properties of left-invariant metrics on a simple Lie group. For more details see for instance \cite{Milnor, Besse, BD, Ba1}. After that we consider the family of deformed metrics on $K$ and determine some of their basic properties, such as volume, Ricci tensor, scalar curvature and Lie derivatives in different directions. The behaviour of these metrics in a dynamical theory on $M\times K$ will be addressed only in section \ref{InternalSymmetryBreaking}, where we study how the unstable modes appear in the higher-dimensional Einstein-Hilbert action.

\subsubsection*{Left-invariant metrics on a group} 

As a vector space, the Lie algebra of a group is the tangent space to the group at the identity element. A vector $v$ in the Lie algebra $\kk= \su$ can be extended to a vector field on the group $K$ in two canonical ways: as a left-invariant vector field $v^\LL$ or as a right-invariant field $v^\RR$.
The explicit flows of these vector fields can be used to show that their Lie bracket is also invariant on $K$ and satisfies
\beq \label{BracketsInvariantFields}
[u^\LL, \, v^\LL ] \ = \ [u, \, v ]^\LL_\kk  \qquad \qquad  [u^\RR, \, v^\RR ] \ = \ - [u, \, v ]^\RR_\mathfrak{k}   \qquad \qquad   [u^\LL, \, v^\RR ] \ = \ 0 \ ,
\eeq
where the bracket $[\, .\, , \, . \,]_\kk$ in the Lie algebra is just the commutator of matrices in the case of matrix Lie groups. Just as with vectors, any tensor in the Lie algebra $\kk$ can be extended to a left or right-invariant tensor field on $\kk$. For example, an inner-product $g$ on $\kk$ can be extended to a left-invariant metric on $K$ by decreeing that the product of left-invariant vector fields has the same value everywhere on $K$ and coincides with $g$ at the identity element of the group. Thus $g(u^\LL, v^\LL) = g(u,v)$. In the opposite direction, every left-invariant metric on $K$ is fully determined by its restriction to the Lie algebra. When a left-invariant metric is applied to right-invariant vector fields, the result is a function on $K$ that is not constant in general, but still simple enough:
\bal \label{ProductInvariantFields}
 g(u^\LL, v^\RR)\, |_h \ =\  g(u, \, \Ad_{h^{-1}} v)   \qquad \qquad    g(u^\RR, v^\RR)\, |_h \ =\  g(\Ad_{h^{-1}} u, \, \Ad_{h^{-1}} v)
\end{align}
for all elements $h \in K$ and all vectors $u, v$ in the Lie algebra. 

The preceding observations imply that right-invariant fields are always Killing for left-invariant metrics on K, since 
\beq \label{LieDerivativeMetric2}
( \Lie_{w^\RRR} g) (u^\LL, v^\LL) \ = \  \Lie_{w^\RRR} \left(  g(u^\LL,v^\LL)  \right)\, -\, g([w^\RR, u^\LL],v^\LL) \,- \,g(u^\LL, [w^\RR, v^\LL]) \ = \ 0 \ .
\eeq
The same is not true for general left-invariant vector fields. The equality
\bal \label{LieDerivativeMetric1}
( \Lie_{w^\LLL} g) (u^\LL, v^\LL) \ &= \  \Lie_{w^\LLL} \left(  g(u^\LL,v^\LL)  \right)\, -\, g([w^\LL, u^\LL],v^\LL)\, - \, g(u^\LL, [w^\LL, v^\LL])    \nonumber  \linebr
 &= \ - \, g([w, u],v) \, - \,g(u, [w, v]) \ 
\end{align}
entails that the Lie derivative $\Lie_{w^\LLL} g$ may be a non-zero, left-invariant 2-tensor on $K$. In the special case when $g$ is an Ad-invariant inner-product on $\kk$, then $g(u^\RR, v^\RR)$ is also a constant function on $K$ and the metric $g$ is both left and right-invariant. In this case left-invariant vector fields are Killing as well. These are the bi-invariant metrics on the group. When the group is simple they coincide with minus the Killing form, up to a constant factor. 

The isometry group of a left-invariant metric $g$ on $K$ is isomorphic to 
\beq \label{IsometriesLeftInvariantMetrics}
\mathrm{Iso}(g) \ = \  ( K \! \times\!  H ) \, /  \, (\, Z(K)\cap H \,) \ ,
\eeq
where $H$ is the closed subgroup of $K$ that preserves the metric under right-translations. The subgroup $H$ can be as large as $K$, for bi-invariant metrics, or as small as the trivial group. The quotient appears because the central elements $x \in Z(K)$ satisfy $x h x^{-1} = h$ for all $h \in K$, so when $x \in Z(K) \cap H$ the element $(x, x^{-1})$ in $K \! \times\!  H$ acts trivially on the manifold $K$. 

Let $H$ be a connected, compact subgroup of $K$. There is a one-to-one correspondence between left-invariant metrics on $K$ preserved by right $H$-translations and inner-products on the Lie algebra $\kk$ that are preserved by $\Ad_h$ for every $h\in H$. This  follows from \eqref{LieDerivativeMetric1}. On the other hand, an $\Ad_H$-invariant inner-product on $\kk$ determines a decompositon $\kk = {\mathfrak{h}} \oplus {\mathfrak{h}}^\perp$ and the orthogonal subspace $ {\mathfrak{h}}^\perp$ is preserved by $\Ad_H$. Since the inner-product restricted to $ {\mathfrak{h}}^\perp$ is still $\Ad_H$-invariant, of course, standard results say that it determines a $K$-invariant metric on the quotient homogenous space $K/H$ \cite[Prop. X.3.1]{KN}. So studying left-invariant metrics on $K$ is an algebraic problem closely related to studying $K$-invariant metrics on $K/H$. Deformations of left-invariant metrics on $K$ can change their isometry group, so provide a sort of continuous, ``upstairs'' path between $K$-invariant geometries in different homogeneous spaces.

The Riemannian volume form $\vol_g$ of a left-invariant metric $g$ is always a left-invariant differential form on the group. In the case of connected, unimodular Lie groups it is also a right-invariant form, even though the metric itself may not be right-invariant.  In this case we always have 
\beq \label{BiinvarianceVolumeForm}
(L_h)^\ast \,\vol_g \ = \ (R_h)^\ast \,  \vol_g \ = \ \vol_g \ ,
\eeq
where $L_h$ denotes the diffeomorphism of $K$ determined by left-multiplication by the group element $h$, and similarly for $R_h$. This bi-invariant volume form of $g$ coincides, up to normalization, with the Haar measure on $K$. The invariance of the volume form together with \eqref{ProductInvariantFields} can be shown to imply that 
\beq \label{ProductInvariantVectors1}
\int_{h\in K}  g(u^\LL, v^\RR)\ \vol_g \  =  \ 0 \ 
\eeq
for all vectors $u$ and $v$ in $\kk$ and for all left-invariant metrics $g$. In contrast, by definition of left-invariant metric, the contraction $g(u^\LL, v^\LL)$ is a constant function on $K$, so we have
\beq \label{ProductInvariantVectors2}
\int_{h\in K}  g(u^\LL, v^\LL)\ \vol_g \  =  \  g(u, v)\ \Vol (K, g)  \ .
\eeq
The integral over $K$ of the product $g(u^\RR, v^\RR)$ is not immediate in general, but it does follow from the second equality in \eqref{ProductInvariantFields} that it is Ad-invariant, and hence proportional to the Cartan-Killing product on a simple algebra $\kk$: 
\beq \label{ProductInvariantVectors5}
\int_{h\in K}  g(u^\RR, v^\RR)\ \vol_g \  =  \  \int_{h\in K}  g(\Ad_{h^{-1}} u, \, \Ad_{h^{-1}} v)\ \vol_g \ \  \propto \   \Tr ({\rm{ad}}_u \, {\rm{ad}}_v) \, \Vol (K, g)  \ .
\eeq
This happens because the second integral above is explicitly averaging the pull-back metric $\Ad^\ast_{h^{-1}} g$ over $K$, and hence is invariant under a change of variable $h \to h' h$ for any fixed group element $h' \in K$.

Finally, the Ricci curvature of a left-invariant metric is also a left-invariant tensor on $K$. This implies that the scalar curvature is constant on the group. Its value can be expressed in terms of a $g$-orthonormal basis $\{e_a\}$ of the Lie algebra $\kk$ through the formula
\beq \label{GeneralScalarCurvature}
R_g \ = \ - \sum_{a,b} \  \frac{1}{4} \, g \left( [e_a, e_b],\, [e_a, e_b] \right)  \,  + \,  \frac{1}{2} \, g([\,e_a, [e_a, e_b]\, ] ,\, e_b) \ .
\eeq
This expression is valid for unimodular Lie groups and is a special case of a well-known formula for the scalar curvature of homogeneous spaces (e.g. see chapter 7 of \cite{Besse}).

\subsubsection*{A particular family of left-invariant metrics on \boldmath$\SU$} 

Think of $\su$ as the space of traceless, anti-hermitian $3\times3$ matrices. It has a natural inner-product
\beq \label{BasicBiinvariantMetric}
\beta_0 (u,v) \ := \ \Tr(u^\dag\, v) \ = \  - \, \frac{1}{6} \,\, B(u,v) \ ,
\eeq
where $B(u,v)$ denotes the Cartan-Killing form $\Tr (\ad_u \circ \ad_v)$. Any matrix in $\su$ can be uniquely written as
\beq \label{AlgebraDecomposition1}
v \ = \ \bmatr - 2 \, v_Y &  - (v'')^\dag \linebr v'' & v_Y  I_2 + v_W \ematr \ ,
\eeq
where $v_Y$ is an imaginary number, $v_W$ is a traceless, anti-hermitian matrix in $\sutwo$ and $v''$ is a vector in $\CC^2$.
This determines a vector space decomposition
\bal \label{AlgebraDecomposition2}
\su \ &= \  \mathfrak{u}(1) \oplus \sutwo \oplus \CC^2   \linebr
v \ &=  \ v_Y\ +\ v_W \ + \ v'' \ .    \nonumber
\end{align}
When acting on vectors in these summands, the Lie bracket of $\su$ satisfies the relations
\bal \label{CommutationRulesSU3}
  [\, \mathfrak{u}(1),  \,  \utwo\,] \ &= \ \{ 0\}  \            &[\,\utwo, \, \CC^2\,] \   &= \ \CC^2    \\
   [\,\sutwo, \, \sutwo\,] \ &= \ \sutwo  \  &[\,\CC^2, \, \CC^2\, ]  \   &=\  \utwo    \nonumber  \ ,
 \end{align}
where of course $\utwo =  \mathfrak{u}(1) \oplus \sutwo$. Identifying $v_Y$, $v_W$ and $v''$ with their image matrices in $\su$, one can define an inner-product $\tbeta$ on $\su$ by
\beq \label{DefinitionTBeta}
\tbeta(u,v) \ := \ \lambda_1 \Tr(u_Y^\dag\, v_Y) \ + \ \lambda_2 \Tr(u_W^\dag\, v_W) \ + \ \lambda_3 \Tr\big[(u'')^\dag \, v'' \big]
\eeq
for positive constants $\lambda_1$, $ \lambda_2$ and $ \lambda_3$. The inner-product $\tbeta$ determines a naturally reductive metric on $\SU$, so essentially a Killing metric with independent rescaling factors in each component of the Lie algebra decomposition. It is clear that decomposition \eqref{AlgebraDecomposition2} is orthogonal with respect to $\tbeta.$

The rescaled inner-product $\tbeta$ is no longer invariant under the adjoint action on $\su$. However, it is invariant under the restriction of this action to a subgroup of $\SU$ isomorphic to $\Utwo$. This subgroup is the image of the homomorphism $\iota: \Utwo \to \SU$ defined by 
\beq
\iota(a) \ = \ \bmatr (\det a)^{-1} &   \\   & a \ematr \ .
\eeq 
Then the restricted adjoint action on $\su$ is given by
\beq \label{AdjointAction2}
\Ad_{\iota(a)} (v) \ = \ \bmatr -\Tr(v_Y) &  - [\,(\det a)\, a\, v'' \,]^\dag \\[0.6em]   (\det a)\, a\, v''  &  v_Y + \Ad_a (v_W) \ematr  \ ,
\eeq
for all matrices $a\in \Utwo$ and vectors $v\in \su$. The product $\tbeta$ is the most general inner-product on $\su$ invariant under such transformations. 

Using the standard correspondence between inner-products on the Lie algebra and left-invariant metrics on the group, it it clear that the left-invariant metric on $K$ determined by $\tbeta$ is not bi-invariant. It is invariant under the diffeomorphisms of $K$ determined by group multiplications on the left. But when it comes to multiplications on the right, it is invariant only under the subgroup $\Utwo$ of $\SU$, unless the constants $\lambda_j$ are all equal. So the isometry group of that left-invariant metric is $\Utwo \times \SU$ or a finite quotient thereof. To be more precise, it follows from \eqref{IsometriesLeftInvariantMetrics} that in the case $K = \SU$ the isometry group of the bi-invariant metric is $(\SU \times \SU) / \mathbb{Z}_3$, while the isometry group of the left-invariant metric determined by $\tbeta$ is
\[
\mathrm{Iso}(\tbeta) \ = \  (\, \SU \times\Utwo \,) /\,  \mathbb{Z}_3 \ = \  ( \, \SU \times \mathrm{SU}(2) \times  \mathrm{U}(1) \, ) /\,  \mathbb{Z}_6 \ .
\]
This coincides with the gauge group of fermionic representations in the Standard Model.

Now let $\{e_a \}$=$\{  u_0, \ldots, u_3, w_1, \ldots, w_4 \}$ be a $\tbeta$-orthonormal basis of $\su$. The vectors are chosen so that $\{w_j\}$ spans the subspace $\CC^2$ of $\su$, the vectors $\{u_1, u_2, u_3\}$ span the subspace $\sutwo$ and $u_0$ is the vector
\beq
u_0 \ = \  \frac{1}{\sqrt{6\, \lambda_1}} \,  \diag(-2i, i, i)  \ ,
\eeq
spanning the subspace $\mathfrak{u}(1)$. This basis is related to a $\beta_0$-orthonormal basis of $\su$ by rescaling the individual vectors with factors of the form $\sqrt{\lambda_a}$. Using this relation it is straightforward to show that the volume forms of $\tbeta$ and $\beta_0$ are related by
\beq \label{V2VolumeForm}
\vol_{\, \tbeta} \ =\ \sqrt{\lambda_1 \, \lambda_2^{\,3} \, \lambda_3^{\, 4}}  \ \vol_{\beta_0} \ .    
\eeq
Using the orthonormal basis in formula \eqref{GeneralScalarCurvature} for the scalar curvature, one calculates that 
\beq \label{ScalarCurvatureDeformation}
R_{\, \tbeta} \ = \ 3 \left( \frac{1}{\lambda_2} +   \frac{4}{\lambda_3} -  \frac{\lambda_1 + \lambda_2}{2\, \lambda_3^2}  \right) \ .         
\eeq
With some more work on the formulae of \cite[ch. 7]{Besse}, in appendix \ref{CalculationsSu(3)} we also calculate that the Ricci tensor of $\tbeta$ is related to the metric itself by
\bal
Ric_{\, \tbeta} \ &= \   \frac{3\, \lambda_1}{2\, \lambda_3^2} \ \tbeta \,  |_{\mathfrak{u}(1)}   \ + \    \left( \frac{1}{\lambda_2}   +  \frac{\lambda_2}{2\, \lambda_3^2}  \right)   \tbeta \, |_{\sutwo}    \ +  \  \frac{3}{4}\, \left( \frac{4}{\lambda_3} -  \frac{\lambda_1 + \lambda_2}{\lambda_3^2}  \right)  \tbeta \, |_{\CC^2} \ \nonumber   \linebr
 &= \     \frac{3\, \lambda_1^2}{2\, \lambda_3^2} \ \beta_0 \,  |_{\mathfrak{u}(1)}   \ + \    \left( 1   +  \frac{\lambda_2^2}{2\, \lambda_3^2}  \right)   \beta_0 \, |_{\sutwo}    \ +  \  \frac{3}{4}\, \left( 4  -  \frac{\lambda_1 + \lambda_2}{\lambda_3}  \right)  \beta_0 \, |_{\CC^2}  \ .
\end{align}
It is clear that $\tbeta$ is Einstein only when $ \lambda_1 =  \lambda_2 =  \lambda_3$. Decomposition \eqref{AlgebraDecomposition2} is orthogonal with respect both to $\tbeta$ and to $Ric_{\, \tbeta}$.

For a left-invariant metric $g$ on $K$ the Lie derivatives $\Lie_{v^\RRR}\, g$ always vanish, as remarked in \eqref{LieDerivativeMetric2}. The derivatives $\Lie_{v^\LLL}\, g$ can be non-zero.
Using \eqref{LieDerivativeMetric1}, the inner-product of these Lie derivatives, as defined in \eqref{DefinitionInnerProduct}, is calculated in appendix \ref{CalculationsSu(3)} to be
\beq \label{InnerProductLieDerivativesBeta}
\big\langle \, \Lie_{u^\LLL}\, \tbeta,  \; \Lie_{v^\LLL}\, \tbeta \,  \big\rangle \ = \ 3\, \Big(\, \frac{1}{\lambda_1} \, +\,  \frac{1}{\lambda_2} \,+ \, \frac{\lambda_1 + \lambda_2}{\lambda_3^2} \, -\,  \frac{4}{\lambda_3} \, \Big)  \ \tbeta (u'', v'')
\eeq
for vectors $u$ and $v$ in $\su$. The double prime, recall, denotes the $\CC^2$-component of the vectors in decomposition \eqref{AlgebraDecomposition2} of $\su$.

\subsubsection*{An unstable deformation of the bi-invariant metric} 

Consider the one-parameter family of left-invariant metrics on $\SU$ obtained from the general deformation \eqref{DefinitionTBeta} by choosing the constants $\lambda_a$ to be
\beq  \label{TTDeformation}
\lambda_1 (s) \ =\  \eta \, e^{2 s}  \qquad  \ \  \lambda_2 (s) \ =\  \eta \, e^{-2 s}  \qquad \ \  \lambda_3 (s) \ =\  \eta \, e^{s}  \ ,
\eeq
where $\eta$ is a positive constant and $s$ a real variable. These metrics will be called $\tbeta_s$. At $s=0$ the metric is bi-invariant and from \eqref{V2VolumeForm} we get that, for any value of $s$,
\beq \label{VolumeDeformation}
\vol_{\tbeta_s} \ = \ \vol_{\tbeta_\SSzero} \ = \ \eta^4 \, \vol_{\beta_0} \ .
\eeq
So $\tbeta_s$ is a deformation of the bi-invariant metric that preserves the volume form. Taking the derivative of $\tbeta_s$ with respect to $s$ and calculating the divergence of the resulting 2-tensor with respect to $\tbeta_s$, one can also verify that
\[
\divergence_{\tbeta_s} \Big( \, \frac{\dd \tbeta_s }{\dd s} \, \Big) \ = \ 0 \ .
\]
So $\tbeta_s$ is a TT-deformation of the bi-invariant metric. The scalar curvature of these metrics can be obtained from \eqref{ScalarCurvatureDeformation}. Denoting it simply by $R(s)$ and taking derivatives with respect to the parameter $s$, we have 
\bal \label{ScalarCurvatureUnstableDeformation}
R(s) \ &= \ \frac{3}{2\, \eta} \, \big(\,    2\, e^{2 s} \, - \, 1 \, +\,  8 \, e^{- s} \,  -\,  e^{-4 s}  \, \big)  \linebr
R'(s) \ &= \ \frac{6}{\eta} \, \big(\,   e^{s} \, - \,  e^{-2 s}  \, \big)^2 \ \nonumber \linebr
R'' (s) \ &= \ \frac{6}{\eta} \, \big(\,   e^{s} \, - \,  e^{-2 s}  \, \big)\,   \big(\,   e^{s} \, + \, 2\,  e^{-2 s}  \, \big) \ \nonumber  \ .
\end{align}
The first derivative $R'(s)$ is always positive except at $s=0$, where it vanishes. So the scalar curvature is a monotone function of $s$ with a saddle point at the Einstein metric $\tbeta_0$. At this point also $R''(0)$ vanishes, but the third derivative is positive. A plot of $R(s)$ looks like this
\begin{figure}[H]
\centering
\includegraphics[scale=0.4]{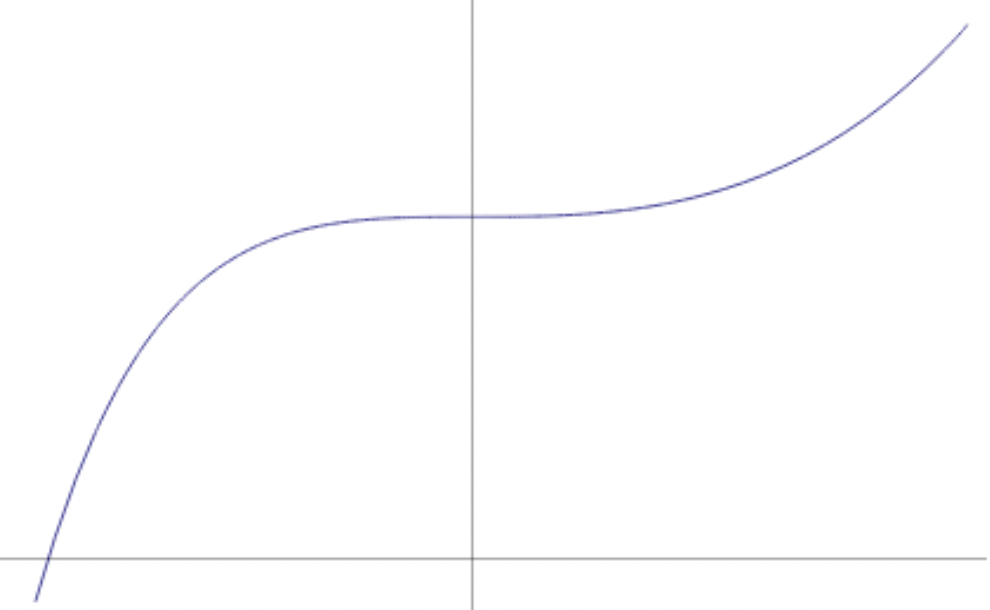}   
\vspace*{-.3cm}
\caption{Scalar curvature $R(s)$. }
\label{fig:CurvatureTTDeformation}
\vspace*{-.4cm}
\end{figure}
Whereas most TT-deformations of the bi-invariant metric on $\SU$ will reduce its scalar curvature, this particular deformation increases it for positive values of the parameter $s$. Since the opposite $ - R_{g_\KKK}$ plays the role of a potential for TT-deformations, we see that the bi-invariant metric is unstable and can unravel in this direction of perturbation.  This family of metrics on $\SU$ was found by Jensen in \cite{Jensen}. 

It is not clear to the author if there are other unstable TT-deformations of the bi-invariant metric on $\SU$, modulo equivalence by diffeomorphisms. The mathematical literature does guarantee that the bi-invariant metric is {\it linearly} stable \cite{Schwahn}. This implies that TT-deformations always have non-positive second derivative $R''(0)$, so a deformation can be unstable only at higher-order. In particular, any TT-instability will always be milder than the rescaling instability of the product metric on $M \times K$ described previously. Recent results also inform that the bi-invariant metric is isolated in the moduli space of Einstein metrics on $\SU$ \cite{BHMW}.

\subsection{Unstable modes and internal symmetry breaking when \texorpdfstring{$K\!=\!\SU$}{KSU}}
\label{InternalSymmetryBreaking}

In this section we study the behaviour of the Einstein-Hilbert action when the internal metric varies within the family of left-invariant metrics on $\SU$  defined previously, in \eqref{DefinitionTBeta} and \eqref{TTDeformation}. These metrics are unstable deformations of the bi-invariant metric. 

We start by deriving the simplest action that controls the dynamics of the deformation fields $\varphi$ and $\phi$.  This is done in \eqref{ActionTTDeformation}. Then we calculate the classical mass of the natural gauge bosons as the bi-invariant metric on $K$ unravels along that unstable perturbation. We find that the bosons' mass changes according to formula \eqref{MassBosonsTTDeformation}. In the next section, using results of the QFT literature, we will argue that this can affect the vacuum energy density of the theory, and hence create an effective potential that could stabilize the unravelling components of the internal metric.

\subsubsection*{Internal symmetry breaking} 

In this section we want to study the dynamics of a two-parameter, unstable deformation of the Einstein product metric $g_P^e = g_M^e + g_K^e$ on $M_4\times K$, where $g_K^e$ is a bi-invariant metric on the internal space $\SU$. Besides the rescaling field $\phi$ studied in section \ref{UnstableModesInflation}, we also consider a TT-deformation of the internal metric parameterized by a field $\varphi$ on $M_4$. Start by writing the internal, bi-invariant Einstein metric as
\[
g_{K}^e \ = \ \frac{15}{2\, \Lambda} \ \beta_0
\]
where $\beta_0$ is the metric defined in \eqref{BasicBiinvariantMetric} and the constant factor was chosen to satisfy the curvature relation \eqref{CurvatureEinsteinProduct2}. Consider the spacetime-dependent, TT-deformation of the Einstein metric
\beq \label{TTDeformedInternalMetric}
\tbeta_K(\varphi) \ :=\  \frac{15}{2\, \Lambda}  \Big[ \, e^{2 \varphi} \;  \beta_0 \  |_{\mathfrak{u}(1)} \ +\  e^{-2 \varphi}  \;  \beta_0 \,  |_{\sutwo} \ + \  e^{\varphi} \;  \beta_0 \,  |_{\CC^2} \, \Big] \ .
\eeq
From \eqref{VolumeDeformation} it is clear that this deformation does not change the volume form on $K$, 
\beq \label{VolumeInvarianceTTDeformation}
\vol_{\tbeta_\KKK(\varphi) } \ = \ \vol_{\tbeta_\KKK(0) }  \ = \  \left( \frac{15}{2\, \Lambda}  \right)^4 \, \vol_{\beta_\SSzero}  \ .
\eeq
In parallel with definition \eqref{NormalizedMetricInflation} in the case of the deformation parameterized by the single field $\phi$,  let us take the normalized internal metric of \eqref{MetricPerturbation2} to be
\beq  \label{NormalizedMetricTTInstability}
\bg_K \ := \ \big( \kappa_P^{-1} \kappa_M \, \Vol_{\tbeta_\KKK(\varphi)} \big) \, \, \tbeta_K(\varphi) \ \  .
\eeq
As in \eqref{MetricPerturbation3}, this leads to a submersive metric on $M\times K$ of the form
\beq  \label{HDMetricTTInstability}
g_P \ = \  \big( \kappa_P^{-1} \kappa_M \, \Vol_{\tbeta_\KKK(\varphi)} \big)^{-1}  \; e^{ \frac{4 \beta \phi}{5}} \; g_M \  +  \ e^{- \frac{\beta \phi}{5}}  \; \tbeta_K(\varphi) \ .
\eeq
This is the higher-dimensional metric that we want to study. It is no longer a warped product metric because $\tbeta_K(\varphi)$ depends on the spacetime coordinates through the field $\varphi$. Our aim is to calculate how the general Lagrangian \eqref{ExpandedLagrangianScalarField} looks like when restricted to this family of higher-dimensional metrics.

Start by observing that the internal volume form associated to $ \tbeta_K(\varphi)$ and $\bg_K$ does not depend on the spacetime coordinates, due to \eqref{VolumeInvarianceTTDeformation}. So the fibres' mean curvature vector determined by the normalized metric $\bg_P$, defined in \eqref{UnscaledMetric}, will vanish. Since the gauge fields are also taken to vanish, we have
\[
N_{\bg_\PPP} \ = \ 0  \ = \  \FF \  .
\]
The kinetic terms for the field $\varphi$ come from the traceless component $\mathring{S}$ of the fibres' second fundamental form. To calculate it, observe that the Lie derivative of $\tbeta_K(\varphi)$ along a vector field $X$ on the base $M_4$ is simply
\beq
\Lie_X \, \big[ \tbeta_K(\varphi) \big] \ =\  \frac{15}{2\, \Lambda}\, \,  \dd \varphi (X) \, \Big[ \ 2 \, e^{2 \varphi} \;  \beta_0 \  |_{\mathfrak{u}(1)} \ - \  2 \, e^{-2 \varphi}  \;  \beta_0 \,  |_{\sutwo} \ + \   e^{\varphi} \;  \beta_0 \,  |_{\CC^2}  \ \Big] \ .
\eeq
These Lie derivatives are symmetric 2-tensors on the internal space. Their inner-product is defined by \eqref{DefinitionInnerProduct} and should be calculated with a $ \tbeta_K(\varphi)$-ortonormal basis $\{e_a\}$ of $TK$. Using the explicit form of $\tbeta_K(\varphi)$  coming from \eqref{TTDeformedInternalMetric} one can compute that
\bal
\left\langle \Lie_{X_\mu}\, \tbeta_K,  \; \Lie_{X_\nu}\, \tbeta_K \right\rangle_{\tbeta_\KKK} \  &= \  \sum\nolimits_{a,b}   \big[ \Lie_{X_\mu}\, g_K \big](e_a, e_b)  \ \big[ \Lie_{X_\nu}\, g_K \big](e_a, e_b)  \nonumber \linebr 
&= \  (\partial_\mu \varphi) (\partial_\nu \varphi) \big[\, 4 \,\dim \mathfrak{u}(1) \ + \ 4\, \dim \sutwo \ + \  \dim \CC^2 \, \big]  \nonumber \linebr 
&= \  20 \,  (\partial_\mu \varphi) (\partial_\nu \varphi)  \ ,
\end{align}
as a function on the base $M_4$. Since a constant rescaling of the internal metric does not affect these inner-products, for the normalized metric $\bg_K$ of \eqref{NormalizedMetricTTInstability} we also have
\[
\left\langle \Lie_{X_\mu}\, \bg_K,  \; \Lie_{X_\nu}\, \bg_K \right\rangle_{\bg_\KKK} \ =\ \left\langle \Lie_{X_\mu}\, \tbeta_K,  \; \Lie_{X_\nu}\, \tbeta_K \right\rangle_{\tbeta_\KKK} \  = \  20 \,  (\partial_\mu \varphi) (\partial_\nu \varphi) \ .
\]
It follows from \eqref{NormT2} that, in the case of vanishing gauge fields, the norm of the fibres' second fundamental form when calculated with the normalized metric $\bg_P$ of \eqref{UnscaledMetric} is just 
\beq
|S_{\bg_\PPP}|^2_{\bg_\PPP} \ = \ |\mathring{S}_{\bg_\PPP}|^2_{\bg_\PPP} \ = \  5 \, g_M^{\mu \nu } \, (\partial_\mu \varphi) (\partial_\nu \varphi) \ .
\eeq
The scalar curvature of $ \tbeta_K(\varphi)$ follows from \eqref{ScalarCurvatureUnstableDeformation} and \eqref{TTDeformedInternalMetric}:
\[
R_{ \tbeta_\KKK(\varphi)}  \ = \  \frac{\Lambda}{5} \, \big(\,    2\, e^{2 \varphi} \, - \, 1 \, +\,  8 \, e^{- \varphi} \,  -\,  e^{-4 \varphi}  \, \big)  \ .
\]
So for the normalized internal metric \eqref{NormalizedMetricTTInstability} we have
\[
R_{ \bg_\KKK(\varphi)}  \ = \  \frac{\Lambda}{5} \, \big( \kappa_P^{-1} \kappa_M \, \Vol_{\tbeta_\KKK(\varphi)} \big)^{-1} \, \big(\,    2\, e^{2 \varphi} \, - \, 1 \, +\,  8 \, e^{- \varphi} \,  -\,  e^{-4 \varphi}  \, \big)  \ .
\]
The remaining constants that appear in the Lagrangian \eqref{ExpandedLagrangianScalarField} are
\bal
\bk \ &= \ \kappa_P^{-1} \kappa_M \,  \Vol_{\bg_\KKK} \ = \ \big( \kappa_P^{-1} \kappa_M \, \Vol_{\tbeta_\KKK (\varphi) } \big)^{5}   \linebr
a_2 \ &= \ a_1 \ = \ \bk^{-1/5} \ = \ \big( \kappa_P^{-1} \kappa_M \, \Vol_{\tbeta_\KKK (\varphi) } \big)^{-1}  \nonumber     \linebr
b_2 \ &= \ 4 \, \beta / 5 \ = \ 4 \, b_1  \ , \nonumber 
\end{align}
where the relation between the volumes of $\bg_K$ and $\tbeta_K (\varphi)$ can be obtained directly from definition \eqref{NormalizedMetricTTInstability}. Bringing together all these formulae, one can finally write down the Lagrangian \eqref{ExpandedLagrangianScalarField} restricted to higher-dimensional metrics $g_P$ of the form \eqref{HDMetricTTInstability}. After integration over the fibre, the result is
\beq \label{ActionTTDeformation}
{\setlength{\fboxsep}{3\fboxsep} \boxed{
\int_M \, \Big[ \, \frac{1}{2\, \kappa_M} \, R_{g_\MMM} \, - \, \frac{1}{2} \, |\dd \phi|^2_{g_\MMM}  \, - \, \frac{5}{2}\, |\dd \varphi|^2_{g_\MMM} \, - \, V(\phi, \varphi) \, \Big]\, \vol_{g_\MMM} \ , 
}}
\eeq
where the potential is 
\bal \label{PotentialTwoDeformations}
V(\phi, \varphi) \ &:= \ \frac{1}{2\, \kappa_M} \, \big[\,  2\, \Lambda\, a_2\, e^{4\beta \phi /5} \, - \,  e^{\beta \phi} \, R_{\bg_\KKK}    \,\big]  \linebr
&= \ \frac{ \kappa_P\, \Lambda}{\kappa_M^2 \ \Vol_{\tbeta_\KKK (\varphi) } } \, e^{4\beta \phi /5} \, \Big[\, 1 \, - \,  \frac{1}{10}\, \big(\,  2\, e^{2 \varphi} \, - \, 1 \, +\,  8 \, e^{- \varphi} \,  -\,  e^{-4 \varphi} \, \big)\, e^{\beta \phi /5}       \,\Big]  \nonumber  \linebr
&= \ \frac{ \kappa_P\, \Lambda^5}{\kappa_M^2 \ \Vol_{\beta_0 } } \, \Big(\frac{2}{15}\Big)^4 \, e^{4\beta \phi /5} \, \Big[\, 1 \, - \,  \frac{1}{10}\, \big(\,  2\, e^{2 \varphi} \, - \, 1 \, +\,  8 \, e^{- \varphi} \,  -\,  e^{-4 \varphi} \, \big)\, e^{\beta \phi /5}       \,\Big]  \ .  \nonumber
\end{align}
The last equality uses the volume identity \eqref{VolumeInvarianceTTDeformation}. When $\varphi =0$ the expression of the potential in the second line manifestly reduces to \eqref{ScalarFieldPotential} in the case $m=4$ and $k=8$, as it should. Omitting a constant factor, the partial derivatives of the potential are
\bal
\frac{\partial V}{\partial \varphi} \ &\propto \  -\,  \frac{1}{10}\, \, e^{\beta \phi} \, \big(\,  2\, e^{\varphi} \, - \, 2 \, e^{-2 \varphi}    \, \big)^2   \linebr
\frac{\partial V}{\partial \phi} \ &\propto \ \frac{4\, \beta}{5}\, \, e^{4\beta \phi /5} \, \Big[\, 1 \, - \,  \frac{1}{8}\, \big(\,  2\, e^{2 \varphi} \, - \, 1 \, +\,  8 \, e^{- \varphi} \,  -\,  e^{-4 \varphi} \, \big)\, e^{\beta \phi /5}       \,\Big]  \nonumber \ .
\end{align}
So the only stationary point occurs at vanishing $\varphi$ and $\phi$, which corresponds to the unscaled, bi-invariant metric on $K$. That is an unstable, saddle point of the potential. Since the second derivative $\frac{\partial^2 V}{\partial \varphi^2}$ also vanishes at this point, while $\frac{\partial^2 V}{\partial \phi^2}$ is negative, the unravelling of the product Einstein metric on $M\times K$ should start by the rescaling perturbation represented by $\phi$, before the slower, TT-perturbation represented by $\varphi$ kicks in to break the isometry group of the internal metric.

\subsubsection*{Mass of the natural gauge bosons} 

In the next few paragraphs we calculate the mass of the gauge bosons associated to the invariant vector fields $e_a^\LL$ and $e_a^\RR$ on $\SU$ when the vacuum metric is $(\phi, \varphi)$-deformed, i.e. when the higher-dimensional metric has the form \eqref{HDMetricTTInstability}. Since a right-invariant field $e_a^\RR$ is Killing for any left-invariant metric on $K$, the bosons associated to $e_a^\RR$ are massless. The squared-mass of the bosons associated to $e_a^\LL$ is given by \eqref{MassBosonsTTDeformation}. These masses are derived from the Lagrangian in the Einstein frame using the formulae of appendix \ref{AppendixBosonsMassEinsteinFrame}.

To start the calculation observe that the vector fields $e_a^\LL$ have vanishing divergence with respect to any left-invariant metric, because the group $\SU$ is unimodular and \eqref{BiinvarianceVolumeForm} applies. So, according to formula \eqref{MassGaugeBosons3}, in the Einstein frame the squared-mass of the gauge boson associated with the field $e_a^\LL$ is given by
\begin{equation} \label{MassGaugeBosons6}
m^2(e_a^\LL) \ = \ e^{\beta \phi} \ \frac{ \int_K \; \left\langle \Lie_{e_a}\, \bg_K,  \; \Lie_{e_a}\, \bg_K \right\rangle_{\bg_K}   \vol_{\bg_\KKK} }{ 2 \int_K   \bg_K (e_a ,  e_a  ) \, \vol_{\bg_\KKK} } \ = \   e^{\beta \phi} \  \frac{ \left\langle \Lie_{e_a}\, \bg_K,  \; \Lie_{e_a}\, \bg_K \right\rangle_{\bg_K}  }{ 2 \,  \bg_K (e_a ,  e_a  )  }   \ .
\eeq
The last equality uses the left-invariance (and hence full constancy) of the integrand functions over $K$. In the calculation of $V(\phi, \varphi)$ we chose a  normalized metric $\bg_K$ related to $\tbeta_K (\varphi)$ by the constant rescaling \eqref{NormalizedMetricTTInstability}. Since the inner-product $\left\langle \Lie_{e_a}\, \bg_K,  \; \Lie_{e_a}\, \bg_K \right\rangle_{\bg_K} $ is invariant under such rescalings, we can write 
\begin{equation} \label{MassGaugeBosons7}
m^2(e_a^\LL) \ = \ \frac{e^{\beta \phi}}{\kappa_P^{-1} \kappa_M \, \Vol_{\tbeta_\KKK(\varphi)}}  \, \,  \frac{ \left\langle\, \Lie_{e_a}\, \tbeta_K(\varphi),  \; \Lie_{e_a}\, \tbeta_K(\varphi) \, \right\rangle_{\tbeta_K(\varphi)}  }{ 2 \,\,  \tbeta_K(\varphi) (e_a ,  e_a  )  }     \ . 
\eeq
The rightmost fraction vanishes when $e_a$ is in the subspace $\utwo$ of $\su$, because $e_a^\LL$ is Killing in this case. When $e_a$ is in the subspace $\CC^2 \subset \su$, the value of the rightmost fraction can be calculated by applying \eqref{InnerProductLieDerivativesBeta} to the metric $\tbeta_K (\varphi)$ as defined in \eqref{TTDeformedInternalMetric}. The result is 
\[
 \frac{ \left\langle\, \Lie_{e_a}\, \tbeta_K(\varphi),  \; \Lie_{e_a}\, \tbeta_K(\varphi) \, \right\rangle_{\tbeta_K(\varphi)}  }{ 2 \,\,  \tbeta_K(\varphi) (e_a ,  e_a  )  }  \ = \ \frac{\Lambda}{5}  \, \Big[ \,      (e^{\varphi} \, - \,  e^{-2 \varphi})^2 \, +\,   (1 \, - \,  e^{- \varphi})^2 \, \Big] \ .
\]
Combining this expression with the volume identity
\[
\Vol_{\tbeta_\KKK(\varphi)} \ = \ \left( \frac{15}{2\, \Lambda}  \right)^4 \, \Vol_{\beta_\SSzero}  \ , 
\]
coming from \eqref{VolumeInvarianceTTDeformation}, we finally obtain that the squared-mass of the gauge field $A^a_\mu$ associated to $e_a^\LL$ is given by
\beq \label{MassBosonsTTDeformation}
{\setlength{\fboxsep}{3\fboxsep} \boxed{
m^2(e_a^\LL)  \ = \  \frac{3}{2} \left( \frac{2\, \Lambda}{15} \right)^5  \frac{e^{\beta \phi}}{\kappa_P^{-1} \kappa_M \, \Vol_{\beta_\SSzero}} \ \Big[ \,      (e^{\varphi} \, - \,  e^{-2 \varphi})^2 \, +\,   (1 \, - \,  e^{- \varphi})^2 \, \Big] 
}}
\eeq
 when $e_a$ is in the subspace $\CC^2$ of $\su$. There are four independent gauge fields in that subspace, and this formula expresses their bosons' mass as a function of the internal deformation fields, $m^2 = m^2(\phi,\varphi)$. The bosons associated with the vector fields $e_a^\LL$ in the subspace $\utwo \subset \su$  remain massless under the $(\phi,\varphi)$-deformations of the internal metric.

\subsection{Stabilizing the internal curvature} 
\label{StabilizingInternalCurvature}

The potential $V(\phi, \varphi)$ described in \eqref{PotentialTwoDeformations} decreases as $\phi$ becomes non-zero and $\varphi$ takes increasingly positive values. So one expects that the initial Einstein product metric on $M\times K$, if it ever represented the system over an ancient region of spacetime, will unravel along the unstable deformations represented by those fields. This means that the internal space would go through a global rescaling and, simultaneously, through a slower TT-deformation that breaks the isometry group and increases the internal scalar curvature. But will these deformations of the internal geometry increase indefinitely, across all orders of magnitude of size and curvature, as suggested by the classical potential? Or at some scale new physics will kick in, physics not contained in the classical Einstein-Hilbert action, to stabilize the metric deformation?

Generally speaking, one may expect that new physics will start to be relevant for sufficiently small internal spaces or for large curvatures. One should not be able to confine quantum particles in arbitrarily small internal directions, nor is GR tested for arbitrarily large curvatures. The question would then be how to go beyond the Einstein-Hilbert action and model mathematically such effects. One can always add to the action new ad hoc terms that depend on the geometry of $K$ and become relevant for big deformations. For example, taking
\[
\int_P \  \big(\, R_{g_\PPP}  \ - \ 2\, \Lambda \ - \ W(g_P) \, \big) \ \vol_{g_\PPP} \ 
\]
and choosing $W = \zeta \, R_{g_\KKK}^2$ to be a quadratic function of the internal scalar curvature, which is a well-defined function on $P$, one can check that for small, positive values of $\zeta$ the effective potential has a local minimum and the rescaling is stabilized once $K$ attains a large curvature. More generally, it would be interesting to investigate the behaviour of $R_{g_\PPP}^2$-gravity or $f(R_{g_\PPP})$-gravity models in this Kaluza-Klein setting. In another line, considering connections with torsion in the internal directions would also be interesting. This is because the Ricci scalar of such a connection will be the traditional scalar curvature plus a term involving the norm of the torsion. This additional term affects the effective potential and may help to counterbalance the runaway scalar curvature under some of its instabilities.

For internal Einstein metrics with no TT-instabilities, where the rescaling represented by $\phi$ is the only instability, one can also follow the tradition of single-field models of inflation and simply experiment with different ad hoc potentials $W(\phi)$, designed to prevent an indefinite rescaling of $K$ and, at the same time, complement the $V(\phi)$ of \eqref{ScalarFieldPotential} to provide a good quantitative model for the initial stages of inflation.

Yet another possibility would be to introduce an effective potential inspired by the QFT vacuum energy density. Well-known QFT calculations suggest that the 4D vacuum energy density depends on the masses of the quantum fields. The contribution of a gauge field of mass $m$ to the vacuum energy density is calculated in \cite{Martin} (after \cite{Ak}) to be  
\beq \label{GaugeBosonsVacuumEnergyDensity2}
\frac{3\, m^4}{64 \pi^2} \, \log\Big(\frac{m^2}{\mu^2} \Big) \ ,
\eeq
where $\mu$ is a new mass scale, not necessary close and presumably smaller than the Planck mass. But in Kaluza-Klein models a gauge field's (classical) mass is given by formulae such as \eqref{MassGaugeBosons} or \eqref{MassGaugeBosons5}, that depend on the vacuum internal metric $g_K^\Szero$. So we have $m^2 = m^2 (g_K^\Szero)$ and a deformation of the internal metric will affect the vacuum energy density. This creates an effective potential $\rho_{\rm vac}^{4D} (g_K)$ that can be added to the 4D action. 

For example, when $K$ is a compact, unimodular Lie group, with its invariant and divergence-free vector fields $\{ e_a^\LL \}$ and $\{ e_a^\RR \}$, one could consider a density  $\rho_{\rm vac}^{4D}(g_K)$ proportional to \eqref{GaugeBosonsVacuumEnergyDensity2} with $m^2$ given by the natural term 
\beq \label{NaturalMassTerm}
 \frac{\kappa_P}{2\, \kappa_M \, (\Vol_{g_\KKK})^2} \, \int_K  \left[ \,  (g_K^\LL)^{ab} \, \big\langle \Lie_{e_a^\LLL}\, g_K,  \, \Lie_{e_b^\LLL}\, g_K \big\rangle \, + \, (g_K^\RR)^{ab} \,  \big\langle \Lie_{e_a^\RRR}\, g_K,  \, \Lie_{e_b^\RRR}\, g_K \big\rangle  \right]  \vol_{g_\KKK}  .
\eeq
Here $(g_K^\LL)^{ab}$ denotes the entries of the inverse matrix of $g_K (e_a^\LL, e_b^\LL)$, and similarly for $(g_K^\RR)^{ab}$. This term coincides with the sum of the squared-masses of the gauge bosons associated to the invariant fields $e_a^\LL$ and $e_a^\RR$, as given by the mass formula \eqref{MassGaugeBosons5}.

In our specific example $K = \SU$, the masses of these gauge bosons were previously calculated for the $(\phi,\varphi)$-deformed, higher-dimensional metric \eqref{HDMetricTTInstability}. The bosons associated to all the fields $e_a^\RR$ and to the fields $e_a^\LL$ in the subspace $\utwo \subset \su$ remain massless under $(\phi,\varphi)$-deformations of the internal metric. The four bosons associated to the $e_a^\LL$ in the subspace $\CC^2$ of $\su$ gain a non-zero squared-mass given by \eqref{MassBosonsTTDeformation}. So the effective potential would be something like
\beq \label{EffectivePotential1}
V_{{\rm eff}} (\phi, \varphi) \ = \ V(\phi, \varphi)  \ + \   \frac{3 \, \zeta}{64\pi^2}\,\, m^4(\varphi, \phi) \, \log \Big[\, \frac{m^2(\varphi, \phi) }{\mu^2} \, \Big]  \ ,
\eeq
where $\zeta$ is a dimensionless, positive constant and there are only four non-zero contributions to the term $m^2(\phi,\varphi)$ of \eqref{NaturalMassTerm}, each one given by \eqref{MassBosonsTTDeformation}. Now the question is whether this effective potential has local minima where the deformation fields can settle, instead of increasing indefinitely. In other words, can an increase of the vacuum energy density compensate the falling potential $V(\phi, \varphi)$ for large values of $\phi$ and $\varphi$? 
The power $m^4(\phi,\varphi)$ grows as $e^{2 \beta \phi} \, e^{4\varphi}$ for large positive values of $\varphi$, as is manifest from \eqref{MassBosonsTTDeformation}. The initial potential $V(\phi, \varphi)$, given by \eqref{PotentialTwoDeformations}, decreases as $- e^{\beta \phi} \, e^{2\varphi}$ for large values of $\varphi$ and $\phi$. So the second term in $V_{{\rm eff}}$ will dominate the initial potential for large, positive values of both $\phi$ and $\varphi$. This means that, in this regime, the unstable $(\phi,\varphi)$-deformations of $g_K$ could be stabilized by contributions coming from the vacuum energy density. A more detailed calculation would be needed to analyze the case where $\phi$ becomes negative when unraveling from the initial value $\phi = 0$.

We also note that, once the internal deformation is stabilized at a value $(\phi_\Szero, \varphi_\Szero)$, the dimensionally-reduced Lagrangian implied by \eqref{ActionTTDeformation} and \eqref{EffectivePotential1} is 
\beq 
\int_M \, \Big[ \, \frac{1}{2\, \kappa_M} \, R_{g_\MMM} \, - \, V_{{\rm eff}} (\phi_\Szero, \varphi_\Szero) \, \Big]\, \vol_{g_\MMM} \ . 
\eeq
So the value of the effective potential at its minimum, more precisely $\kappa_M \, V_{{\rm eff}} (\phi_\Szero, \varphi_\Szero)$, acts as the de facto 4D cosmological constant. In particular, the unravelling of the primordial Einstein metric along instabilities, by lowering the potential energy of the fields, does indeed contribute to reduce the value of that constant. Having in mind the shape of the rescaling potential of figure \ref{fig:HillTopPotential}, it seems that an unraveling of the rescaling instability in the $\phi < 0$ direction could more easily produce a positive, very small value of the cosmological constant after the corrections contained in $V_{{\rm eff}}$ are considered.

The stabilization of the metric deformations at values $(\phi_\Szero, \varphi_\Szero)$ would fix the present-time, internal vacuum metric $g_K^\Szero$. This metric determines all the different gauge couplings and gauge bosons' masses in the model, as described before. The masses of the Higgs particles, in turn, will be determined by the second derivatives of the effective potential $V_{{\rm eff}} (g_K)$ at its minimum $g_K^\Szero$, as in the usual Higgs mechanism.


Finally, we stress that the arguments in this section merely pretend to suggest that it may be possible to stabilize the deformations of the internal metric using physics not contained in the Einstein-Hilbert action. For instance, using the QFT vacuum energy density. These are not fully justified calculations. The usual notion of mass and formalism leading to the QFT formula \eqref{GaugeBosonsVacuumEnergyDensity2}, for example, are developed on a Minkowski background, not on a general, or even de Sitter, four-dimensional background. It also uses the boson's renormalized mass, not the classical bare mass. Even within our Kaluza-Klein framework, in this section we have considered only the contribution of the most basic gauge fields---those associated with the invariant vector fields $e_a^\LL$ and $e_a^\RR$ on $K$---to the vacuum energy density, whereas in a general submersion presumably all massive gauge and scalar fields could contribute. Also fermions should presumably contribute. Lastly, we note that the stabilization of the TT-deformation field $\varphi$ would follow from both mass formulae \eqref{MassGaugeBosons} or \eqref{MassGaugeBosons3}, while the stabilization of the rescaling field $\phi$ in this simple setting only follows in the case of the Einstein frame mass formula \eqref{MassGaugeBosons3}.

\subsubsection*{Electroweak symmetry breaking}
 
 In the last few sections we saw how the bi-invariant metric on $\SU$ is naturally unstable. In a dynamical Kaluza-Klein model, it should unravel along a TT-deformation that breaks the isometry group down to $( \SU \mkern-1mu \times \mkern-1mu \mathrm{SU}(2) \mkern-1mu  \times \mkern-1mu \mathrm{U}(1)  ) /  \mathbb{Z}_6$. Then we argued that it may be possible to stabilize this deformation using an ad hoc potential, as is done in models of cosmological inflation, or using results from quantum physics, such as the dependency of the vacuum energy density on the bosons' mass and hence on the internal geometry. Under the stabilized internal metric, the four gauge bosons associated with the broken internal symmetries are massive.

 The unavoidable next question is, of course, how to go farther and account for the electroweak symmetry breaking at a lighter mass scale. It is easy enough to find a second TT-deformation of the internal metric, represented by a new scalar field $\sigma$, 
 \[
\tbeta_K = \tbeta_K (\varphi, \sigma) \ ,
\] 
that further breaks the isometry group down to $\SU \mkern-1mu \times \mkern-1mu  \mathrm{U}(1)$. An explicit example is given in \cite{Ba1}, formula 5.6. More generally, see the parameterization of inequivalent left-invariant metrics on $\SU$ given in \cite{Coq} for different isometry groups.
But will any such deformation increase the internal scalar curvature, and hence decrease the potential $V(\phi, \varphi, \sigma)$, to allow for a spontaneous metric unravelling in this direction? The answer to this question should follow from a mechanical but long calculation, using formula \eqref{GeneralScalarCurvature} for the scalar curvature on $K$. 
If the deformation $\sigma$ does decrease the potential, the next question is how can the newly broken symmetries correspond to light bosons, with a mass much smaller than the scale \eqref{ScaleGaugeCouplings} of the gauge couplings? In other words, how can the deformation field $\sigma$ be stabilized at a very small value $\sigma_\Szero$, so that some Lie derivatives $\Lie_{e_a^\LL} \,\, \tbeta_K (\varphi_\Szero, \sigma_\Szero)$ appearing in \eqref{MassGaugeBosons5} will be non-zero but still have a norm many orders of magnitude below the unity? 

The answer to this last question, if it exists within the Kaluza-Klein framework, should probably rely again on physics beyond the classical Einstein-Hilbert action. The addition of new terms to the action, on the way to an effective potential, entails the introduction of new constants, such as the numbers $\zeta$ or $\mu$ appearing in the vacuum energy density in \eqref{EffectivePotential1}. These constants can {\it a priori} be very different from the Planck constant, since they represent the scale of the new physics. They will appear in the extended equations of motion and, in favourable conditions, can affect the classical field configurations that minimize the effective potential. Thus, when we go beyond the Einstein-Hilbert action, the classical vacuum metric $g_K^\Szero$ can in principle have small ``wrinkles'' produced by components in more than one length scale, so may be able to generate gauge bosons with very different masses.

\newpage

\section{Comments on fermions}
\label{CommentsFermions}

The purpose of this section is to briefly discuss some of the implications for fermions of the Kaluza-Klein picture described previously in these notes. Having gauge fields associated to non-isometric diffeomorphisms of the internal space creates both opportunities---such as the possibility to generate mass for the gauge bosons within the Kaluza-Klein framework, or the possibility to evade the Atiyah-Hirzebruch theorem---and additional theoretical challenges---such as understanding how fermions should transform under diffeomorphisms that do not preserve the metric. This challenge exists in any gravity theory, not just in Kaluza-Klein.

We start the section discussing possible new ways to circumvent the traditional no-go arguments against chiral fermions in Kaluza-Klein. Then we spend most of the time describing three possible geometric approaches to the transformation of fermionic fields under non-isometric diffeomorphisms of the underlying space. Some of these approaches already exist in the literature, in different forms, but do not seem to be too widely explored. They imply thinking about the definition of spinor itself, and how to extend it to objects that are not tied down to a fixed background metric.


\subsection{No-go arguments against chiral fermions}
\label{NoGoArguments}

In the standard Kaluza-Klein framework, as in \cite{Witten83}, fermions are represented by spinors on $M\times K$ satisfying a Dirac-like, higher-dimensional equation of motion. The observed fermions with their light masses (when compared to the Planck mass) are associated to zero modes of the internal Dirac operator $\sD^K$, or to zero modes of a deformation $\sM^K$ of that operator generally called the internal mass operator. It is necessary to consider deformations because the Schr{\"o}dinger-Lichnerowicz formula implies that $\sD^K$ does not have zero modes on an internal space with positive scalar curvature, such as our $K$ with the metrics $g^e_K$ or $g^\Szero_K$, but a deformation $\sM^K$ may well have them. If the mass operator on $K$ still anticommutes with the internal chirality operator, as $\sD^K$ does, then it must have an equal number of zero modes with positive and negative internal chirality,  i.e. it must have a vanishing index. This follows from the fact that $\sD^K$ has vanishing index (because it has no zero modes), together with the invariance of the index under continuous operator deformations that preserve the anticommutation relation with the chirality operator. This property of the mass operator does not contradict experiment, even assuming that the internal chirality is correlated with spacetime chirality, because there is a equal number of left-handed and right-handed fermions in the Standard Model.

Let us denote by $H$ the isometry group of the internal vacuum metric $g^\Szero_K$. The Dirac operator commutes with isometries, and so should the mass operator if we assume it to be natural. So there is a natural $H$-representation on the finite-dimensional space of zero modes of $\sM^K$. The (Kosmann) Lie derivative of a spinor along a Killing vector field commutes with the chirality operator. So there are separate $H$-representations on the spaces of positive and negative chirality zero modes of $\sM^K$. In general these are reducible representations. Now fix an irreducible $H$-representation $\rho$. The number of independent, positive chirality zero modes of $\sM^K$ that transform according to $\rho$ minus the number negative chirality zero modes that transform according to the same $\rho$ is called the $\rho$-index of the operator $\sM^K$. A result of Atiyah and Hirzebruch says that the $\rho$-index of $\sD^K$ vanishes when $K$ is connected, even-dimensional and the group $H$ is compact connected \cite{AH}. By invariance of the $\rho$-index under deformations of the operator that preserve its $H$-invariance, also the $\rho$-index of $\sM^K$ should vanish under the same conditions. As Witten famously pointed out in \cite{Witten83}, this means that one cannot obtain realistic Kaluza-Klein models based on a connected, even-dimensional $K$ and an isometry group $H=G_{\text{SM}}$ acting on the zero modes of $\sM^K$ with spacetime chirality correlated to internal chirality. In such a model the right-handed fermions would transform exactly as the left-handed fermions do under the Standard Model group $G_{\text {SM}}$, and this goes against the observed, chiral nature of fermions.

This argument raises a significant difficulty in the traditional Kaluza-Klein framework. As discussed in the Introduction, to address that difficulty in these notes we suggest that the Standard Model group $G_{\text {SM}}$ should not be identified with the exact isometry group $H$ of $g^\Szero_K$, but should instead be identified with the larger group $G'$ mentioned in the beginning of the section. This seems quite natural because, in the Kaluza-Klein framework, the mass of a gauge boson is proportional to the Lie derivative of $g^\Szero_K$ along associated internal symmetry. So only the massless bosons should be associated with exact isometries of the vacuum metric. Associating the weak force to nearly-Killing, but not exactly-Killing vector fields on the internal space would have the following advantages:
\begin{itemize}
\item[i)] the weak bosons would gain a non-zero mass within the Kaluza-Klein framework;
\item[ii)] the internal symmetries associated to the weak force would not commute with $\sD^K$ or $\sM^K$, so the weak field would be able to mix fermions with different masses, as happens in the Standard Model;
\item[iii)] The restrictions dictated by the Atiyah-Hirzebruch theorem only apply to representations of the isometry group of $g^\Szero_K$, so {\it a priori} the weak gauge fields would be able to couple differently to the positive and negative chirality zero modes of $\sM^K$, potentially circumventing the no-go arguments of \cite{Witten83}.
\end{itemize}
Besides the transversal argument based on the Atiyah-Hirzebruch theorem, there are other arguments in the literature that rule out the existence of chiral fermions in specific dimensions, for example restricting realistic Kaluza-Klein theories to internal spaces of dimension $k=4n+2$. These arguments are based on a study of the spinor representations in different dimensions and their interplay with the chirality operator and complex conjugation \cite{Witten81, Wetterich83, Witten83}. Once again, these arguments do not directly apply to weak gauge fields associated with non-Killing vector fields on the internal space, essentially because it is unclear how fermionic fields should transform under non-isometric diffeomorphisms of internal space.  The usual definitions of spin structure, spin bundle and spinor all start with a manifold equipped with a fixed background metric. The spin groups themselves are double covers of the special orthogonal groups in different dimensions and signatures. When the background metric is allowed to change under full diffeomorphism symmetries, as in genuine gravity or Kaluza-Klein theories, the transformation laws of spinors and fermionic fields are not established as a settled matter.

This brings us to the next part of these comments, which turns around the question of how fermions should transform under non-isometric diffeomorphisms. We describe three possible geometric approaches to address this question. Two of these approaches already exist in the literature, in different forms, but do not seem to be too widely explored. It would be interesting to further study them while having the Kaluza-Klein framework in mind. The purpose would be to investigate if they can be useful to model any properties of the interactions of the weak force with fermions.

Before finishing this section, however, we should also point out that the classical Kaluza-Klein literature contains several alternative proposals to circumvent the no-go arguments that rule out the existence of chiral fermions. Those proposals include using non-compact internal spaces \cite{Wetterich84}, adding gauge fields to the higher-dimensional theory \cite{CM, CS} or generalizing the assumptions of Riemannian geometry \cite{Weinberg84}, each having its own advantages and drawbacks. More references can be found in the reviews \cite{Bailin, CFD}.

\subsection{Spinors and \texorpdfstring{$\widetilde{\GL}_k$}{tGL} representations}
\label{InfiniteSpinorRepresentations}

One approach to discussing the transformation of fermions under general diffeomorphisms is to work with $\tGLp_k$ groups and representations, instead of working with the traditional spin groups. The next few paragraphs give a quick outline of the geometrical setting, although we do not study the $\tGLp_k$-representations themselves. In the physics literature this approach is generally considered in the case of four-dimensional spinors and Lorentzian geometry (e.g. \cite{Neeman, Mi, DP, NS}). Due to our Kaluza-Klein setting, here we will mostly be thinking about the Riemannian geometry of spinors in the internal space. 

For $k \geq 3$ the identity component of the general linear group, denoted by $\GL^+_k$,  has a universal double cover that satisfies the commutative diagram
\beq \label{TildeGL}
\begin{tikzcd}
\Spin_k ( \mathbb{R}) \ \arrow[hookrightarrow]{r} \arrow[swap]{d}{\chi} &\  \tGLp_k( \mathbb{R}) \arrow{d}{\chi} \\%
\SO_k ( \mathbb{R}) \ \arrow[hookrightarrow]{r}   & \ \GL^+_k ( \mathbb{R}) 
\end{tikzcd}
\eeq
Here the vertical arrows are double covers and the horizontal arrows are inclusions of groups. There is an infinite number of inequivalent representations of $\Spin_k ( \mathbb{R})$ on finite-dimensional vector spaces, the usual spinor representations. In contrast, it is well-known that the enlarged group $\tGLp_k( \mathbb{R})$ has no faithful finite-dimensional representations \cite[ch. II.5]{LM}. All  its finite-dimensional representations factor through $\GL^+_k ( \mathbb{R})$ and hence are not fermionic in nature. There are, however, genuine $\tGLp_k( \mathbb{R})$-representations on  infinite-dimensional vector spaces. Consider one such representation
\[
\sigma: \  \tGLp_k(\mathbb{R}) \times \Delta \ \longrightarrow \ \Delta  \ .
\]
Restricting $\sigma$ to the maximal compact subgroup $\Spin_k ( \mathbb{R})$ inside $\tGLp_k( \mathbb{R})$, the infinite-dimensional space $\Delta$ decomposes as sum
\beq \label{DecompositionDelta}
\Delta \ = \ \bigoplus_{m=1}^\infty \, \Delta_m  \ ,
\eeq
where each $\Delta_m$ is a finite-dimensional, irreducible representation space of $\Spin_k ( \mathbb{R})$. The spin action preserves the subspaces $\Delta_m$ but the $\tGLp_k$-action will in general mix them, and so will the action of any subgroup of $\tGLp_k$ not contained in the spin group. Exactly which inequivalent spin representations appear in decomposition \eqref{DecompositionDelta}, as well as their multiplicities, depends of course on the initial $\tGLp_k$-representation $\sigma$. It is not clear to the author which collections of irreducible spin representations $\{ \Delta_m: m\in \mathbb{N} \}$ can be obtained by restricting some infinite-dimensional $\tGLp_k$-representation. For example, is there a $\sigma$ such that all its components $\Delta_m$ are equivalent to the basic half-spin representations?

The next step is to go from representations to spinors on manifolds, so one needs to talk about spin structures. Let $F^+  (K) \rightarrow K$ denote the bundle of oriented frames in the tangent bundle $TK$. A topological spin structure on $K$ is a double cover $\theta: \widetilde{F}^+ (K) \rightarrow F^+  (K)$ equipped with a $\tGLp_k$-action on the right that satisfies 
\beq \label{EquivarianceSpinStructure}
\theta (p\cdot h) \ = \ \theta (p)\cdot \chi (h)
\eeq
for all spinorial frames $p \in \widetilde{F}^+ (K)$ and all group elements $h$ in $\tGLp_k ( \mathbb{R})$ \cite[ch. 2]{Bourguignon}. Here $\chi$ is the double cover map of diagram \eqref{TildeGL}.
When the manifold $K$ has a metric $g$ there is a canonical relation between this notion of spin structure and the standard one. With a metric one can pick the orthonormal frames of $TK$ among all the oriented ones, and so define the sub-bundle $F_\sSO (K, g)$ of orthonormal oriented frames inside $F^+  (K)$. Restricting the double cover $\theta$ to the inverse image of $F_\sSO (K, g)$, we get a submanifold $F_\sSpin (K, g) \subset \widetilde{F}^+ (K)$ that is a double cover of $F_\sSO (K, g)$. Due to the commutation of \eqref{TildeGL}, it satisfies the equivariance property \eqref{EquivarianceSpinStructure} for any $h$ in $\Spin_k ( \mathbb{R})$. In other words, for each metric $g$ on $K$ the topological spin structure $\theta$ restricts to a standard spin structure $F_\sSpin (K, g)$ that fits in the commutative diagram of principal bundles over $K$:
\beq \label{RestrictionSpinStructure}
\begin{tikzcd} 
F_\Spin (K, g) \ \arrow[hookrightarrow]{r} \arrow[swap]{d}{\theta} &\  \widetilde{F}^+ (K) \arrow{d}{\theta} \\%
F_\SO (K,g) \ \arrow[hookrightarrow]{r}   & \ F^+ (K)
\end{tikzcd} 
\eeq
Here the vertical arrows are double covers and the horizontal arrows are inclusions. The spinor bundle of $S^\sigma$ is defined as the associated vector bundle 
\[
S^\sigma  := \widetilde{F}^+(K) \times_\sigma \Delta \longrightarrow \ K \ . 
\]
Its fibres are isomorphic to the infinite-dimensional representation space $\Delta$. Each metric $g$ on $K$ defines a different sub-bundle $F_\Spin (K, g) \subset \widetilde{F}^+ (K)$ and, from \eqref{DecompositionDelta}, a different decomposition 
\beq  \label{DecompositionSDelta}
S^\sigma  \ = \ \bigoplus_{m=1}^\infty  \ S^\sigma_m (g)  \ .
\eeq
The components $S^\sigma_m (g)$ are finite-dimensional vector bundles over $K$. For two different metrics on $K$ the respective decompositions are related by an automorphism of $S^\sigma$, in other words by a $\tGLp_k$-gauge transformation, so there are isomorphisms $S^\sigma_m (g) \simeq S^\sigma_m (g')$ between the component vector bundles. 

Now let $\rho$ denote a left-action of a connected Lie group $G$ on the manifold $K$. Using the derivative maps, $\rho$ has a lift to a $G$-action on the tangent bundle $TK$ that is linear on the fibres. Since a frame is just a collection of tangent vectors, $\rho$ also has a lift to a $G$-action on the frame bundle $F^+  (K)$. This lifted left-action, denoted by $\rho$ as well, commutes with the standard right-action of $\GL^+_k ( \mathbb{R})$ on $F^+ (K)$. 

Since the topological spin structure $\widetilde{F}^+ (K) \rightarrow F^+ (K)$ is a double cover and $G$ is connected, standard results guarantee that the action $G \times F^+ (K)  \rightarrow F^+ (K)$ lifts to a $G$-action on $\widetilde{F}^+ (K)$ or, at worst, to an action of a double cover group $\tilde{G}$ of $G$ (e.g. \cite[Th. I.9.1]{Brendon}). This lifted action, denoted by $\trho$, fits into the commutative diagram of left-actions
\beq \label{DiagramLeftActions}
\begin{tikzcd}
\tilde{G} \times \widetilde{F}^+ (K)   \arrow[swap]{r}{\trho}  \arrow[swap]{d} &  \widetilde{F}^+ (K)  \arrow{d} \\%
G \times F^+ (K)  \arrow[swap]{r}{\rho}     \arrow[swap]{d} & F^+ (K)  \arrow{d}  \\
G \times  K  \arrow[swap]{r}{\rho}   &  K
\end{tikzcd}
\eeq
Since the left-action $\rho$ commutes with the right-action of $GL^+_k ( \mathbb{R})$ on $F^+ (K)$, it follows that the lift $\trho$ commutes with the right $\tGLp_k (\mathbb{R})$-action on $\widetilde{F}^+ (K)$. This implies that $\trho$ induces a left-action on the total space of the associated bundle $\widetilde{F}^+(K) \times_\sigma \Delta$, which of course is the spinor bundle $S^\sigma$. This action is linear on the fibres and fits into the commutative diagram
\beq \label{DiagramLeftActionsSpinors}
\begin{tikzcd}
\tilde{G} \times S^\sigma  \arrow[swap]{r}{\trho}  \arrow[swap]{d} &  S^\sigma \arrow{d} \\%
G \times K  \arrow[swap]{r}{\rho}   &  K
\end{tikzcd}
\eeq
If the manifold $K$ is equipped with a metric $g$ and the group $G$ acts through isometries, then the lifted action $\trho$ preserves the spinor sub-bundles $S^\sigma_m (g)$ in decomposition \eqref{DecompositionSDelta}. In this case the $G$-action does not mix spinors with values in different representation spaces $\Delta_m$, even though some of these spaces may just be copies of each other. In contrast, if $G$ does not act through isometries and only a subgroup $G' \subset G$ does, then $G'$ preserves the representation spaces but another part of $G$ can mix them. 

Coming back to our proposed Kaluza-Klein models, where a gauge group $G_{\rm SM}$ acts on the internal space $K$ but only a subgroup $G'$ preserves the vacuum metric, that interplay of the group representations with the different $\Delta_m$ is not too dissimilar from the way that the distinct parts of  $G_{\rm SM}$ preserve or mix the different fermionic generations.

In the eight-dimensional example $K=\SU$, the smallest non-trivial representation spaces of $\Spin_8 ( \mathbb{R})$ are of course the eight-dimensional representation spaces $\Delta_v$, $\Delta_+$ and $\Delta_-$, related to each other by the triality outer-automorphisms of $\Spin_8 ( \mathbb{R})$. It would be interesting to understand how different infinite-dimensional representations of $\tGLp_8( \mathbb{R})$ mix these three subspaces when restricted to compact subgroups not entirely contained in $\Spin_8 ( \mathbb{R})$. For instance when restricted to $\Utwo$-subgroups with $U(2) \cap \Spin_8 ( \mathbb{R}) = U(1)$.

The existence of the lifted action $\trho$ on the spinor bundle $S^\sigma$, depicted in diagram \eqref{DiagramLeftActionsSpinors}, also guarantees that there is a well-defined notion of Lie derivative $\Lie_X \psi$ of a $\tGLp$-spinor along any vector field $X$ on $K$. This derivative can be defined by a formula analogous to \eqref{DefnLieDerivativeSpinors}. It satisfies the standard identity $[\Lie_X, \, \Lie_Y] \psi = \Lie_{[X, Y]} \psi$ for all fields $X$ and $Y$. In contrast, the usual Kosmann-Lichnerowicz derivative of traditional spinors on $S^\sigma(g,K)$ satisfies this identity only when $X$ or $Y$ are Killing fields on $(K, g)$ \cite{Kosmann}.

\subsection{Extended spinors}
\label{Extended spinors}

A second approach to modeling the transformation of fermions under non-isometric diffeomorphisms would be to work with what we call extended spinors, a simpler version of universal spinors. These fields have values on a spin representation space $\Delta$, for example the usual, finite-dimensional, spin-$1/2$ representation spaces, but instead of being defined on the manifold $K$, the fields are now defined on the total space of the inner-product bundle $\MT (K)$ over $K$. For our purposes, the main advantage of working with extended spinors is that they have a natural transformation rule under non-isometric diffeomorphisms of $K$, unlike the traditional spinors. They exist whenever $K$ has a spin structure and restrict to the traditional spinors once a metric on $K$ is chosen. We now describe their construction, which we have not been able to find in the literature.

The oriented frame bundle $F^+ (K) \rightarrow K$ has a right-action of $\GL_k^+ ( \mathbb{R})$ mixing the vectors that compose the frame.  It is a free and transitive action. Identifying frames related by the subgroup of $\SO_k ( \mathbb{R})$-transformations, we can consider the quotient space
\beq \label{DefInnerProductBundle}
\MT (K) \ := \ F^+ (K) \, / \, \SO_k ( \mathbb{R}) \ = \ \widetilde{F}^+ (K) \, / \, \Spin_k ( \mathbb{R}) \ .
\eeq
Then $\MT (K)$ is a bundle over $K$ with fibres isomorphic to $GL_k^+  / \SO_k $, so of real dimension $k(k+1)/2$. We will call it the inner-product bundle of $TK$. Each point in $\MT (K)$ over a base point $x\in K$ represents a $\SO_k$-equivalence class of frames on the tangent space $T_x K$, and hence determines a unique inner-product on $T_xK$. In other words, $\MT (K)$ can be thought of as the space of pairs $(x, h_x)$, where $x$ is a point in $K$ and $h_x$ is an inner-product on $T_x K$. The bundle $\MT (K)$ is the open subset of the symmetric bundle $\Sym^2 (T^\ast K)$ defined by picking only the positive-definite products on the fibres. A section of $\MT (K) \rightarrow K$ is just a metric on the tangent bundle of $K$. The inner-product bundle fits in the commutative diagram of bundles over $K$:
\beq \label{DiagramProductBundle}
\begin{tikzcd}
\widetilde{F}^+ (K)   \arrow[swap]{r}  \arrow[swap]{rd} &  F^+ (K) \arrow{d}  \\
  & \MT (K)  
\end{tikzcd}
\eeq
Here the horizontal arrow is a double cover, the vertical arrow is a principal $\SO_k ( \mathbb{R})$-bundle and the diagonal arrow is a principal $\Spin_k ( \mathbb{R})$-bundle. We are of course assuming the existence of a topological spin structure $\widetilde{F}^+ (K)$ on $K$, as described near \eqref{EquivarianceSpinStructure}. For a given spin representation $\sigma: \Spin_k ( \mathbb{R}) \times \Delta \rightarrow \Delta$, which now can be finite-dimensional, the extended spinor bundle of $\ST^\sigma$ is defined as the associated vector bundle 
\beq \label{DefExtSpinorBundle}
\ST^\sigma (K) \ := \ \widetilde{F}^+ (K) \times_{\sigma}  \Delta \ \longrightarrow \  \MT (K) \ .
\eeq
So it is a bundle with fibre $\Delta$ over the larger base $\MT (K)$. In particular, when $\Delta$ is the usual spin-$1/2$ representation space, with its Clifford multiplication of spinors by $\SO_k$-vectors, there will also be a canonical multiplication of extended spinors by vectors tangent to $K$, as happens with the usual spinors. A section of $\ST^\sigma (K)$ is locally represented by a $\Delta$-valued map $\psi (x, h_x)$ on the total space of the inner-product bundle.

The relation between the extended spinor bundle and the usual spinor bundles over $K$ can be described as follows.
 A metric $g$ on $K$ is the same thing as a section of the inner-product bundle $\MT (K) \rightarrow K$, so defines a map
\[
g: \  K\ \longrightarrow \ \MT (K) \ .
\]
The pullback by this map of the extended spinor bundle $\ST^\sigma (K) \rightarrow \MT (K)$ is a vector bundle over $K$ with fibre $\Delta$. This pullback bundle is then isomorphic to the usual spinor bundle $\ST^\sigma (K, g) \rightarrow K$ determined by the representation $\sigma$, the metric $g$ and the spin structure on $K$ obtained by restriction of the topological spin structure, as in \eqref{RestrictionSpinStructure}. So for every metric $g$ on $K$ we can write simply
\[
 g^\ast(\ST^\sigma (K)) \ \simeq \ \ST^\sigma (K, g) \ \longrightarrow \ K \ 
\]
as vector bundles over $K$. Thus, in a sense, the extended spinor bundle over $\MT (K)$ contains all the usual spinor bundles over $K$ determined by the different metrics on that manifold.

There is a second natural construction of the extended spinor bundle, which we now describe. Take the principal $\SO_k ( \mathbb{R})$-bundle $F^+ (K) \rightarrow \MT (K)$ depicted in the commutative diagram \eqref{DiagramProductBundle}. Using the fundamental representation of this group, we can consider the associated bundle
\beq \label{AssociatedBundle2}
E(K) \ :=  \  F^+ (K) \times_{\SO_k}  \mathbb{R}^k \ \longrightarrow \  \MT (K) \ .
\eeq
It is a bundle with typical fibre $\mathbb{R}^k $ over the base $\MT (K)$. It is isomorphic to the pullback bundle
\[
E(K) \ \simeq  \ \pi^\ast  (TK) \ , 
\]
where $\pi: \MT(K) \rightarrow K$ denotes the natural projection. Since a point $q$ in $\MT(K)$ represents an inner-product on the tangent space $T_{\pi(q)} K$, it follows that the vector bundle $E(K) \rightarrow \MT(K)$ has a canonical inner-product on its fibres, defined simply by 
\[
\langle u, v \rangle_{q} \ := \ q(u, v) \ ,
\]
where $u$ and $v$ are vectors in the fibre $E_q \simeq T_{\pi(q)} K$. Moreover, since $E(K)$ is a pullback of the tangent bundle $TK$, the topological spin structure on $K$ determines a topological spin structure on the vector bundle of oriented frames of $E(K)$, which can then be restricted to the $\langle \cdot , \cdot \rangle$-orthogonal frames. So for a given spin representation $\sigma$ we have another natural spinor bundle over $\MT (K)$, namely the spinor bundle $S^\sigma (E)$ associated to $E$, to $\langle \cdot , \cdot \rangle$ and that spin structure \cite{LM}. There is, however, an isomorphism of vector bundles
\[
S^\sigma (E) \ \simeq  \  \ST^\sigma(K) \ \longrightarrow \  \MT (K) \ ,
\]
so this is just a second description of the same extended spinor bundle \eqref{DefExtSpinorBundle}.

The main advantage of working with the extended spinor bundle over the restricted spinor bundles is that it has a natural action of general diffeomorphisms of $K$, even when they are not isometries. We will now describe how this comes about.
Let $\rho$ denote a left-action of a connected Lie group $G$ on the manifold $K$. This group can be as large as the identity component of the diffeomorphism group of $K$, denoted by $\Diff^\Szero(K)$. As noted in diagram \eqref{DiagramLeftActions}, $\rho$ can be lifted to left $G$-actions on the tangent bundle $TK$ and on the oriented frame bundle $F^+  (K)$. The latter commutes with the right $\GL_k$-action on that bundle, so it follows from definition \eqref{DefInnerProductBundle} that it descends to a $G$-action on the inner-product bundle $\MT (K)$. As also noted near diagram \eqref{DiagramLeftActions}, after passing to a double cover $\tilde{G}$ of $G$ if necessary, standard results (e.g. \cite[Th. I.9.1]{Brendon}) guarantee the action $\rho$ has another lift $\trho$ to the double cover $\widetilde{F}^+  (K)$ that commutes with the right $\tGLp_k$-action on that cover. So there are induced actions on the associated bundles \eqref{AssociatedBundle2} and  \eqref{DefExtSpinorBundle}  that fit in the commutative diagrams
\beq \label{DiffActionESpinors}
\begin{tikzcd}
G \times E (K)  \arrow[swap]{r}{\rho}  \arrow[swap]{d} &  E(K) \arrow{d} &   & \tilde{G} \times \ST^\sigma (K)  \arrow[swap]{r}{\trho}  \arrow[swap]{d} &  \ST^\sigma (K) \arrow{d}  \\%
G \times \MT (K)  \arrow[swap]{r}{\rho} \arrow[swap]{d}  & \MT (K)  \arrow{d}  &  & G \times \MT (K)  \arrow[swap]{r}{\rho} \arrow[swap]{d}  & \MT (K)  \arrow{d}  \\
G \times K  \arrow[swap]{r}{\rho}   & K  & &  G \times K  \arrow[swap]{r}{\rho}   & K
\end{tikzcd} 
\eeq
Extended spinors are defined to be the sections of $ \ST^\sigma (K)$. Therefore, as desired, there is a natural $\widetilde{\Diff}{\vphantom{\Diff}}^\Szero(K)$-action on extended spinors. 

Let us write down these actions in somewhat more detail. For any group element $f \in G$ the transformation $\rho_f$ is a diffeomorphism of $K$. It acts on points in $K$; through the derivative map it acts on vectors tangent to $K$; and through pullback it acts on inner-products on the tangent spaces to $K$. The inner-product bundle $\MT (K)$ can be thought of as the space of pairs $(x, h_x)$, where $h_x$ is an inner-product on the tangent space $T_xK$. So the induced left-action on $\MT(K)$ can be written as
\[
(x, \, h_x)  \ \longmapsto \ \big(\, \rho_f(x), \,(\rho_f^{-1})^\ast (h_x) \,  \big) \ .
\]
A point in the associated bundle $E(K)$, as defined in \eqref{AssociatedBundle2}, can be written as an equivalence class $[ x, \, h_x, \, \{ e_a\} , \, v]$, where $v$ is a vector in $\mathbb{R}^k$, $x$ is point in $K$, $h_x$ is an inner-product on $T_x K$ and $\{ e_a\}$ is an oriented, $h_x$-orthonormal frame of $T_x K$. Then the lifted $G$-action on $E(K)$ is determined by the  transformation rule
\[
\big[ \, x, \, h_x, \, \{ e_a\} , \, v \, \big]  \ \longmapsto \ \big[\, \rho_f(x), \,(\rho_f^{-1})^\ast (h_x) , \,\{ (\rho_f)_{\ast} \, e_a \} , \, v \,  \big]  \ ,
\]
where $(\rho_f)_{\ast} \, e_a$ denotes the pushforward of the vector $e_a$ in the frame. The action is well-defined because $(\rho_f)_{\ast}$ commutes with the right $\SO_k$-action that mixes the vectors $e_a$ of the $h_x$-orthonormal frame. Note also that the base $\{ (\rho_f)_{\ast} \, e_a \}$ is still orthonormal with respect to the pullback inner-product $(\rho_f^{-1})^\ast (h_x)$ on the tangent space $T_{\rho_f(x)}K$.

The lifted $\tilde{G}$-action on the extended spinor bundle $\ST^\sigma(K)$ can be written down in similar terms. A point in this bundle is an equivalence class $[ x, \, h_x, \,\{ \tilde{e}_a\} , \, s]$, where $s$ is a vector in $\Delta$ and $\{ \tilde{e}_a\}$ is a lift to $\widetilde{F}^+ (K)$ of an oriented, $h_x$-orthonormal frame $\{ e_a \}$ of $T_xK$. The action of $\tilde{f} \in \tilde{G}$ on the total space of $\ST^\sigma(K)$ is then determined by the transformation rule
\[
[ \,x, \, h_x, \,\{ \tilde{e}_a\} , \, s \, ]  \ \longmapsto \ \big[\, \rho_f(x), \,(\rho_f^{-1})^\ast (h_x) , \, \{  \tilde{\rho}_{\tilde{f}} \, ( \tilde{e}_a) \} , s \,  \big] \ .
\]
The spin frame $ \{ \tilde{\rho}_{\tilde{f}} \, ( \tilde{e}_a) \}$ is a lift to  $\widetilde{F}^+ (K)$ of the $(\rho_f^{-1})^\ast (h_x)$-orthonormal frame $\{ (\rho_f)_{\ast} \, e_a \}$ of the tangent space $T_{\rho_f(x)}K$. 

Extended spinors have an operation of Clifford multiplication and a natural inner-product. They can also be differentiated using Lie derivatives and a compatible covariant derivative. These derivations are described in section \ref{LieDerivativesSpinors} for universal spinors but also apply to the special case of extended spinors.

\subsection{Universal spinors}
\label{Universal spinors}

A third approach to modeling the transformation of fermions under non-isometric diffeomorphisms of internal space is to work with universal spinors, a sort of spinors that are defined for all metrics simultaneously. We now describe this approach. Universal spinors have been considered before in the mathematical literature, although rarely, it seems, and sometimes in slightly different guises (e.g. \cite{Swift, Ammann}). The extended spinors of the previous section can be regarded as a special, finite-dimensional case of universal spinors, a relation that we also describe below.

Let $K$ be a connected manifold and let $\MM$ be the space of Riemannian metrics on it. Universal spinors are sections $\psi(g,x)$ of a universal spinor bundle $S \rightarrow \MM \times K$. This vector bundle restricts to the usual $S(g,K)$ over each slice $\{g\} \times K$, has an operation of Clifford multiplication, has a natural inner-product on the fibres, and also comes equipped with a compatible connection. Unlike the restricted spinor bundles, the universal bundle has an action of the double cover of the diffeomorphism group of $K$. This  means that universal spinors have a natural transformation rule under general diffeomorphisms of $K$.

The construction of the universal spinor bundle is very similar to the construction of the extended spinor bundle described in the last section. The reader may want to skip the details. We start be considering a sort of universal tangent bundle, i.e. a vector bundle $E \rightarrow \MM \times K$ whose fibres are isomorphic to the tangent spaces to $K$. More precisely, take the projection $\pi_2: \MM \times K \rightarrow K$ and use the pullback of bundles to define
\beq \label{UniversalTangentBundle}
E := \pi_2^\ast (TK) \ \longrightarrow \ \MM \times K \ .
\eeq
So $E$ is a vector bundle over $\MM \times K$ with fibres $E_{(g,x)} \simeq T_x K$. This vector bundle has a natural inner-product $\langle \cdot , \cdot \rangle$ on the fibres. Given two vectors $u$ and $v$ in a fibre $E_{(g,x)}$, it is defined simply by 
\beq \label{NaturalProductE}
\langle u, v \rangle_{(g,x)} \ := \ g_x(u, v) \ .
\eeq
The diffeomorphism group of $K$ acts both on $K$ and, by pullback, on the metrics in $\MM$. So there is a natural left-action 
\[
\rho:\  \Diff (K) \times \MM \times K \ \longrightarrow \ \MM \times K \qquad \quad  \rho_f (g, x) \ :=  \left( (f^{-1})^\ast g, \, f(x) \right) \ .
\]
This action lifts to the total space of the vector bundle $E \rightarrow \MM \times K$. Given a diffeomorphism $f$ and a vector $v$ in the fibre $E_{(g,x)}\simeq T_x K$, the derivative map of $f$ allows one to define 
\beq \label{ActionE}
\rho_f (v)\ :=\  (\dd f)_x (v)  \quad \in \ T_{f(x)}K \simeq E_{\rho_f (g, x)} \ .
\eeq
Using the definition of $\langle \cdot , \cdot \rangle$, it is clear that the inner-product of any two sections of $E$ satisfies
\beq \label{ActionPreservesProduct}
\langle\,  \rho_f (u), \, \rho_f (v)\, \rangle \circ \rho_f \ = \    \langle u,\, v \rangle 
\eeq
as a function on $\MM \times K$. So the lifted action preserves the natural inner-product on $E$.

To talk about universal spinors we also need a spin structure on the vector bundle $E$. So let us take the usual path and denote by $F_\sSO  (E) \rightarrow \MM \times K$ the bundle of oriented frames on $E$ that are orthonormal with respect to $\langle \cdot , \cdot \rangle$. It is a principal bundle with a natural right-action of $\SO_k ( \mathbb{R})$. The restriction of $F_\sSO  (E)$ to a slice $\{g\} \times K$ is isomorphic to the bundle $F_\sSO (K,g)$ of orthonormal frames on $(TK, g)$. A spin structure on $E$ is a double cover 
\[
\theta:\ F_\sSpin (E)\ \longrightarrow \ F_\sSO  (E)
\]
equipped with a $\Spin_k ( \mathbb{R})$-action on the right that satisfies
\beq \label{EquivarianceSpinStructure2}
\theta (p\cdot h) \ = \ \theta (p)\cdot \chi (h)
\eeq
for all $p \in F_\sSpin (E)$ and all group elements $h$ in $\Spin_k ( \mathbb{R})$. Here $\chi$ denotes the standard double cover of groups $\Spin_k ( \mathbb{R}) \rightarrow \SO_k ( \mathbb{R})$. 
Since $E$ is a pullback of the tangent bundle $TK$, a topological spin structure on $K$, as defined in \eqref{EquivarianceSpinStructure}, determines a topological spin structure on the bundle of oriented frames of $E$, which can then be restricted to the $\langle \cdot , \cdot \rangle$-orthogonal frames. So any topological spin structure on $K$ determines a spin structure on the bundle $E \rightarrow \MM \times K$, i.e. the cover $F_\sSpin (E)$ exists whenever $K$ is spin.

Given this fact, the universal spinor bundles are defined in the usual way. For a fixed representation $\sigma: \Spin_k ( \mathbb{R}) \times \Delta \rightarrow \Delta$, the universal spinor bundle $S^\sigma (K)$ is just the associated bundle
\[
S^\sigma (K) \ := \ F_\sSpin (E) \times_\sigma \Delta \ \longrightarrow \ \MM \times K \ .
\]
Since the restriction of the principal bundle $F_\sSpin (E)$ to a slice $\{g\} \times K$ is isomorphic to the bundle $F_\sSpin (K, g)$ over $K$ determined by the same topological spin structure on $K$, it follows that the restriction of $S^\sigma (K)$ to $\{g\} \times K$ is isomorphic to the classical spinor bundle $S^\sigma(K, g) \rightarrow K$ determined by $F_\sSpin (K, g)$.

We have already seen that the left-action $\rho$ of $\Diff (K)$ on the product $\MM \times K$ lifts to an action on $E$ that is linear on the fibres and preserves the inner-product $\langle \cdot , \cdot \rangle$. Since a frame on $E$ is just a collection of vectors, $\rho$ also induces an action on $F_\sSO  (E)$ of the identity component $\Diff^\Szero(K)$ of the diffeomorphism group. The restriction to the identity component guarantees the transformed frames, besides being orthonormal, are oriented as well. This induced left-action, also denoted by $\rho$, commutes with the right-action of $\SO_k (\mathbb{R})$ on $F_\sSO  (E)$.

Now let $G$ denote the identity component $\Diff^\Szero(K)$ or a connected subgroup of it. Since $F_\sSpin (E) \rightarrow F_\sSO  (E)$ is a double cover and $G$ is connected, standard results guarantee that the action $G \times F_\sSO (E)  \rightarrow F_\sSO(E)$ lifts to an action of $G$ on $F_\sSpin (E)$ or, at worst, to an action of a double cover group $\tilde{G} \rightarrow G$ (e.g. \cite[Th. I.9.1]{Brendon}). This lifted action, denoted by $\trho$, therefore fits into the commutative diagram of left-actions
\[ 
\begin{tikzcd}
\tilde{G} \times F_\sSpin(E)  \arrow[swap]{r}{\trho}  \arrow[swap]{d} &  F_\sSpin (E) \arrow{d} \\%
G \times F_\sSO (E)  \arrow[swap]{r}{\rho}     \arrow[swap]{d} & F_\sSO(E)   \arrow{d}  \\
G \times \MM \times K  \arrow[swap]{r}{\rho}   & \MM \times K
\end{tikzcd}
\]
Since the left-action $\rho$ commutes with the right-action of $\SO_k (\mathbb{R})$ on $F_\sSO  (E)$, it follows that the lift $\trho$ commutes with the right $\Spin_k (\mathbb{R})$-action on $F_\sSpin (E)$. This implies that $\trho$ induces a left-action on the total space of associated bundles $ F_\sSpin (E) \times_\sigma \Delta$, which of course are the spinor bundles $S^\sigma(K)$. This action is linear on the fibres and preserves the inner-product on $S^\sigma(K)$. So we have another commutative diagram
\beq \label{DiffActionUSpinors}
\begin{tikzcd}
\tilde{G} \times S^\sigma(K)  \arrow[swap]{r}{\trho}  \arrow[swap]{d} &  S^\sigma (K) \arrow{d} \\%
G \times \MM \times K  \arrow[swap]{r}{\rho}   & \MM \times K
\end{tikzcd}
\eeq
Universal spinors are defined to be the sections of the universal bundle $S^\sigma(K)$. Therefore, as desired, there is a natural $\widetilde{\Diff}{\vphantom{\Diff}}^\Szero(K)$-action on universal spinors, but not on the usual restricted spinors, defined for a fixed background metric $g$ on $K$.

When $\sigma$ is the fundamental spin representation, the lifts of $\rho$ to the spinor bundle $S^\sigma (K)$ and to the universal tangent bundle $E$ defined in \eqref{UniversalTangentBundle} are compatible with each other under Clifford multiplication. This means that for every spin diffeomorphism $\tilde{f}$ in $\tilde{G}$ we have
\beq \label{ActionCliffordMult}
\trho_{\tilde{f}} (v\cdot \psi) \ = \ \rho_{\chi(\tilde{f})} (v)\, \cdot\, \trho_{\tilde{f}} (\psi)  
\eeq
as sections of $S^\sigma (K)$. Here $\chi$ denotes the double cover map $\tilde{G} \rightarrow G$, $v$ is any section of $E$ and $\psi$ is any universal spinor.

\subsubsection*{Relation between universal and extended spinors}

Universal spinors are $\Delta$-valued maps $\psi (g, x)$ defined on the product space $\MM\times K$. 
Physically, it seems natural to introduce the following locality conditions on these fields: 
\begin{itemize}
\item[{\bf L1)}] $\psi (g, x)  = \psi (g', x)$ whenever the metrics $g$ and $g'$ coincide in a neighbourhood of the point $x\in K$;
\item[{\bf L2)}] $\psi (g, x)  = \psi (g', x)$ whenever the metrics $g$ and $g'$ coincide at the point $x\in K$.
\end{itemize}
Both conditions are diffeomorphism-invariant. If they are satisfied by a given universal spinor $\psi$, then they are satisfied by all the transformed spinors $\trho_h (\psi)$, where $h \in \widetilde{\Diff}{\vphantom{\Diff}}^\Szero(K)$ and $\trho$ is the action defined in \eqref{DiffActionUSpinors}. A universal spinor $\psi$ that satisfies L2 will automatically satisfy L1, so the second condition is more restrictive. It is equivalent to saying that $\psi$ descends to a field on the inner-product bundle $P(K)$, defined in the previous section. In this sense, the extended spinors of the previous section can be regarded as the special family of universal spinors satisfying L2. In contrast, condition L1 is equivalent to saying that the universal spinor $\psi$ descends to a field on the total space of germs of metrics over $K$, denoted here by $\mathcal{P}(K)$. 
Now consider the natural projection maps
\[
\MM \times K  \ \stackrel{\xi_1}{\longrightarrow} \ \mathcal{P}(K)  \  \stackrel{\xi_2}{\longrightarrow} \ P(K) \ \qquad \quad (g, \, x)\  \longmapsto\  ([g_x], \,x)\  \longmapsto\  (g_x,\, x) \ .
\]
It is clear that the universal bundles $E$ and $S^\sigma(K)$, as defined in this section, are isomorphic to the pullbacks by $\xi_2 \circ \xi_1$ of the analog bundles defined in the previous section for extended spinors:
\[
E_{\mathrm{uni}} (K) \ \simeq \ (\xi_2 \circ \xi_1)^\ast \,  E_{\mathrm{ext}}(K) \qquad \quad \ \  S_{\mathrm{uni}}^\sigma(K) \ \simeq \ (\xi_2 \circ \xi_1)^\ast \, \, \ST_{\mathrm{ext}}^\sigma (K) \ .
\]
It also clear that by considering the pullback bundles $\xi_2^\ast \,  E_{\mathrm{ext}}(K)$ and $\xi_2^\ast \,  \ST_{\mathrm{ext}}^\sigma (K)$ over the space of germs $\mathcal{P}(K)$, one can define a third notion of spinors, larger than the extended spinors of the previous section and smaller than the universal spinors of this section. Those spinors will be $\Delta$-valued maps defined on the total space of $\mathcal{P}(K)$. They will also have an action of the spin diffeomorphism group $\widetilde{\Diff}{\vphantom{\Diff}}^\Szero(K)$.

\subsubsection*{Fermions and universal spinors}

The relevant question for Kaluza-Klein is whether physical fermions in a gravity theory can be modeled by any of these enlarged notions of spinor. In this case fermions would not be represented by simple fields $\psi(x)$ on spacetime, but would also depend explicitly on the background metric. They would be modeled by fields $\psi (g, x)$ on the product space $\MM \times K$, in the case of universal spinors, or by fields $\psi (x, h_x)$ on the total space of the inner-product bundle $\MT (K)$, in the case of extended spinors. 

One can also conceive that physical fermions could be modeled by objects that depend only on the spacetime coordinates and on the vacuum metric. In this case, denoting by $\MM^\Szero \subset \MM$ the subset of vacuum metrics on $K$, one would restrict the universal spinor bundle to the domain $\MM^\Szero \times K$ and consider sections of that bundle. So fermions would be modeled by $\Delta$-valued maps $\psi (g, x)$ defined on $\MM^\Szero \times K$. These maps still have an action of the group $\widetilde{\Diff}{\vphantom{\Diff}}^\Szero(K)$ provided that $\MM^\Szero$ is preserved by diffeomorphisms, which will be true for any diffeomorphism-invariant action for the metric. Note that the space $\MM^\Szero$ may no longer be contractible to a point, as happens for $\MM$ in the case of Riemannian metrics, so may be a space with non-trivial topology. In the limit, one can even try to model fermions with universal spinors restricted to the product $\mathcal{O}_{g^\SSzero} \times K$, where $\mathcal{O}_{g^\SSzero} \subset \MM^\Szero$ denotes the orbit of the chosen vacuum metric under the diffeomorphisms of $K$. These restricted universal spinors would still transform naturally under non-isometric diffeomorphisms of the vacuum and would still generalize the usual spinors, which are defined only over $\{ g^\Szero\} \times K$ for a fixed vacuum metric.

A finite-dimensional version of this setting can be studied when $K$ is a Lie group and, instead of considering all metrics and all diffeomorphisms, one considers only left-invariant metrics on $K$ and the canonical diffeomorphisms determined by left and right-multiplication on the group. In this case the space $\MM_\LL$ of left-invariant metrics is finite-dimensional and the vacuum submanifold $\MM^\Szero_\LL$ can be compact. If one restricts universal spinors $\psi (g, x)$  to the domain $\MM^\Szero_\LL \times K$, then a harmonic expansion of $\psi$ can be obtained from the tensor product of harmonic expansions in the $K$-component and in the $\MM^\Szero_\LL$-component. This would increase the multiplicity that each mode of the Dirac operator on $K$ appears in the overall harmonic expansion, in a phenomenon reminiscent of the fermionic generations. Only the non-isometric diffeomorphisms of $K$ coming from right-multiplication act non-trivially on the component $\MM^\Szero_\LL$, so only these diffeomorphisms would be able to mix $K$-modes associated to different $\MM^\Szero_\LL$-harmonics in the tensor product expansion.

\subsection{Lie and covariant derivatives of universal spinors}
\label{LieDerivativesSpinors}

The next few paragraphs describe the existence of natural derivations of universal spinors. Their main properties are stated below and will be justified in more detail elsewhere. Although we work with the more general universal spinors, analog derivations with similar properties also exist for the extended spinors of section \ref{Extended spinors}.

We have seen before that the universal spinor bundle $S^\sigma (K) \rightarrow \MM \times K$ has an action $\trho$ of the double cover of the identity component of the diffeomorphism group of $K$. This covers the natural $\Diff^\Szero (K)$-action $\rho$ on the product $\MM \times K$, as in diagram \eqref{DiffActionUSpinors}. These actions induce a natural Lie derivative $\Lie_X \psi$ of universal spinors through the definition
\beq \label{DefnLieDerivativeSpinors}
(\Lie_X \psi)(g,x)  \ := \ -\, \frac{\dd}{\dd t}\ \,  \trho_{\Phi_t^X} \, \circ \,  \psi \, \circ \,  \rho^{-1}_{\Phi_t^X} (g,x) \ \  |_{t=0} \ .
\eeq
Here $X$ is any vector field on $K$ and $\Phi_t^X$ denotes its flux, so an element of  $\Diff^\Szero (K)$. This definition of Lie derivative is quite general. It can be applied to sections of any vector bundle with an action that is linear on the fibres. In particular one can use it to define Lie derivatives of extended spinors or of the spinors of section \ref{InfiniteSpinorRepresentations}.

Definition \eqref{DefnLieDerivativeSpinors} implies the standard identities:
\begin{align} \label{NaturalnessLieDerivative}
\Lie_X \Lie_Y \psi \, -\,  \Lie_Y \Lie_X \psi \ &=\ \Lie_{[X,Y]} \psi   \\
\Lie_X (\psi_1\, +\,  f \psi_2) \ &=\ \Lie_X \psi_1 \, + \, f\, \Lie_X \psi_2 \, + \, \dd f(X_{\MM\times K}) \, \psi_2  \nonumber
\end{align}
for all vector fields $X$ and $Y$ on $K$ and for all scalar functions $f$ on $\MM \times K$. Here $X_{\MM\times K}$ denotes the vector field on $\MM\times K$ induced by $X$ and the action $\rho$, i.e.
\beq \label{InducedVectorField}
X_{\MM\times K} \ |_{(g,x)} \ := \  \frac{\dd}{\dd t}\ \,  \rho_{\Phi_t^X} (g,x) \ \, |_{t=0} \ = \ -\,  \Lie_X g \ + \ X \, |_x
\eeq
as a tangent vector in $T_{(g,x)}(\MM \times K)$, which is isomorphic to $\Sym^2(T^\ast K) \oplus T_xK$.
In contrast, note that the standard Kosmann-Lichnerowicz derivative of the usual spinors on $S(K, g) \rightarrow K$ satisfies the first identity in \eqref{NaturalnessLieDerivative} only when $X$ or $Y$ are Killing fields on $(K,g)$ \cite{Kosmann}. 

The action $\trho$ preserves the natural inner-product on the fibres of $S^\sigma (K)$, which are isomorphic to the spin representation space $\Delta$. This property implies that the induced Lie derivative satisfies
\beq  \label{LieDerivativeProduct2}
\langle \Lie_X \psi_1 ,\, \psi_2 \rangle \ + \ \langle \psi_1 ,\,  \Lie_X \psi_2 \rangle \ = \  \left[ \dd \langle \psi_1 ,\, \psi_2 \rangle \right] (X_{\MM\times K}) \ , 
\eeq
as functions on $\MM \times K$, for all vector fields $X$ on $K$.

The vector bundle $E \rightarrow \MM \times K$, defined in \eqref{UniversalTangentBundle}, also has a left-action of the identity component of the diffeomorphism group of $K$, as described in \eqref{ActionE}. So a formula analogous to \eqref{DefnLieDerivativeSpinors} can be used to define a natural Lie derivative $\Lie_X v$ of the sections of $E$. This guarantees that the derivative will have properties analogous to \eqref{NaturalnessLieDerivative}. Since this $\Diff^\Szero (K)$-action preserves the natural inner-product on $E$, as described in \eqref{ActionPreservesProduct}, the corresponding Lie derivative will also satisfy a formula analogous to \eqref{LieDerivativeProduct2}. 

Because $E$ is just the pullback bundle $\pi_2^\ast (TK)$, a section $v$ of $E$ can also be viewed as a vector field on the product manifold $\MM \times K$. It is a vector field that projects to zero under the pushforward map $(\pi_1)_\ast$, where $\pi_1 : \MM \times K \rightarrow \MM$ denotes the natural projection. Using the explicit form of the action on $E$, described in \eqref{ActionE}, one can then derive the simple formula
\[
\Lie_X v \ = \ [X_{\MM\times K}, \, v] \ .
\]
On the right-hand side we are taking the Lie bracket of vector fields on $\MM\times K$. Note that $X_{\MM\times K}$ is a field on $\MM\times K$ projectable to $\MM$. In fact we have
\[
(\pi_1)_\ast (X_{\MM\times K}) \ |_g  \  = \ -\, \Lie_X g
\] 
by formula \eqref{InducedVectorField}. Then standard identities of submersions \cite[ch. 9.C]{Besse} imply that 
\[
(\pi_1)_\ast \left( [X_{\MM\times K}, \, v] \right)  \ =\   \left[ (\pi_1)_\ast \left(X_{\MM\times K} \right) , \, (\pi_1)_\ast\left( v  \right) \right] \ = \ 0 \ ,
\]
because $v$ projects to zero. So the bracket $[X_{\MM\times K}, \, v]$ projects to zero under $(\pi_1)_\ast$, and hence is a well-defined section of $E$.

When $\sigma$ is the fundamental spin representation, it follows from \eqref{ActionCliffordMult} and definition \eqref{DefnLieDerivativeSpinors} that the Lie derivatives $ \Lie_X  v$ and $\Lie_X \psi$ are compatible with each other under Clifford multiplication, in the sense that 
\beq
\Lie_X ( v \cdot \psi) \ = \ ( \Lie_X  v ) \cdot \psi \, + \, v\cdot (\Lie_X \psi)
\eeq
for all vector fields $X$ on $K$. 

The universal spinor bundle $S^\sigma (K) \rightarrow \MM \times K$ has a natural connection $\nabla$. This connection restricts to the Levi-Civita connection for derivations along vector fields on $K$. For derivations along vector fields on $\MM$, it restricts to the connection described by Bourguignon and Gauduchon, defining the parallel transport of spinors under a change of metric \cite{BG}. The connection $\nabla$ is compatible with the natural inner-product of universal spinors, in the sense that 
\beq \label{CovariantDerivativeProduct}
\langle \nabla_W \psi_1 ,\, \psi_2 \rangle \ + \ \langle \psi_1 ,\,  \nabla_W \psi_2 \rangle   \ = \ \left[ \dd \langle \psi_1 ,\, \psi_2 \rangle \right] (W)
\eeq
for all vector fields $W$ on $\MM \times K$. It is also compatible with the Lie derivative of universal spinors along vectors fields on $K$, in the sense that
\beq  \label{LieDerivativeConnection}
\Lie_X (\nabla_W \psi) \ = \ \nabla_{\Lie_X W}\, \psi \ + \ \nabla_W (\Lie_X \psi)
 \eeq
for all vector fields $X$ on $K$ and all fields $W$ on $\MM \times K$. The connection $\nabla$ is not flat. Its curvature has non-zero components when restricted to the $K$-factor in the product $\MM \times K$, also when restricted to the $\MM$-factor, and also for the mixed components with one leg in each factor. 

The vector bundle $E \rightarrow \MM \times K$ also has a connection $\nabla$ compatible with the natural inner-product on its fibres, defined in \eqref{NaturalProductE}. It satisfies identities analogous to \eqref{CovariantDerivativeProduct} and \eqref{LieDerivativeConnection}, but applied to vector sections $v$ instead of spinor sections $\psi$. When $\sigma$ is the fundamental spin representation, the natural connections on $E$ and on $S^\sigma (K)$ are compatible with Clifford multiplication, in the sense that  
\beq
\nabla_W ( v \cdot \psi) \ = \ ( \nabla_W  v ) \cdot \psi \, + \, v\cdot (\nabla_W \psi)
\eeq
for all vector fields $W$ on $\MM \times K$.  

One can also write down a formula expressing the Lie derivative of universal spinors in terms of the connection $\nabla$ on the vector bundles $E$ and $S^\sigma (K)$. It is
\beq  \label{FormulaLieDerivativeSpinors}
\Lie_X \psi \ = \  \nabla_{X_{\MM} + X} \, \psi \ - \ \frac{1}{8} \, \sum_{a ,\, b}\, \Big(  \langle \nabla_{e_a} X, \, e_b \rangle \, - \,  \langle \nabla_{e_b}X, \, e_a \rangle   \Big) \, (\dd \sigma)(e_a \cdot e_b) \, \psi \ .
\eeq
Here $\{ e_a\}$ denotes a local trivialization of $E \rightarrow \MM \times K$ that is orthonormal with respect to $\langle \cdot, \cdot \rangle$; the linear map $\dd\sigma: {\mathfrak{spin}}_k \rightarrow {\rm{End}}(\Delta)$ is the derivative of the spin representation $\sigma$; the vector field $X$ on $K$ can be viewed as a field on $\MM \times K$ or as a section of $E$ that does not depend on $\MM$; and finally $X_{\MM}$ denotes the vector field on $\MM$ induced by $X$ and the action of $\Diff^\Szero (K)$ on $\MM$, as in \eqref{InducedVectorField}, so that
\[
X_{\MM} \ |_{g} \ = \ -\,  \Lie_X g 
\]
as an element of $T_{g}\MM \simeq \Sym^2(T^\ast K)$. This local expression for the Lie derivative of universal spinors is very similar to the standard Kosmann-Lichnerowicz formula in the case of regular spinors with fixed background metric \cite{Kosmann}. The only difference is the appearance of the field $X_{\MM}$ on $\MM$ and the fact that $\nabla$ is a connection on the universal bundle $S^\sigma (K) \rightarrow \MM \times K$, not on the restricted spinor bundles $S^\sigma (K, g) \rightarrow K$. It follows that the restriction of $\Lie_X \psi$ to a slice $\{g\} \times K$ inside $\MM \times K$ coincides with the Kosmann-Lichnerowicz derivative only when $X_{\MM} \ |_{g} \, = \, -\,  \Lie_X g \, = \, 0 $, i.e. only when $X$ is Killing for the metric $g$.

These canonical derivations can be used to construct natural operators on universal spinors. For example, if $\{ e_a\}$ denotes a local, $\langle \cdot, \cdot \rangle$-orthonormal trivialization of the bundle $E \rightarrow \MM \times K$, one can define
\beq \label{UniversalDiracOperator}
\sD_K \, \psi \ := \ \sum\nolimits_a \, e_a \cdot \nabla_{e_a} \psi  \ .
\eeq
Since the connection $\nabla$ on the universal bundle restricts to the Levi-Civita connection $\nabla^g$ over each slice $\{ g\} \times K$ inside $\MM \times K$, the operator $\sD_K$ restricts to the standard Dirac operator $\sD^g_K$ over those slices, of course.

\subsubsection*{Operators on universal spinors on a Lie group}

Let us now consider the case where the internal space $K$ is a Lie group and restrict universal spinors to the product $\MM_L \times K$, where $\MM_L$ denotes the finite-dimensional space of left-invariant metrics on $K$. Then the tangent bundle $TK$ and the universal bundles $E$ and $S^\sigma (K)$ are all trivial bundles. Let us pick a global, oriented, $\langle \cdot, \cdot \rangle$-orthonormal trivialization $\{ e_a\}$  of the bundle $E \rightarrow \MM \times K$ such that the restriction of $e_a$ to the slices $\{ g\} \times K$ are left-invariant vector fields on $K$. Any such two trivializations are related by a transformation $B: \MM_L \rightarrow {\rm{SO}}_k$ such that 
\beq \label{TransformationTrivialization}
e'_a \, |_{(g,x)} \ = \ \sum\nolimits_b \ B_a^b(g) \,  \, e_b\,  |_{(g,x)} \ . 
\eeq
The components $B_a^b(g)$ do not depend on the point $x\in K$ because both $e'_a$ and $e_a$ are left-invariant vector fields. Now denote by $(e_a)_{\MM}$ the vector field on $\MM_L$ induced by the section $e_a$ of $E$, i.e.
\[
(e_a)_{\MM} \ |_g \ := \ \ - \,\Lie_{e_a (g, \cdot)} g \ , 
\]
where $e_a (g, \cdot)$ denotes the restriction of $e_a$ to the slice $\{ g\} \times K$. The right-hand side is a left-invariant, symmetric 2-tensor on $K$, so an element of $T_g \MM_L$. Since the right-hand side is also $C^\infty(\MM_L)$-linear on the $e_a$ entry and the transformation components $B_a^b(g)$ do not depend on $x$, it follows from \eqref{TransformationTrivialization} that also for the induced fields 
\beq \label{TransformationTrivialization2}
(e'_a)_\MM \, |_{g} \ = \ \sum\nolimits_b \ B_a^b(g) \,  \, (e_b)_\MM \,  |_{g} \   .
\eeq
This transformation rule makes it clear that one can define another operator on universal spinors on a Lie group, besides \eqref{UniversalDiracOperator}, by 
\beq \label{UniversalOperators1}
\sD_\MM \, \psi \ := \ \sum\nolimits_a \, e_a \cdot \nabla_{(e_a)_\MM} \psi   \ .
\eeq
Using the Lie derivative of universal spinors, as written in formula \eqref{FormulaLieDerivativeSpinors}, one can check that also the operator 
\beq \label{UniversalOperators2}
\slLie \, \psi \ := \ \sum\nolimits_a \, e_a \cdot \Lie_{e_a} \psi    
\eeq
is well-defined and independent of the choice of left-invariant trivialization $\{ e_a\}$ of $E$. Using the Kosmann-Lichnerowicz derivative 
\beq  \label{KosmannDerivativeSpinors}
L_X \, \psi \ := \  \nabla_{X} \, \psi \ - \ \frac{1}{8} \, \sum_{a ,\, b}\, \Big(  \langle \nabla_{e_a} X, \, e_b \rangle \, - \,  \langle \nabla_{e_b}X, \, e_a \rangle   \Big) \, (\dd \sigma)(e_a \cdot e_b) \, \psi \ \ , 
\eeq
which is similar to $\Lie_X\psi$ but does not satisfy the first identity in \eqref{NaturalnessLieDerivative}, one can also consider the operator on universal spinors
\beq \label{UniversalOperators3}
\sL \, \psi \ := \ \sum\nolimits_a \, e_a \cdot L_{e_a} \, \psi     \  \ . 
\eeq
This operator does not involve any derivation along the $\MM$-component in the cartesian product $\MM \times K$. So it is also well-defined for traditional spinors on a Lie group equipped with a fixed left-invariant metric. From formula \eqref{FormulaLieDerivativeSpinors} it is clear that $\slLie = \sL + \sD_\MM$.

All these operators are invariant under the $\widetilde{\Diff}{\vphantom{\Diff}}^\Szero(K)$-action on the universal spinor bundle. In fact, one can check that 
\[
[\Lie_X, \,\sD_K ] \ = \ [\Lie_X, \,\sD_\MM] \ = \ [\Lie_X, \,\slLie] \ = \ [\Lie_X, \,\sL ] \ =  \ 0
\]
as operators on universal spinors for all vector fields $X$ on $K$. These commutation relations are not valid in general if the Kosmann-Lichnerowicz derivative $L_X$ is used instead of the Lie derivative $\Lie_X$. For example, $[L_X, \,\sL ]$ vanishes only when $X$ is Killing. If fermions in Kaluza-Klein were to be modeled by universal (or extended) spinors on a Lie group, it would be natural to look for combinations of these operators as models for the internal mass operator.

\vspace*{2cm}

\subsection*{Acknowledgements}

It is a pleasure to thank Kirill Krasnov and Nick Manton for helpful comments on an earlier version of these notes.

\newpage

\vspace{.3cm}

\appendix

\section{Appendices}

\subsection{Weyl rescalings in Riemannian submersions}
\label{AppendixWeylRescalings}

Let $\pi: (P, g_P) \to (M, g_M)$ be a Riemannian submersion with fibre $K$. In this appendix we consider Weyl rescalings of the higher-dimensional metric that act differently on its horizontal and vertical parts, while preserving the submersive structure. As described near \eqref{MetricDecomposition}, a submersive metric $g_P$ is equivalent to a triple $(g_M, A, g_K)$ formed by a metric on the base $M$, the gauge fields $A$ and a family of metrics $g_K$ on $K$, one for each fibre. Thus, given any pair of smooth functions $f$ and $\omega$ on the base $M$ (which can also be regarded as functions on $P$ that are constant along the fibres), one can define a new higher-dimensional metric $\tilde{g}_P$ through the rescaled data
\beq  \label{B.0}
 (\tilde{g}_M,  \,A, \,\tilde{g}_K) \ := \ (e^{2 \omega}\, g_M, \, A, \, e^{2f} \, g_K) \ .
\eeq
Then the projection $\pi: (P, \tilde{g}_P) \to (M, \tilde{g}_M)$ is still a Riemannian submersion. Denote by $m$ and $k$ the real dimensions of the manifolds $M$ and $K$. The volume forms transform according to
\beq  \label{B.1}
\vol_{\tg_\PPP}  \ = \  e^{m\omega+k f} \, \vol_{g_\PPP}    \qquad \qquad  \vol_{\tg_\KKK}  \ = \  e^{k f}  \, \vol_{g_\KKK}  \qquad \qquad  \vol_{\tg_\MMM}  \ = \  e^{m\omega} \, \vol_{g_\MMM} \ .
\eeq
Since the function $f$ is constant along the fibres, the scalar curvature of the fibre metrics rescales simply as 
\beq \label{B.2}
R_{\tg_\KKK} \ = \ e^{-2 f} \,   R_{g_\KKK} \ .
\eeq
For the metric on the base, a well-known formula for the transformation of the scalar curvature under rescalings says that \cite{Wald}
\beq \label{B.3}
R_{\tg_\MMM} \ = \ e^{-2 \omega}\, \big[ \,  R_{g_\MMM} \, - \, 2(m-1)\, \DAlembert_{g_\MMM} \omega  \, - \, (m-1)(m-2)\,   |\dd  \omega |^2_{g_\MMM}  \, \big] \ ,
\eeq
where $\DAlembert_{g_\MMM} = g_M^{\mu \nu}\, \nabla_\mu \nabla_\nu$ denotes the negative Laplacian acting on functions on $M$. Since the gauge fields (the one-form $A$ on $M$ with values on the vertical vector fields on $P$) are not rescaled in \eqref{B.0}, the tensor $\FF$ of \eqref{tensorF} also remains unchanged. The squared-norm $|\FF|^2$, however, depends explicitly on the metrics, so it follows from \eqref{NormF} that it changes as
\beq \label{B.4}
\left|\, \FF\, \right|^2_{\tg_\PPP} \ = \ e^{2(f-2\omega)} \, \left|\, \FF\, \right|^2_{g_\PPP}  \ .
\eeq
The transformation rule for the second fundamental form of the fibres can be deduced from \eqref{ExplicitTensorS2}. Using the properties of the Lie derivative of a 2-tensor and the fact that $f$ is constant along the fibres, one first calculates that 
\bal
(\Lie_{e_a} \tg_P ) (U,V) \ &= \ e^{2f} \, (\Lie_{e_a} g_P ) (U,V)  \nonumber  \linebr
(\Lie_X \tg_P ) (U,V) \ &= \ e^{2f} \, \left[ \, (\Lie_X  g_P ) (U,V)   \ + \ 2\, \dd f(X) \, \, g_P (U,V) \,  \right]  \nonumber  \ .
\end{align}
Then from \eqref{ExplicitTensorS2} one can derive that
\beq \label{B.5}
\tilde{S}_U V \ = \ e^{2(f-\omega)}\, \big[ \, S_U V \ - \  g_P(U, V) \, (\grad_{g_\PPP} f) \, \big]  \ 
\eeq
for any vertical vector fields $U$ and $V$ on $P$. Taking the metric trace of this expression over the vertical sub-bundle produces the transformation rule of the mean curvature vector of the fibres: 
\beq \label{B.6}
\tilde{N} \ = \ e^{-2 \omega}\, \big[ N \ - \ k \, (\grad_{g_\PPP} f) \, \big] \ .
\eeq
This is an equality of horizontal vector fields on $P$. Combining the previous two expressions, it is easy to check that the traceless component of the fibres' second fundamental form, defined by 
\[
\mathring{S}_U V \ := \ S_U V \; - \; \frac{1}{k} \, g_P(U, V) \, N \  \ ,
\]
transforms under the rescalings as
\beq \label{B.12}
\mathring{\tilde{S}}_{U} V \ = \ e^{2(f-\omega)}\, \mathring{S}_{U} V \ .
\eeq
The squared-norms of these tensors, as defined in \eqref{NormsTensors}, transform as
\bal \label{B.7}
\big| \mathring{\tilde{S}} \big|^2_{\tg_\PPP} \ &= \  e^{-2 \omega} \, \big| \mathring{S} \big|^2_{g_\PPP}  \, \nonumber \linebr
| \tilde{N} |^2_{\tg_\PPP} \ &= \  e^{2 \omega}\, | \tilde{N} |^2_{g_\PPP} \ = \ e^{- 2 \omega}\, \big| \, N \, - \,  k\,  \grad_{g_\PPP} f   \, \big|^2_{g_\PPP} \ .
\end{align}
To compute the transformation rule of $\divergence_{g_\PPP} N$, start by observing that it depends on the metric both through $N$ and through the divergence operator. For a fixed vector field $Z$ on $P$, it follows from the general relation $\Lie_Z \vol_g = (\divergence_g Z) \vol_g $ and the rescaling rule for volume forms that the divergence of $Z$ transforms under Weyl rescalings as
\beq \label{B.8}
\divergence_{\tg_\PPP} Z \ = \ \divergence_{g_\PPP} Z \ + \ m\, \dd \omega (Z) \ + \ k\, \dd f (Z)  \ .
\eeq
Combining \eqref{B.8} with expression \eqref{B.6} for $\tilde{N}$, a short calculation then shows that
\beq \label{B.9}
\divergence_{\tg_\PPP} (\tilde{N})  \ = \  e^{-2\omega} \, \Big\{  \divergence_{g_\PPP} (N) \, + \,  \dd \big[ k f  +  (m-2) \omega \big] ( \, N \, - \,  k\,  \grad_{g_\PPP} f  \,)  \ - \ k  \, \DAlembert_{g_\PPP} f   \, \Big\}
\eeq
as a real function on $P$. For reference, the transformation rules of the components of $R_{g_\PPP}$ that appear in the action are written below after multiplication by the volume form: 
\bal \label{B.10}
R_{\tg_\KKK} \; \vol_{\tg_\PPP} \ &= \  e^{m\omega + (k - 2)f} \, R_{g_\KKK}  \; \vol_{g_\PPP} \nonumber \linebr
R_{\tg_\MMM} \; \vol_{\tg_\PPP} \ &= \  e^{(m-2)\omega + kf} \, \big[ \,  R_{g_\MMM} -  2(m-1)\, \DAlembert_{g_\MMM} \omega   -  (m-1)(m-2)\,   |\dd  \omega |^2_{g_\MMM}  \, \big]   \; \vol_{g_\PPP} \nonumber \linebr
| \FF |^2_{\tg_\PPP} \; \vol_{\tg_\PPP} \ &= \  e^{(m-4)\omega + (k+ 2)f} \, | \FF |^2_{g_\PPP}  \; \vol_{g_\PPP} \nonumber \linebr
| \mathring{\tilde{S}} |^2_{\tg_\PPP} \; \vol_{\tg_\PPP} \ &= \  e^{(m-2)\omega + kf} \, | \mathring{S} |^2_{g_\PPP}   \; \vol_{g_\PPP}  \nonumber \linebr
| \tilde{N} |^2_{\tg_\PPP} \; \vol_{\tg_\PPP} \ &= \ e^{(m- 2) \omega + k f}\, \big|   N \, - \,  k\,  \grad_{g_\PPP} f    \big|^2_{g_\PPP} \; \vol_{g_\PPP}   \ \ .
\end{align}
In the particular case where $e^f = e^\omega =: \Omega$, the higher-dimensional metric as a whole transforms as
\beq \label{SimpleRescaling}
\tg_P \ = \ \Omega^2 \, g_P \ .
\eeq 
One can then check that the transformation rules listed above, when applied to the different components of decomposition \eqref{DecompositionScalarCurvature} and \eqref{DivergenceN}, imply that the scalar curvature $R_{g_\PPP}$ as a whole transforms as 
\beq \label{B.11}
R_{\tg_\PPP} \ = \ \Omega^{-2} \, \big[ \,  R_{g_\PPP} \, - \, 2(n-1)\, \DAlembert_{g_\PPP} ( \log \Omega)  \, - \, (n-1)(n-2)\,   |\dd  \log \Omega |^2_{g_\PPP}  \, \big] \ ,
\eeq
where $n= m+k$ is the total dimension of $P$. This of course coincides with the usual transformation of the scalar curvature under a simple Weyl rescaling \cite{Wald}. When $f=\omega = \log \Omega$ the rules above also imply that 
\beq
| \tilde{N} |^2_{\tg_\PPP} \ = \ \Omega^{-2}  \left\{  \, |N |^2_{g_\PPP}  \, + \, k^2 \,  |\dd \log \Omega |^2_{g_\PPP}  \, - \, 2\, k \,  (\dd \log \Omega)(N) \, \right\} \ , \nonumber
\eeq
and that 
\[
\divergence_{\tg_\PPP} (\tilde{N})  \ = \  \Omega^{-2} \, \Big[ \, \divergence_{g_\PPP} (N) \, -\, k\, \DAlembert_{g_\PPP} (\log \Omega) \linebr  + \, (n-2) (\dd \log \Omega) (N) \, + \, k(2-n) \big|\dd \log \Omega \big|^2_{g_\PPP} \, \Big] 
\]
as real functions on $P$. Taking these formulae for $R_{\tg_\PPP}$, $| \tilde{N} |^2_{\tg_\PPP}$ and $\divergence_{\tg_\PPP} (\tilde{N})$, one can look for constants $\alpha_1$ and $\alpha_2$ such that
\beq
R_{\tg_\PPP} \ + \ \alpha_1\,  | \tilde{N} |^2_{\tg_\PPP} \ + \ \alpha_2 \, \divergence_{\tg_\PPP} (\tilde{N})  \ = \ \Omega^{-2} \, \Big[ \,   R_{g_\PPP} \ + \ \alpha_1\,  | N |^2_{g_\PPP} \ + \ \alpha_2 \, \divergence_{g_\PPP} (N)  \, \Big]
\eeq
for every rescaling function $\Omega$. This defines a system of linear equations for $\alpha_1$ and $\alpha_2$ that has a solution for
\beq
\alpha_1 \ = \ - \, \frac{(n-1)(n-2)}{k^2}   \qquad \qquad \quad   \alpha_2 \ = \ -\, 2\  \frac{n-1}{k} \  .
\eeq
Therefore the function on $P$
\beq
W_{g_\PPP} \ :=  \  R_{g_\PPP} \ - \ \frac{(n-1)(n-2)}{k^2} \; \big| N \big|^2_{g_\PPP} \, - \, 2\, \frac{(n-1)}{k} \ \divergence_{g_\PPP}  N 
\eeq
transforms simply as $W_{\tg_\PPP} \, = \, \Omega^{-2} \, W_{g_\PPP}$ under the rescaling \eqref{SimpleRescaling}.

\newpage

\subsection{Bosons' mass in the Einstein frame}
\label{AppendixBosonsMassEinsteinFrame}

The purpose of this short section is to register how the gauge bosons' mass formula \eqref{MassGaugeBosons} gets adjusted when we use the Einstein frame Lagrangian \eqref{ExpandedLagrangianScalarField} instead of the Jordan frame Lagrangian in \eqref{ActionP}. Replicating the calculation of section \ref{SectionMassGaugeBosons}, one works in the approximation where the gauge fields $A^a_\mu$ are small, close to their vanishing ``vacuum'' value, and where the internal metric is constant and equal to $g_K^\Szero$ as one moves across the fibres. After integrating over the fibre, the terms of Lagrangian \eqref{ExpandedLagrangianScalarField} that depend on $A^a_\mu$ are proportional to
\bal  \label{GaugeTerms2}
\frac{1}{4} \; g_M^{\mu \nu} \, g_M^{\sigma \rho}   \ (F^a_{A})_{\mu \sigma} \ (F^b_{A})_{\nu \rho}  \ B_{ab} \ + \ g_M^{\mu \nu} \ A^a_\mu \; D_{\nu a}     \ + \ 
 \ g_M^{\mu \nu} \ A^a_\mu \  A^b_\nu   \ C_{ab}  \ ,
\end{align}
but in the case of the Einstein frame the coefficients $B_{ab}$, $C_{ab}$ and $D_{\nu a}$ are now given by
\bal \label{MassCalculationCoefficients2}
B_{ab}  \ &:= \  e^{-\beta \phi_\SSzero}\, \int_K   \bg_K^\Szero (e_a ,  e_b  ) \ \vol_{\bg_\KKK^\SSzero}  \nonumber \linebr
D_{\nu a} \ &:= \   \frac{1}{2} \; \int_K  \left[\; \left\langle \Lie_{e_a}\, \bg_K^\Szero,  \ \Lie_{X_\nu}\, \bg_P^\Szero \right\rangle \ +\ 4 \, (1-\lambda)\; \Lie_{X_\nu}  (\divergence_{\bg_\KKK^\SSzero} e_a)  \; \right]  \, \vol_{\bg_\KKK^\SSzero}  \nonumber \linebr
C_{ab}  \ &:= \   \frac{1}{4} \; \int_K \left[ \; \left\langle \Lie_{e_a}\, \bg_K^\Szero,  \ \Lie_{e_b}\, \bg_K^\Szero \right\rangle  \ + \ 4\, (\lambda - 1)\, ( \divergence_{\bg_\KKK^\SSzero} e_a) \,   (\divergence_{\bg_\KKK^\SSzero} e_b) \; \right]  \, \vol_{\bg_\KKK^\SSzero}  \ ,
\end{align}
instead of \eqref{MassCalculationCoefficients}. Following the argument described in section \ref{SectionMassGaugeBosons}, choosing vector fields $e_a$ on $K$ that simultaneously diagonalize the quadratic forms $B_{ab}$ and $C_{ab}$, the four-dimensional equations of motion for the gauge fields are then
\beq \label{EquationMotionGaugeFields2}
g_M^{\mu \nu} \; (\nabla^M_\nu F^a_{A})_{\mu \sigma} \ - \ 2 \; \frac{C_{aa} }{B_{aa}}\; A^a_\sigma  \ = \ 0 \ .
\eeq
The coefficient $2\,C_{aa}/B_{aa}$ will be called the squared-mass of the gauge field. It is now given by
\begin{equation} \label{MassGaugeBosons3}
\left(\text{Mass} \, A_\mu^a \right)^2 \ = \ e^{\beta \phi_\SSzero} \ \frac{ \int_K \big[ \; \left\langle \Lie_{e_a}\, \bg_K^\Szero,  \; \Lie_{e_a}\, \bg_K^\Szero \right\rangle_{\bg_K^\SSzero}  \, + \, 4\, (\lambda -1)\, (\divergence_{\bg_\KKK^\SSzero} e_a)^2 \; \big] \vol_{\bg_\KKK^\SSzero} }{ 2 \int_K   \bg_K^\Szero (e_a ,  e_a  ) \, \vol_{\bg_\KKK^\SSzero} } \ ,
\eeq
instead of \eqref{MassGaugeBosons}. So the difference is the appearance of the factor $e^{\beta \phi_\SSzero}$ and the use of the normalized version $\bg_K^\Szero$ of the internal vacuum metric instead of the simple vacuum metric $g_K^\Szero$. The symbol $\left\langle \cdot \, , \, \cdot \right\rangle_{\bg_K^\SSzero}$ on the right-hand side means that the internal product of $2$-tensors defined in \eqref{DefinitionInnerProduct} should be taken with respect to the metric $\bg_K^\Szero$.

We would prefer to write the mass formula in terms of the proper vacuum metric $g_K^\Szero$, instead of the normalized metric $\bg_K^\Szero$ and the vacuum value of the rescaling field $\phi$. The defining relation for $\phi$ was, in \eqref{MetricPerturbation2} and \eqref{ActionConstants2},
\[
g_K \ =\  a_1 \, e^{- b_1 \phi} \, \bg_K \  = \ \bk^{ \frac{-2}{m+k-2}} \,  e^{- \frac{(m-2)  \beta \phi}{m+k-2} }\,   \bg_K \ .
\]
Taking the volume of the metrics on both sides of the equation and using \eqref{ActionConstants3} we get
\[
\big(\, \kappa_P^{-1} \kappa_M \Vol_{g_K}  \big)^{\frac{2}{k}} \ = \ \bk^{ \frac{-2}{m+k-2} +  \frac{2}{k} }  \  e^{- \frac{(m-2)  \beta \phi}{m+k-2} } \ ,
\]
and so
\beq \label{RelationPhiVolume}
e^{\beta \phi}  \ = \   \bk^{\frac{2}{k} } \  \big(\, \kappa_P^{-1} \kappa_M \Vol_{g_K}  \big)^{- \frac{2(m+k-2)}{k(m-2)}}  \ = \  \big(\, \kappa_P^{-1} \kappa_M \Vol_{\bg_K}  \big)^{\frac{2}{k} } \  \big(\, \kappa_P^{-1} \kappa_M \Vol_{g_K}  \big)^{- \frac{2(m+k-2)}{k(m-2)}}   \ .
\eeq
On the other hand the inner-product $\left\langle \Lie_{e_a}\, g_K,  \; \Lie_{e_a}\, g_K \right\rangle_{g_\KKK}$ and the divergence $(\divergence_{g_\KKK} e_a)$ are both invariant under constant rescalings of the internal metric $g_K$, so take the same value when calculated with respect to $g_K$ or to $\bg_K$. This implies that the quotient
\[
\frac{ \int_K \left[ \; \left\langle \Lie_{e_a}\, g_K,  \; \Lie_{e_a}\, g_K \right\rangle  \, + \, 4\, (\lambda -1)\, (\divergence_{g_\KKK} e_a)^2 \; \right] \vol_{g_\KKK} }{  2\, (\kappa_P^{-1} \kappa_M \Vol_{g_\KKK})^{-2/k} \, \int_K   g_K (e_a ,  e_a  ) \, \vol_{g_\KKK} }
\]
is also invariant under constant rescalings of $g_K$. Combining this fact with \eqref{RelationPhiVolume} one recognizes that the right-hand side of \eqref{MassGaugeBosons3} can be written as
\begin{equation} \label{MassGaugeBosons4}
\frac{1}{ (\kappa_P^{-1} \kappa_M \Vol_{g^\SSSzero_\KKK})^{\frac{2(m+k-2)}{k(m-2)}}} \ \frac{ \int_K \left[ \; \left\langle \Lie_{e_a}\, g_K^\Szero,  \; \Lie_{e_a}\, g_K^\Szero \right\rangle  \, + \, 4\, (\lambda -1) (\divergence_{g_\KKK^\SSzero} e_a)^2 \; \right] \vol_{g_\KKK^\SSzero} }{ 2\, (\kappa_P^{-1} \kappa_M \Vol_{g^\SSSzero_\KKK})^{-2/k} \, \int_K   g_K^\Szero (e_a ,  e_a  ) \, \vol_{g_\KKK^\SSzero} } \ . \nonumber
\eeq
Here the rightmost fraction is invariant under constant rescalings of $g_K^\Szero$. Simplifying the denominators we finally get
\begin{equation} \label{MassGaugeBosons5}
{\setlength{\fboxsep}{3\fboxsep} \boxed{
\left(\text{Mass} \, A_\mu^a \right)^2 \ = \ \frac{ \int_K \big[  \, \left\langle \, \Lie_{e_a}\, g_K^\Szero,  \, \Lie_{e_a}\, g_K^\Szero \, \right\rangle_{g_\KKK^\SSzero} \, + \,  4 (\lambda -1) (\divergence_{g_\KKK^\SSzero} e_a)^2 \,  \big] \, \vol_{g_\KKK^\SSzero} }{ 2\ \big(\, \kappa_P^{-1} \kappa_M \Vol_{g^\SSSzero_\KKK}\, \big)^{2/(m-2)} \, \int_K  \, g_K^\Szero (e_a ,  e_a  ) \, \vol_{g_\KKK^\SSzero} } \, . 
}}
\eeq
This formula describes how the classical mass of the gauge bosons depends on the vacuum geometry of the internal space, determined by the metric $g_K^\Szero$. In the physical case $m=4$, it indicates that all squared-masses should scale as $(\Vol_{g^\SSSzero_\KKK})^{-1 - \frac{2}{k}}$ when the internal space changes in size. Importantly, when the calculation is performed in the Einstein frame, the scale of the bosons' masses also depends on the free parameter $\kappa_P$ appearing in the higher-dimensional action \eqref{ActionP}. So that scale can be very different from $\kappa_M^{-1}$. In other words, the Kaluza-Klein framework does not force masses to be in the Planck scale.

As a word of caution, however, note that the notion of mass of a Klein-Gordon field is not particularly clear outside of the flat Minkowski case. The physical interpretation of the coefficient $2 \, C_{aa} / B_{aa}$ that appears in the equation of motion \eqref{EquationMotionGaugeFields2}, written in the vacuum approximation, should require additional care on curved backgrounds.

\newpage

\subsection{Quadratic forms on \texorpdfstring{$\su$}{su3}}
\label{CalculationsSu(3)}

Let $g$ be a left-invariant metric on $K= \SU$ and let $u$ and $v$ be vectors in the Lie algebra $\su$. Identifying these vectors with left-invariant vector fields $u^\LL$ and $v^\LL$ on the group manifold, the product $g (u^\LL, v^\LL)$ is a constant function on $K$. So the Lie derivative of the metric becomes 
\[
(\Lie_{w^\LLL} g)(u^\LL, v^\LL) \ = \ -\, g([w, u], \, v) \, - \, g(u, \, [w, v])
\] 
for any vector $w$ in $\su$, as in \eqref{LieDerivativeMetric1}. Choosing a $g$-orthonormal basis $\{ e_a\}$ of the Lie algebra, the inner-product of Lie derivatives can be expressed as
\bal \label{LieDerivativeProduct}
\big\langle \, \Lie_{u^\LLL}\, g \, ,  \; \Lie_{u^\LLL}\, g \,  \big\rangle \ &= \   \sum_{a,\, b} \, (\Lie_{u^\LLL} g)(e_a, e_b) \  (\Lie_{u^\LLL} g)(e_a, e_b) \nonumber \linebr
&= \  2\,  \sum_{a, \, b} \,  g([u, e_a], \, e_b)^2  \, +\,   g([u, e_a], \, e_b)\, g([u, e_b], \, e_a)   \nonumber \linebr
&= \ 2\, \sum_{a} \,  g( [u, e_a], \,  [u, e_a]) \, + \, g( [u, [u, e_a]], \,  e_a) \  \nonumber \linebr
&= \ 2\, A_{g} (u,u) \, + \, 2 \, B(u,u) \ .
\end{align}
In the last equality we have denoted by $A$ the quadratic form on the Lie algebra
\beq
A_{g} (u,v) \ := \ \sum\nolimits_a  \,  g( [u, e_a], \,  [v, e_a]) 
\eeq
and by $B(u,v)$ the Killing form $\Tr (\ad_u \circ \ad_v)$. 

Now let $g$ be the special metric $\tbeta$ defined in \eqref{DefinitionTBeta}. The commutation rules \eqref{CommutationRulesSU3} and the $\tbeta$-orthogonality of decomposition \eqref{AlgebraDecomposition2} imply that 
\[
A_{\tbeta} \big(\, \utwo, \, \CC^2\, \big) \ = \ \{  0 \} \ .
\]
The $\Ad_{\Utwo}$-invariance of the metric $\tbeta$ implies that $A_{\tbeta}$ is also $\Ad_{\Utwo}$-invariant. Since any element $u \in \sutwo$ can be taken to its opposite by an $\Ad_{\Utwo}$-transformation and the element $u_0 \in \mathfrak{u}(1)$ is invariant under the same transformation, we also have that
\[
A_{\tbeta} (u_0, u) \ = \ A_{\tbeta} (u_0, -u) \ = \ 0 \ .
\]
Hence the decomposition $\su =   \mathfrak{u}(1) \oplus \sutwo \oplus \CC^2$ is fully orthogonal with respect to $A_{\tbeta}$. Since the $\Ad_{\Utwo}$-transformations are transitive on the maximal spheres inside each component of the decomposition, the $\Ad_{\Utwo}$-invariance of both $A_{\tbeta}$ and $\tbeta$ implies that these quadratic forms are proportional to each other inside each component, i.e.
\[
A_{\tbeta} \ = \ \mu_1\, \tbeta\, |_{\mathfrak{u}(1)} \ + \  \mu_2\, \tbeta\, |_{\sutwo} \ + \ \mu_3\, \tbeta\, |_{\CC^2}  \ 
\] 
for positive constants $\mu_k$. The value of these constants can be deduced from explicit calculations using the $\tbeta$-orthonormal basis $\{  u_0, \ldots, u_3, w_1, \ldots, w_4 \}$ of $\su$. For example, one calculates that
\bal \label{ExplicitSums}
\sum_{j,k =1}^4 \, \tbeta \big(\, [w_j, w_k], \, [w_j, w_k] \, \big) \ &= \ 6\, (\lambda_1 + \lambda_2) \, \lambda_3^{-2}     &\sum_{j=1}^3 \, \tbeta \big(\, [u_0, u_j], \, [u_0, u_j] \, \big) \ &= \  0 \nonumber \linebr
\sum_{j=1}^4 \, \tbeta \big(\, [u_0, w_j], \, [u_0, w_j] \, \big) \ &= \  6\, \lambda_1^{-1}   &\sum_{j,k=1}^3 \, \tbeta \big(\, [u_j, u_k], \, [u_j, u_k] \, \big) \ &= \  12\, \lambda_2^{-1} \nonumber \linebr
\sum_{j=1}^4 \, \tbeta \big(\, [u_k, w_j], \, [u_k, w_j] \, \big) \ &= \  2\, \lambda_2^{-1} \ \ \    {\rm for} \  k=1,2,3 .&  
\end{align}
Taking these results and their sums, one can extract the values of the $\mu_k$ to conclude that
\beq
A_{\tbeta} \ = \  \frac{6}{\lambda_1}\,\, \tbeta \, |_{\mathfrak{u}(1)} \ + \  \frac{6}{\lambda_2}\,\, \tbeta \, |_{\sutwo} \ + \ \frac{3}{2} \, \Big(\, \frac{\lambda_1 + \lambda_2}{\lambda_3^2} \, + \, \frac{1}{\lambda_1} \, + \, \frac{1}{\lambda_2} \, \Big) \, \tbeta \,  \, |_{\CC^2} \ .
\eeq
Substituting this result for $A_{\tbeta}$ into expression \eqref{LieDerivativeProduct} and using the relation between the Killing form and the product $\tbeta$, as in \eqref{BasicBiinvariantMetric} and \eqref{DefinitionTBeta}, we obtain
\beq \label{ProductLieDerivatives2}
\big\langle \, \Lie_{u}\, \tbeta,  \; \Lie_{u}\, \tbeta \,  \big\rangle \ = \ 3\, \Big(\, \frac{1}{\lambda_1} +  \frac{1}{\lambda_2} +  \frac{\lambda_1 + \lambda_2}{\lambda_3^2} -  \frac{4}{\lambda_3} \, \Big)  \ \tbeta (u'', u'') \ , 
\eeq
where $u''$ denotes the component of $u$ in the subspace $\CC^2$ of $\su$. Now consider the second quadratic form on the Lie algebra
\beq
C_{\tbeta} (u,v) \ := \ \sum\nolimits_{a,b}  \,  \tbeta( [e_a, e_b] , \, u) \ \tbeta( [e_a, e_b] , \, v)  \ .
\eeq
A similar reasoning, using the $\Ad_{\Utwo}$-invariance of this form and the explicit results \eqref{ExplicitSums}, allows one to calculate that 
\beq
C_{\tbeta} \ = \  \frac{6 \, \lambda_1}{\lambda_3^2}\,\, \tbeta \, |_{\mathfrak{u}(1)} \ + \  \frac{2}{\lambda_2}\, \Big( \,   2 \,+\,  \frac{\lambda_2^2}{\lambda_3^2}    \, \Big) \, \tbeta \, |_{\sutwo} \ + \  \Big(\, \frac{3}{\lambda_1} \, + \, \frac{3}{\lambda_2} \, \Big) \, \tbeta \,  \, |_{\CC^2} \ .
\eeq
A well-known formula for the Ricci tensor of left-invariant metrics on unimodular Lie groups \cite[ch. 7]{Besse} then says that 
\[
Ric_{g} \ = \    - \, \frac{1}{2} \, A_{g} \ - \ \frac{1}{2} \, B \ + \ \frac{1}{4} \,C_{g} \ , 
\]
where $B$ stands for the Killing form on the Lie algebra. Substituting the results for $A_{\tbeta}$ and $C_{\tbeta}$ into this expression and using the relation between the Killing form and the products $\beta_0$ and $\tbeta$, as in \eqref{BasicBiinvariantMetric} and \eqref{DefinitionTBeta}, one can write
\bal
Ric_{\, \tbeta} \ &= \   \frac{3\, \lambda_1}{2\, \lambda_3^2} \ \tbeta \,  |_{\mathfrak{u}(1)}   \ + \    \left( \frac{1}{\lambda_2}   +  \frac{\lambda_2}{2\, \lambda_3^2}  \right)   \tbeta \, |_{\sutwo}    \ +  \  \frac{3}{4}\, \left( \frac{4}{\lambda_3} -  \frac{\lambda_1 + \lambda_2}{\lambda_3^2}  \right)  \tbeta \, |_{\CC^2} \ \nonumber    \ .
\end{align}


\renewcommand{\baselinestretch}{1.2}

\vspace{1cm}

\end{document}